%% file: ms.tex
\documentclass[useAMS,usegraphicx,usenatbib]{mn2e}

\usepackage{longtable}
\usepackage{subfigure}
\usepackage{ulem}

\title[Radio sources with ultrahigh polarization]{Radio sources with ultrahigh polarization}

\author[H.~Shi et al.]{H.~Shi,$^1$
  H.~Liang,$^2$\thanks{E-mail:Haida.Liang@ntu.ac.uk}
    J. L.~Han$^1$ and R. W.~Hunstead$^3$\\
$^1$National Astronomical Observatories, Chinese Academy of Science,
  Jia20 Datun Road, Chaoyang District, Beijing 100012,
  China\\ 
$^2$School of Science and Technology, Nottingham Trent
  University, Nottingham NG11 8NS, UK\\ 
$^3$Sydney Institute for Astronomy, School of Physics, University 
of Sydney, NSW 2006, Australia}
\begin{document}

\date{}

\pagerange{\pageref{firstpage}--\pageref{lastpage}} \pubyear{2009}

\maketitle

\label{firstpage}

\begin{abstract}

A sample of 129 unresolved radio sources with ultrahigh linear
polarization ($>30$ per cent) has been selected from the NRAO VLA Sky
Survey. Such high average linear polarization is unusual in
extragalactic sources. Higher resolution Australia Telescope Compact
Array and Very Large Array observations confirm the high average
polarization but find that most of these sources are extended. The
Sloan Digital Sky Survey spectroscopy, where available, shows that the
optical counterparts are elliptical galaxies with no detectable
emission lines. The optical spectra, radio luminosity, linear size and
spectral index of these sources are typical of radio-loud active
galactic nuclei. Galaxy counts within a 1\,Mpc radius of the radio
sources show that these highly polarized sources are in environments
similar to their low polarization ($<2$ per cent)
counterparts. Similarly, the line-of-sight environments of the
ultrahigh polarization sources are on average indistinguishable from
those of the low-polarization sources. We conclude that the
extraordinarily high average polarization must be due to intrinsic
properties of the sources, such as an extremely ordered source
magnetic field, low internal thermal plasma density or a preferential
orientation of the source magnetic field perpendicular to the line of
sight.
\end{abstract}


\section{Introduction}
Polarized radio emission is an important astrophysical diagnostic of
the physical conditions in radio sources. Synchrotron radiation is the
mechanism responsible for the radio emission from extragalactic
sources. In a uniform magnetic field, the fractional linear
polarization of synchrotron radiation is given by
\begin{equation}
\Pi=\frac{P}{I}=\frac{3-3\alpha}{5-3\alpha}
\end{equation}
where $P=\sqrt{Q^{2}+U^{2}}$ is the linear polarization flux density,
$Q$ and $U$ are the Stokes flux densities corresponding to the two
orthogonal components of linear polarization, $I$ is the Stokes
integrated flux density and $\alpha$ is the spectral index defined by
$S\propto \nu^{\alpha}$. Hence, synchrotron radiation is intrinsically
linearly polarized at $\sim$60 to 80 per cent for spectral indices
$\alpha$ in the range of $0$ to $-1.5$.  However, owing to
depolarization effects, most extragalactic radio sources are observed
to have only a few per cent net linear polarization across the entire
source and rarely have percentage linear polarization $>25$ per cent
even on the smallest angular scales (Saikia \& Salter 1988; Kronberg
1994). The frequency-independent depolarization effects are mainly due
to the tangling of the magnetic field on scales smaller than the size
of the observing beam. The frequency-dependent depolarization effects
are mainly caused by Faraday rotation. Faraday rotation, either within
the synchrotron emitting region or in any magnetoionic media along the
line of sight, can cause depolarization (e.g. Burn 1966; Ruzmaikin,
Sokoloff \& Shukurov 1988; Sokoloff et al. 1998; Fletcher et
al. 2004). Faraday depth is defined as
\begin{equation}
\phi=0.81 \int n_{e}{\bf B_{||}} dl \ {\rm rad\,m}^{-2},
\end{equation} 
where $n_e$ is the electron density in cm$^{-3}$, ${\bf B_{||}}$ is
the line-of-sight component of the magnetic field strength in $\mu$G
and $l$ is the path length along the line of sight (Burn 1966). The
rotation measure is the observed Faraday depth, defined as the slope
of the polarization position angle $\chi$ versus $\lambda^{2}$:
\begin{equation}
\chi(\lambda^{2})=\chi_{0}+ \phi \lambda^{2},
\end{equation}
where 
\begin{equation}
\chi=\frac{1}{2}\tan^{-1}\frac{U}{Q},
\end{equation}
and $\chi_{0}$ is the intrinsic polarization position angle. Internal
Faraday rotation occurs when the radio-emitting plasma is mixed with
thermal plasma. Internal Faraday rotation always causes
depolarization, even if the magnetic field is ordered and the thermal
plasma is uniform. A foreground thermal plasma that has an ordered
magnetic field and uniform electron density can cause Faraday rotation
but not depolarization. In the absence of internal Faraday rotation
and beam depolarization, the Faraday depth towards a source is equal
to its rotation measure.

Depolarization in general increases at lower resolution, wider
observing bandwidth and longer wavelength. An understanding of the
polarization/depolarization properties of radio sources has important
implications for the study of the evolution of radio galaxy and quasar
environments (Goodlet \& Kaiser 2005; Bernet et al. 2008). Very little
is known about the origin and growth of magnetic fields in galaxies
and clusters of galaxies. Recently, magnetic fields have been detected
for the first time in $z>0.5$ galaxies (Bernet et al. 2008; Wolfe et
al. 2008).

\begin{figure*}
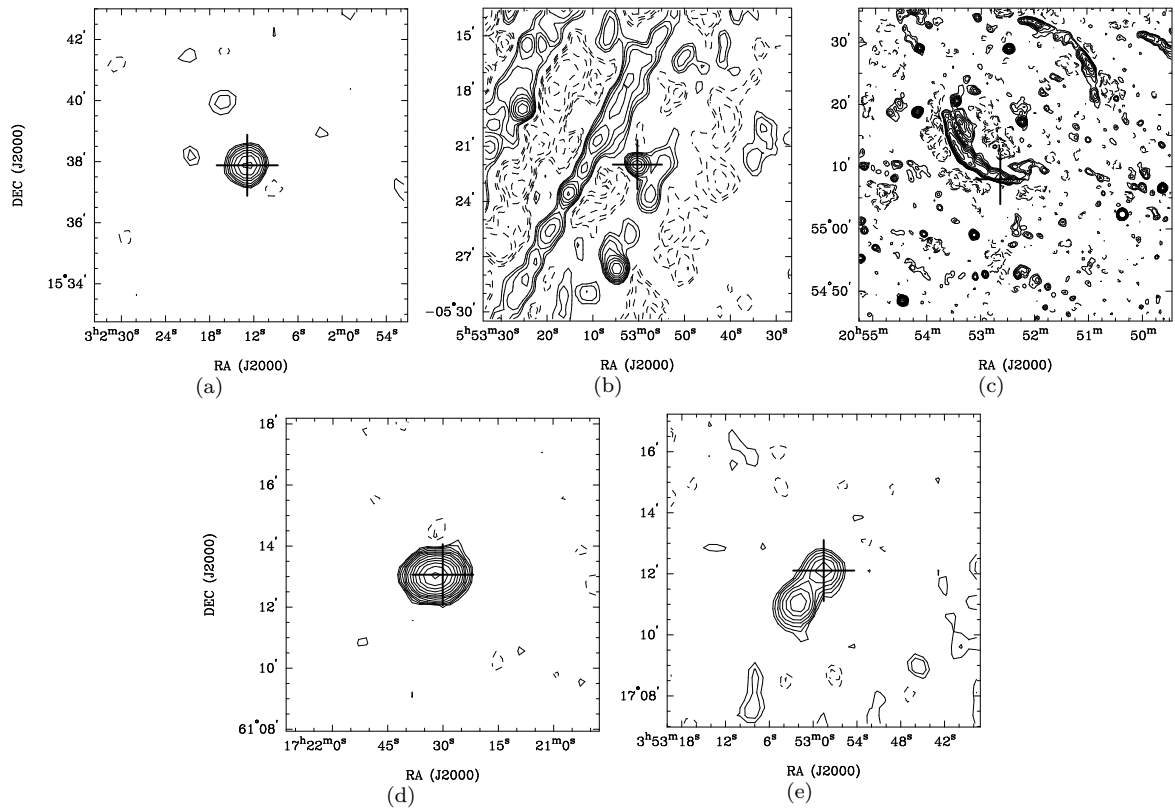

    \centering
    \subfigure[]{\includegraphics[height=5.3cm,angle=-90.0]{J030212+153752.ps}
                \label{fig:good}}
    \subfigure[]{\includegraphics[bb=44 145 583 688,height=4.9cm,angle=-90.0,clip]{bad.ps}
                \label{fig:bad}}
    \subfigure[]{\includegraphics[bb=44 145 583 688,height=4.9cm,angle=-90.0,clip]{diffuse.ps}
                \label{fig:diffuse}}
    \subfigure[]{\includegraphics[height=5.3cm,angle=-90.0]{offset.ps}
                \label{fig:offset}}
    \subfigure[]{\includegraphics[bb=44 145 583 688,height=4.9cm,angle=-90.0,clip]{double.ps}
                \label{fig:double}}
    \caption{(a) Example of a source included in the final list of
      ultrahigh polarization sources: J030212+153752. Examples of
      different types of sources eliminated after visual inspection
      are: (b) a field with strong side-lobes as a result of a nearby
      strong source; (c) a source that is part of a diffuse emission,
      such as a supernova remnant; (d) a source fitted to the
      residuals of a Gaussian fit to a nearby peak; and (e) a source
      with a neighbouring source within 90\,arcsec. The contours of
      the sources are $\pm 2^{n/2}$\,mJy~beam$^{-1}$, where $n=0, 1,
      2, 3,$ \ldots\ .  The NVSS catalogue position of the selected
      sources are marked by `+'. }
    \label{fig:example}
\end{figure*}

A first indication of the existence of ultrahigh average polarization
radio sources came from a study of the radio properties of the Bullet
cluster (1E0657$-$56; Liang et al. 2000). Liang et al. (2001),
serendipitously, found a peculiar, extended
(15\,arcsec\,$\times$\,4.5\,arcsec), ultra-steep spectrum ($\alpha\sim
-1.5$) radio source J0658.7$-$5559. In projection, it lies 2\,arcmin
from the cluster centre and was found with an average percentage
linear polarization of $\sim 55$ per cent at 8.8\,GHz. Such a high
average linear polarization is extremely unusual for radio
sources. While the nature of this peculiar source is still an open
question, it is almost certainly extragalactic (Liang et
al. 2001). The only radio sources known to have such high fractional
polarization are pulsars, which are point sources, unlike
J0658.7$-$5559, which is extended. Pulsar astronomers have used this
unique property to search for pulsars (e.g. Crawford et al. 2000; Han
et al. 2004). The polarization properties of radio sources were
summarised by Han \& Tian (1999), where the distribution of the
fractional linear polarization of known extragalactic sources, such as
radio galaxies, quasars and BL Lac objects, were compared with those
of known pulsars identified in the NRAO VLA Sky Survey (NVSS). The
average linear polarization of extragalactic radio sources rarely
exceeds 25 per cent, whereas $\sim50$ per cent of pulsars have linear
polarization $>25$ per cent (fig. 2 in Han \& Tian 1999).

The NVSS (Condon et al. 1998) mapped 82 per cent of the celestial
sphere at 1.4~GHz and detected $\sim 1.8$ million radio sources with
flux densities greater than 2.5\,mJy. It is the first large survey
with full polarization information. A sample of unresolved radio
sources was originally selected from the NVSS with percentage linear
polarization greater than 30 per cent to search for new
pulsars. However, follow-up observations of these sources with Parkes
and Jodrell Bank telescopes failed to find any new pulsars (Han et
al. 2004). Subsequent observations found that most of the sources were
extended and extragalactic.

This paper reports the existence of a subset of extragalactic radio
sources with extremely high average linear polarization and explores
the nature of these radio sources. Details of sample selection are
given in Section 2. Section 3 gives the high resolution polarization
measurements with the Very Large Array (VLA) and Australia Telescope
Compact Array (ATCA) to confirm the high linear polarization results
from the NVSS. Optical identification, radio luminosity, linear size
and spectral index measurements are described in Sections 4, 5 and
6. Section 7 discusses the nature of these highly polarized radio
sources. The conclusions are given in Section 8.

Cosmological parameters of $\Omega_{m}=0.3$, $\Omega_{\lambda}=0.7$
and $H_{0}=70$\,km\,s$^{-1}$\,Mpc$^{-1}$ are assumed throughout this
paper. Unless otherwise stated, all errors quoted are 1$\sigma$.

\begin{figure*}
  \centering
  \includegraphics[angle=-90.0, width=80mm]{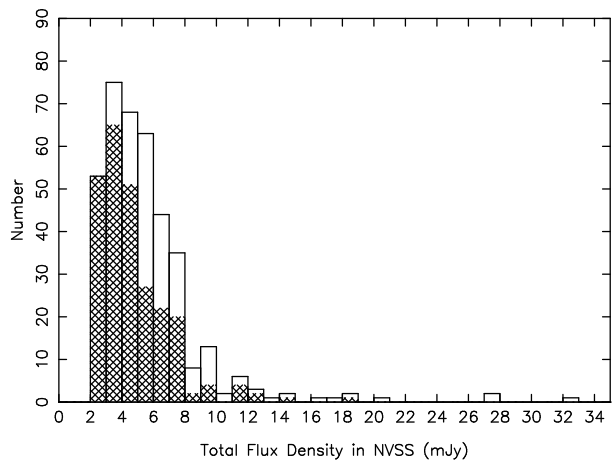}
\caption{Distribution of flux densities for the 381 sources selected
  by the initial four selection criteria. The cross-hatched histogram
  depicts the 252 sources excluded after detailed visual
  inspection. The fraction of sources excluded is greatest at the
  lowest flux densities. }
  \label{fig:381vs252}
\end{figure*}

\begin{figure*}
  \centering
  \includegraphics[angle=-90.0, width=100mm]{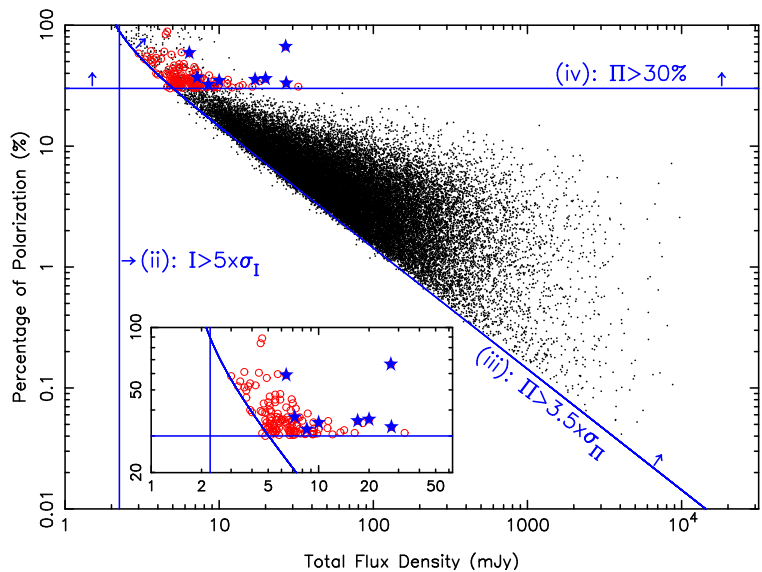}
\caption{Percentage polarization versus the total 1.4-GHz flux density
  of the sources that satisfy the first three of the selection
  criteria (38454 sources marked as black dots). The final sample of
  129 isolated, unresolved and highly polarized sources with
  $\Pi\ge30$ per cent are marked by the red circles, except where the
  sources are identified with known pulsars (blue stars). The blue
  curves define the selection criteria (ii)--(iv) (Section 2) by
  assuming typical values for rms noise in the Stokes $I$, $Q$ and $U$
  images ($\sigma_I=0.45$, $\sigma_Q=0.29$ and $\sigma_U=0.29$
  mJy\,beam$^{-1}$). The inset gives an enlarged view of the relative
  positions of the selected sources in the graph.}
  \label{fig:pp-I}
\end{figure*}

\section{Sample selection}\label{sec:samp}
An all-sky sample of ultrahigh polarization radio sources is selected
from the NVSS through the following criteria:
   \begin{enumerate}
   \item ~~Unresolved sources;
   \item ~Total flux density $I\ge5\sigma_I$;
   \item Percentage linear polarization $\Pi\ge3.5\sigma_{\Pi}$;
   \item Percentage linear polarization $\Pi\ge30$ per cent.
   \end{enumerate}

The NVSS has a full width at half-maximum (FWHM) beam size of
45\,arcsec. Elliptical Gaussians were fitted to sources to produce the
NVSS catalogue.  In cases where the source is unresolved, the
catalogue lists the 98 per cent confidence upper limit to the source
size.  We select only those unresolved NVSS sources (marked by `$<$'
for both the deconvolved major and minor axes in the NVSS catalogue)
that do not have a large residual after the Gaussian fit (i.e. those
not marked by `P*' and `S*' in the NVSS catalogue). For the NVSS, the
average rms noise in a Stokes $I$ image is $\sigma_{I}\sim
0.45$\,mJy\,beam$^{-1}$ and the average rms noise in a Stokes $Q$ or
$U$ image is $\sigma_{Q}=\sigma_{U}\sim 0.29$\,mJy\,beam$^{-1}$ which
implies an average rms noise in a linear polarization $P$ image of
$\sigma_{P}\sim 0.41$\,mJy\,beam$^{-1}$. Since a simple error
propagation gives the rms fractional linear polarization error as
\begin{equation}
     \sigma_{\Pi} = \sqrt{ \left(\frac{\sigma_I}{I}\right)^2 +
                \left(\frac{\sigma_P}{P}\right)^2}~\Pi,
\end{equation}
criterion (iii) implies that $P\ge 3.5\sigma_{P}$ is always true.

Similarly, criterion (iii) defines an implicit relationship between
the fractional linear polarization $\Pi$ and the Stokes $I$ flux
density of a source through
\begin{equation}
     I \ge 3.5\times\sqrt{ \sigma_{I}^{2} + \frac{\sigma_P^{2}}{\Pi^{2}}}
\end{equation}
For the average rms values of $\sigma_{I}$ and $\sigma_{P}$ given
above, this implies that on average if $I\ge 5$\,mJy, then criterion
(iii) ensures that criterion (iv) is always true. In retrospect,
criterion (ii) is very weak as its removal does not change the final
source list. However, criteria (i), (iii) and (iv) together do not
necessarily guarantee that criterion (ii) is always satisfied.

A total of 38454 sources in the NVSS satisfied the first three
conditions. Only 381 sources satisfied all four criteria. However, a
detailed visual inspection of the actual NVSS images in a $1\degr
\times 1\degr$ field found that not all of the 381 sources were
reliable. The following types of sources were eliminated: (1) 190
sources in fields with significant CLEAN artifacts which can include
regions with strong sources or extended diffuse emission
(e.g. Fig.~\ref{fig:bad} and Fig.~\ref{fig:diffuse}); (2) 18 sources
found in the residuals of a Gaussian fit to a nearby peak (e.g.
Fig.~\ref{fig:offset}); NVSS fits elliptical Gaussians to peaks in a
field and if the residual in the area covered by a single Gaussian fit
is too high, then multiple Gaussians were fitted simultaneously; (3)
42 sources that have one or more neighbouring sources within a radius
of 90 arcsec (e.g. Fig.~\ref{fig:double}; given the NVSS source
density, the probability of finding two unrelated sources within 90
arcsec is less than 10 per cent, Condon et al. 1998); and (4) two
sources on the edge of the survey field near Dec. $-40\degr$.

\begin{figure*}
  \centering
  \includegraphics[angle=-90.0, width=80mm]{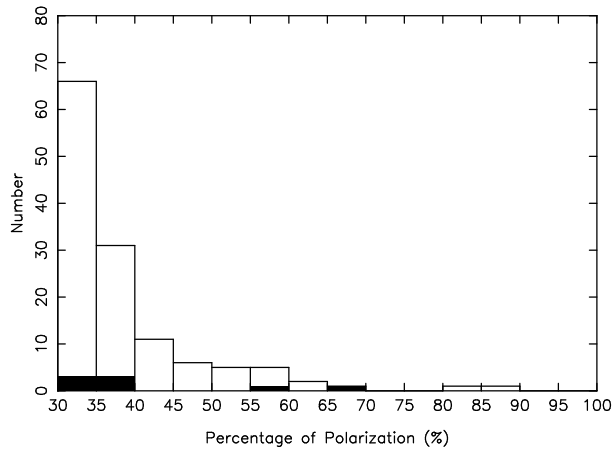}
\caption{Distribution of percentage polarization for the final sample
  of 129 sources, where the eight known pulsars are marked in solid
  black.}
  \label{fig:pp}
\end{figure*}

The final sample of 129 isolated, unresolved, highly polarized NVSS
sources that satisfy all the selection criteria is listed in
Table~\ref{tab:129nvss}. An example of such a source is shown in
Fig.~\ref{fig:good}. There are eight known pulsars in this sample. The
flux density distribution of the 381 sources selected according to the
initial four selection criteria is shown in Figure~\ref{fig:381vs252}
along with the distribution of the 252 sources excluded after the
above visual inspection. As expected, the fraction of sources excluded
after the detailed visual inspection decreases as the flux density
increases.  Figure~\ref{fig:pp-I} plots the percentage polarization
versus the total flux density for all the sources that satisfy the
first three criteria.  The selection criteria (ii) -- (iv) plotted are
based on typical values for the rms noise in Stokes $I$, $Q$ and $U$
images as given above.  The final list of sources are marked by red
circles, with the known pulsars marked by blue stars. The distribution
of fractional polarization of the sources is shown in
Fig.~\ref{fig:pp} where the majority have percentage polarization
between 30 and 40 per cent. There appears to be an increase in the
number of sources with high linear polarization towards lower flux
densities. This may be an effect of there being more sources at low
flux densities. Without detailed statistical tests, it is difficult to
conclude whether the low flux density sources are preferentially more
polarized as have been previously reported (Taylor et al. 2007).

\begin{figure*}
  \centering
  \includegraphics[angle=-90.0, width=90mm]{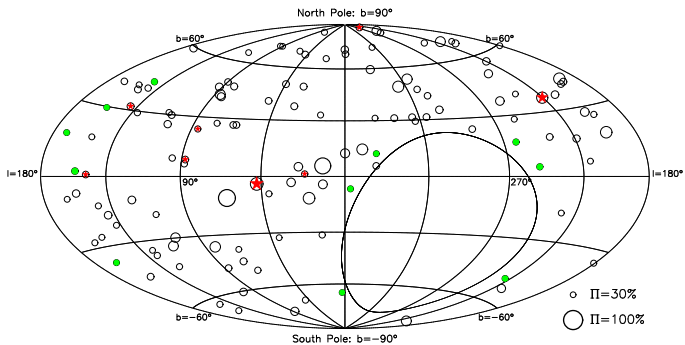}
\caption{Distribution on the sky in Galactic coordinates of the final
  sample of 129 isolated, unresolved and highly polarized sources. The
  empty southern region below Dec. of $-40\degr$ corresponds to the
  area inaccessible to the VLA. The sizes of the circles correspond to
  their fractional linear polarization. The red stars indicate known
  pulsars in the sample and the green filled circles correspond to
  sources with high-resolution polarization observations. }
  \label{fig:sky}
\end{figure*}

The distribution of the final sample on the sky in Galactic
coordinates is given in Fig.~\ref{fig:sky}, and shows no correlation
with the Galactic latitude.

\section{High resolution polarization measurements}\label{sec:highres}

In order to confirm the high polarization of these sources, higher
resolution radio observations of selected sources were obtained with
the ATCA and the VLA at 1.4\,GHz.
Tables~\ref{tab:highres}~\&~\ref{tab:obs_nvss} give the results of
these radio observations. Sources from the final sample in the
Dec. range $-40\degr<\delta<-30\degr$ and RA range from 16$^h$ to
10$^h$ were selected for observation with the ATCA. In addition, five
out of the 10 sources in the final sample with percentage linear
polarization uncertainty less than 5 per cent (excluding pulsars) were
observed with the VLA.

\begin{figure*}
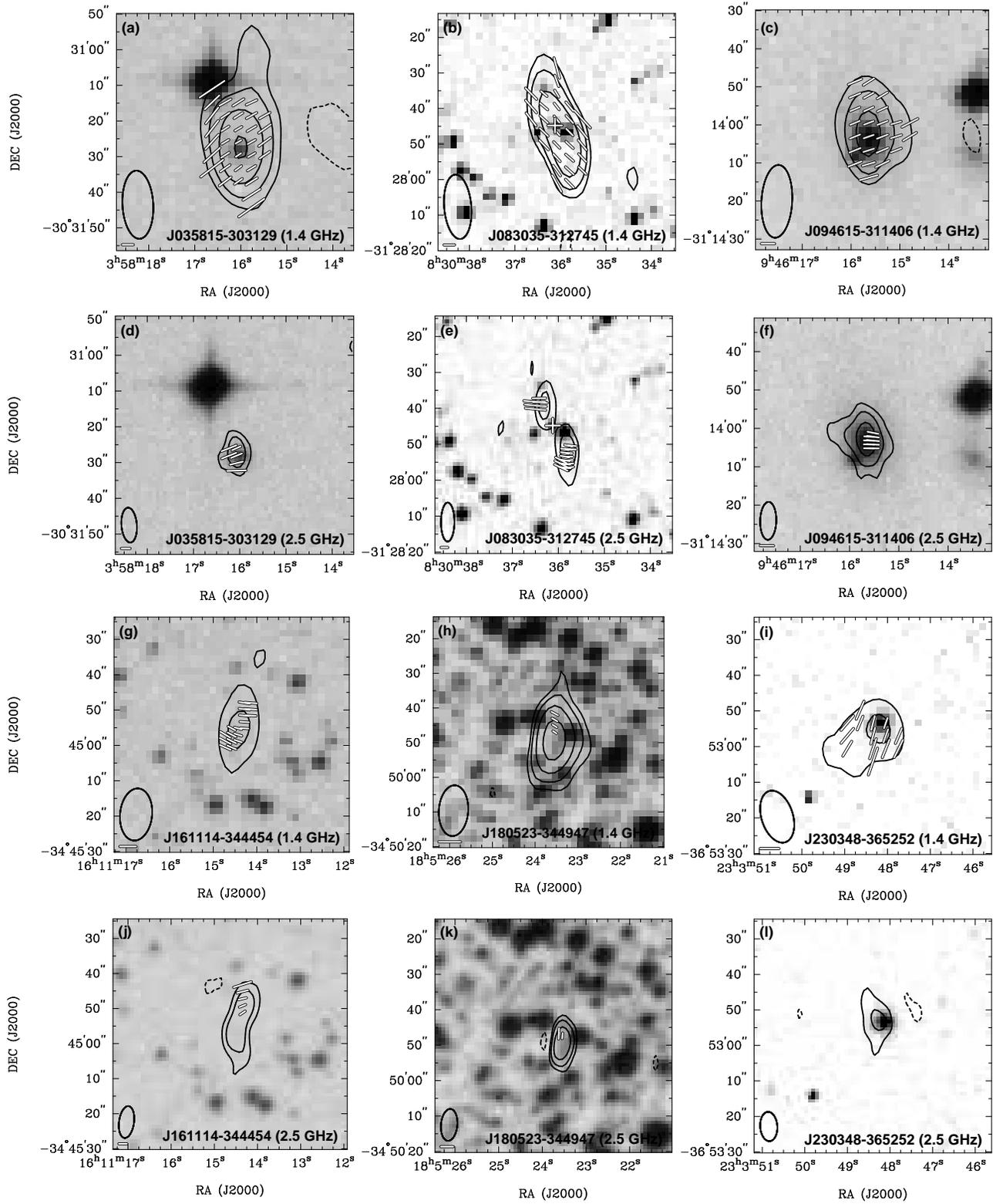

  \centering
  \includegraphics[bb=36 28 576 653,height=6.0cm,angle=-90.0,clip]{J0358-3031_pol_opt_1384.ps}
    \label{fig:J0358a} 
  \includegraphics[bb=36 86 571 648,height=5.4cm,angle=-90.0,clip]{J0830-3127_pol_opt_1384.ps}
    \label{fig:J0830a}
  \includegraphics[bb=36 86 576 648,height=5.3cm,angle=-90.0,clip]{J0946-3114_pol_opt_1384.ps}
    \label{fig:J0946a} \\
  \includegraphics[bb=36 26 571 648,height=6.0cm,angle=-90.0,clip]{J0358-3031_pol_opt_2496.ps}
    \label{fig:J0358b}
  \includegraphics[bb=36 86 571 648,height=5.4cm,angle=-90.0,clip]{J0830-3127_pol_opt_2496.ps}
    \label{fig:J0830b}
  \includegraphics[bb=36 86 567 648,height=5.4cm,angle=-90.0,clip]{J0946-3114_pol_opt_2496.ps}
    \label{fig:J0946b} \\
  \includegraphics[bb=36 26 570 656,height=6.0cm,angle=-90.0,clip]{J1611-3444_pol_opt_1384.ps}
    \label{fig:J1611a}
  \includegraphics[bb=36 86 571 664,height=5.4cm,angle=-90.0,clip]{J1805-3449_pol_opt_1384.ps}
    \label{fig:J1805a}
  \includegraphics[bb=36 86 571 648,height=5.4cm,angle=-90.0,clip]{J2303-3652_pol_opt_1384.ps}
    \label{fig:J2303a} \\
  \includegraphics[bb=36 27 568 660,height=6.0cm,angle=-90.0,clip]{J1611-3444_pol_opt_2496.ps}
    \label{fig:J1611b}
  \includegraphics[bb=36 86 567 649,height=5.4cm,angle=-90.0,clip]{J1805-3449_pol_opt_2496.ps}
    \label{fig:J1805b}
  \includegraphics[bb=36 86 568 648,height=5.4cm,angle=-90.0,clip]{J2303-3652_pol_opt_2496.ps}
    \label{fig:J2303b} \\
\caption{ATCA radio contours and polarization E-vectors at 1.4\,GHz
  and 2.5\,GHz overlaid on SuperCOSMOS {\it R}-band images for all
  sources except for J083035$-$312745 where the {\it B}-band image was
  used. The optical identification of J083035$-$312745 is marked with
  the white symbol '+'. The contours are $\pm 3\sigma \times 2^{n}$
  where $n=0, 1, 2, 3,$ \ldots\ and $\sigma= 0.15, 0.19, 0.35, 0.16,
  0.17, 0.17, 0.46, 0.22, 0.38, 0.22, 0.19, 0.27$\,mJy~beam$^{-1}$ for
  panels (a), (b), (c), (d), (e), (f), (g), (h), (i), (j), (k) and
  (l), respectively. The beam size is shown at the lower left-hand
  corner of the images and 30 per cent linear polarization is
  indicated by the length of line segment below the beam. Those
  without vector overlays are sources with no detectable polarization
  at that radio frequency.}
  \label{fig:polmap2}
\end{figure*}

\begin{figure*}
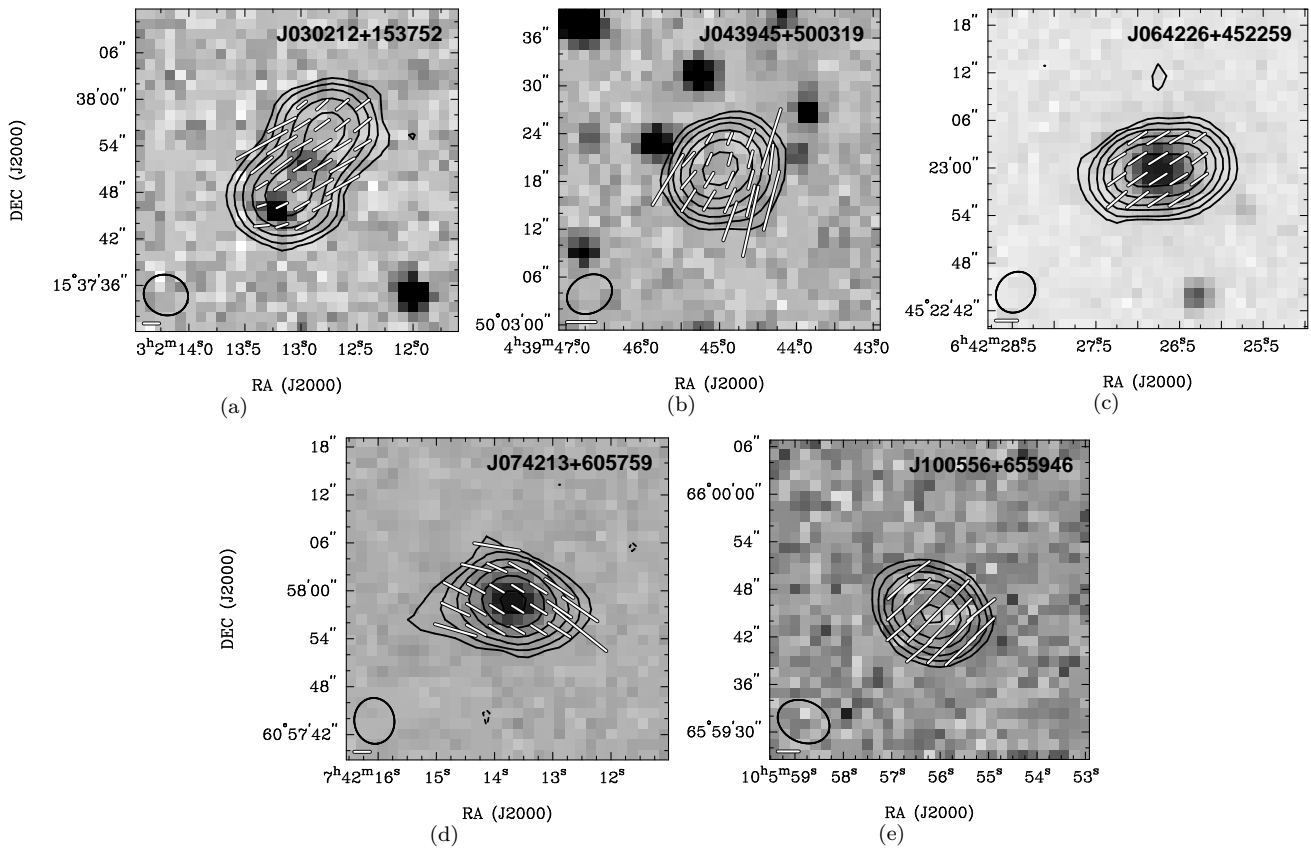

  \centering
  \subfigure[]{\includegraphics[bb=47 27 580 648,height=6.0cm,angle=-90.0,clip]{J0302+1537_pol_opt.ps}
    \label{fig:J0302+1537}}
  \subfigure[]{\includegraphics[bb=47 86 580 658,height=5.5cm,angle=-90.0,clip]{J0439+5003_pol_opt.ps}
    \label{fig:J0439+5003}}
  \subfigure[]{\includegraphics[bb=47 85 580 649,height=5.4cm,angle=-90.0,clip]{J0642+4522_pol_opt.ps}
    \label{fig:J0642_1460}}
  \subfigure[]{\includegraphics[bb=44 27 580 649,height=6.0cm,angle=-90.0,clip]{J0742+6057_pol_opt.ps}
    \label{fig:J0742_6057}}
  \subfigure[]{\includegraphics[bb=41 85 580 660,height=5.5cm,angle=-90.0,clip]{J1005+6559_pol_opt.ps}
    \label{fig:J1005+6559}}
\caption{VLA radio contours and polarization E-vectors at 1.4\,GHz
  overlaid on SuperCOSMOS {\it R}-band images.  The contour levels are
  $\pm 3\sigma \times 2^{n}$, where $n=0, 1, 2, 3,$ \ldots\ and
  $\sigma= 0.05, 0.07, 0.05, 0.05, 0.08$\,mJy~beam$^{-1}$ for panels
  (a), (b), (c), (d) and (e), respectively. The beam size is shown at
  the lower left-hand corner of the images and 30 per cent linear
  polarization is indicated by the length of line segment below the
  beam. The optical images are scaled to the same pixel size as the
  corresponding radio images. }
  \label{fig:polmap}
\end{figure*}

The ATCA has orthogonal linearly polarized feeds in each of the two
observing frequencies. In continuum mode, the 128-MHz bandwidth at
each observing frequency is divided into 32 channels. ATCA
observations were conducted on 2001 January 18 and 19.  The 6C array
configuration was used and observations were conducted in snapshot
mode simultaneously at 1384\,MHz and 2496\,MHz. Data calibration
followed standard procedures in MIRIAD. The images were formed with
bandwidth synthesis using the individual channels. Natural weighting
was used to maximise the detection sensitivity. Since ATCA snapshot
observations suffer from very sparse sampling of the uv plane, care
was taken to minimise clean bias. The images were first cleaned
unconstrained to identify the sources.  The dirty images were then
cleaned with box constraints on only those sources that also appear in
the NVSS catalogue within the primary beam. The total clean iterations
performed were between 1400 and 5500 for the 1.4-GHz data and between
300 and 1000 for the 2.5-GHz data.  The Q and U images were cleaned
with only a few hundred iterations.  Polarization bias was corrected
using estimates from the Q and U image rms noise, which was also very
close to the V image rms noise. The percentage linear polarization
from NVSS was confirmed within error margins for all but one source
(J180523$-$344947), which was found to be polarized at 8 per cent.
Figure~\ref{fig:polmap2} shows the sources observed by the ATCA at
both 1384\,MHz and 2496\,MHz. Pixels with polarization intensity less
than $3\sigma$ and polarization position angle error greater than
10\degr\ were clipped from the final polarization image. There appears
to be little Faraday depolarization judging from the small difference
in average polarization between the two frequencies, as shown in
Table~\ref{tab:obs_nvss}.

VLA follow-up observations were conducted on 2001 February 25 and 28
in a snapshot mode using the B array configuration.  The observations
were obtained in continuum mode with full polarization measurements at
1435\,MHz and 1485\,MHz. The VLA has circularly polarized feeds, with
a single channel of 50-MHz bandwidth at each frequency. Data
calibration, image formation and deconvolution followed standard
calibration procedures in AIPS. Figure~\ref{fig:polmap} shows the
deconvolved radio images derived from natural weighting of the uv
data. Polarization intensity was corrected for Ricean bias based on
the rms noise in the Q and U images which were similar to the noise in
the V images. The high linear polarization was confirmed in all cases
by the VLA observations.

\begin{figure*}
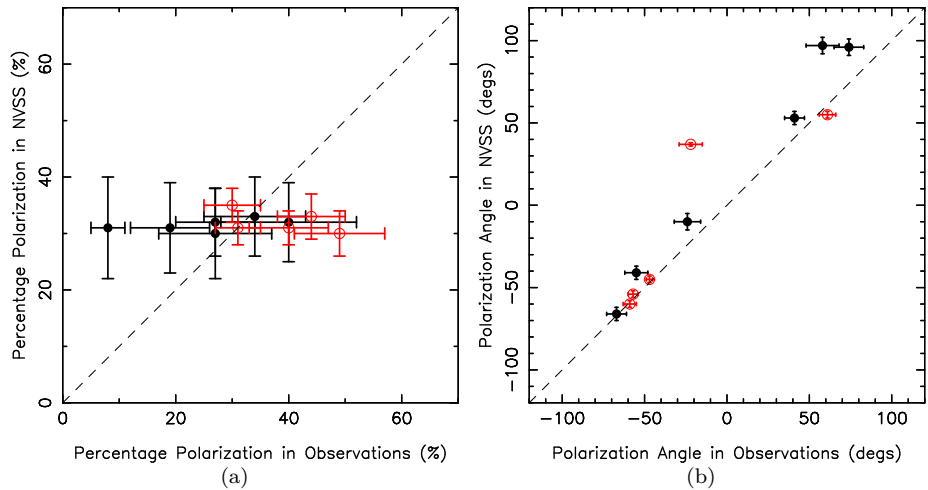

    \centering
    \subfigure[]{\includegraphics[height=6cm,angle=-90.0]{pp_obs.ps}
                \label{fig:pp_obs}}
    \subfigure[]{\includegraphics[height=6cm,angle=-90.0]{pa_obs.ps}
                \label{fig:pa_obs}}
\caption{(a) NVSS percentage linear polarization versus that measured
  at higher resolution with the ATCA (black points) and VLA (red
  circles) at 1.4\,GHz; and (b) NVSS polarization angle (position
  angle of the E-vector) versus those measured at higher resolution
  with the ATCA (black points) and VLA (red circles).}
  \label{fig:fpol}
\end{figure*}

Figure~\ref{fig:pp_obs} shows the NVSS fractional linear polarization
versus those measured at higher resolution at the ATCA and VLA. High
polarization was confirmed in all but one source.
Figure~\ref{fig:pa_obs} shows that the polarization position angles
measured by the high-resolution observations also agree with the NVSS
values.

\begin{figure*}
  \centering
  \includegraphics[angle=-90.0, width=80mm]{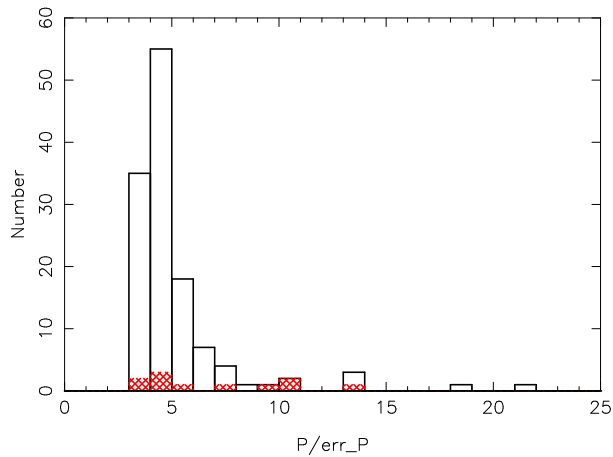}
\caption{Distribution of the NVSS polarization intensity
  signal-to-noise for the sample of 129 high-polarization sources. The
  cross-hatched histogram shows the 11 sources that were selected for
  observation at high resolution with ATCA and VLA.}
  \label{fig:p_err}
\end{figure*}

Figure~\ref{fig:p_err} shows that the distribution of the polarization
intensity signal-to-noise of the 11 sources selected for
high-resolution observations span a similar range in polarization
signal-to-noise ratio as the whole sample. Therefore, in so far as
they are representative of the whole sample, the higher resolution
ATCA and VLA measurements confirm that the high linear polarizations
measured by the NVSS are reliable.

\section{Optical identification} 
Faint Images of the Radio Sky at Twenty-centimeters (FIRST, Becker,
White \& Helfand 1995) has a much higher resolution (5\,arcsec
compared to 45\,arcsec) and higher point source sensitivity (0.15\,mJy
rms noise compared to 0.45\,mJy rms) than the NVSS, and hence is more
reliable for optical identifications. It covers $9000$\,deg$^{2}$ at
the north Galactic cap and a $\sim 2.5\degr$ wide strip along the
celestial equator, but did not measure linear polarization.  It was
designed to cover the same region of sky as the Sloan Digital Sky
Survey (SDSS, York et al. 2000). SDSS contains broad-band photometric
measurements in five bands, ${\it u} (\sim 355\,{\rm nm})$, ${\it g}
(\sim 469\,{\rm nm})$, ${\it r} (\sim 617\,{\rm nm})$, ${\it i} (\sim
748\,{\rm nm})$ and ${\it z} (\sim 893{\rm nm})$ (Fukugita et
al. 1996) and spectroscopic observations for the brighter objects
covering the wavelength range $3800-9200$\,\AA.

There are 45 sources in our final sample that fall within the FIRST
survey region, of which 36 were detected in the FIRST catalogue within
10\,arcsec of the NVSS position (see Table~\ref{tab:highres}). The
undetected sources in the FIRST survey region are most likely to be
faint extended sources, since the rms noise levels in the images are
all less than $0.4$\,mJy\,beam$^{-1}$. Correlations with the SDSS
(Data Release 7) and Two Micron All Sky Survey (2MASS) catalogues
found that 22 of the FIRST-detected sources have an optical
counterpart (see Table~\ref{tab:SDSS}). For slightly extended FIRST
sources, optical identifications are found within 3\,arcsec of the
radio peak position. For sources that are well resolved (e.g. those
that are resolved into doubles), optical identifications were sought
along the principal axis of the source. Figure~\ref{fig:optid} shows
radio contours overlaid on SDSS $i$-band images for all the
FIRST-detected sources with optical
identifications. Table~\ref{tab:SDSS} lists the SDSS position of the
optical identification, the SDSS {\it g}- and {\it i}-band magnitudes,
the 2MASS $K$ magnitude, redshift (either spectroscopically determined
or estimated from SDSS multiband photometry) and the classification of
the identification based on the optical spectrum. All are extended in
the optical.  Out of 11 sources with SDSS spectra, 10 are associated
with elliptical galaxies with typical absorption lines (Balmer lines,
strong Mg I line, distinct $4000$-\AA\ break and Ca II H \& K) and no
emission lines.  The one exception, J163221+155147, showed emission
lines of [O\,II], H$\beta$, [O\,III], [O\,I], H$\alpha$, [N\,II] and
[S\,II], typical of a LINER that has undergone recent star formation;
it is classified as an ultraluminous infrared galaxy and an OH
megamaser (Darling \& Giovanelli 2000).

In addition, of the 11 sources with high-resolution ATCA and VLA
observations, seven were found to have optical identifications in
either SuperCOSMOS or 2MASS within 3\,arcsec of the central position
of the radio source.  The SuperCOSMOS Sky Survey (SSS) is a digitized
all-sky survey using three sets of photographic survey plates from the
UK Schmidt, ESO Schmidt and Palomar Schmidt telescopes.
Table~\ref{tab:cosmos} gives the position of the optical
identification from SuperCOSMOS, the optical magnitudes, $B_J$ and
$I_N$, in the photographic $B$ (IIIaJ emulsion) and $I$ bands (IV-N
emulsion) respectively, the 2MASS $K$ magnitude and a
spectroscopically determined redshift if it could be measured.
Spectra were obtained for two of the optical identifications using the
Australian National University's (ANU) 2.3-m telescope and Double Beam
Spectrograph, with a combined wavelength range $\sim 3000-9000$\AA.
Both spectra are identified with elliptical galaxies with no emission
lines.

It appears that the majority of these isolated, compact high
polarization radio sources are elliptical galaxies with no detectable
emission lines.  The 13 optical identifications with measured spectra
(see Table \ref{tab:SDSS} and \ref{tab:cosmos}) are all extragalactic,
with redshifts ranging from $0.05$ to $0.46$.  Only J163221+155147 in
the sample of 129 sources given in Table~\ref{tab:129nvss} was
detected in the IRAS catalogue, indicating that few of these sources
are luminous star-forming galaxies.

\input sdss_final_22.tex

\section{Radio luminosity and morphology}

\begin{figure*}
  \centering
  \includegraphics[angle=-90.0, width=80mm]{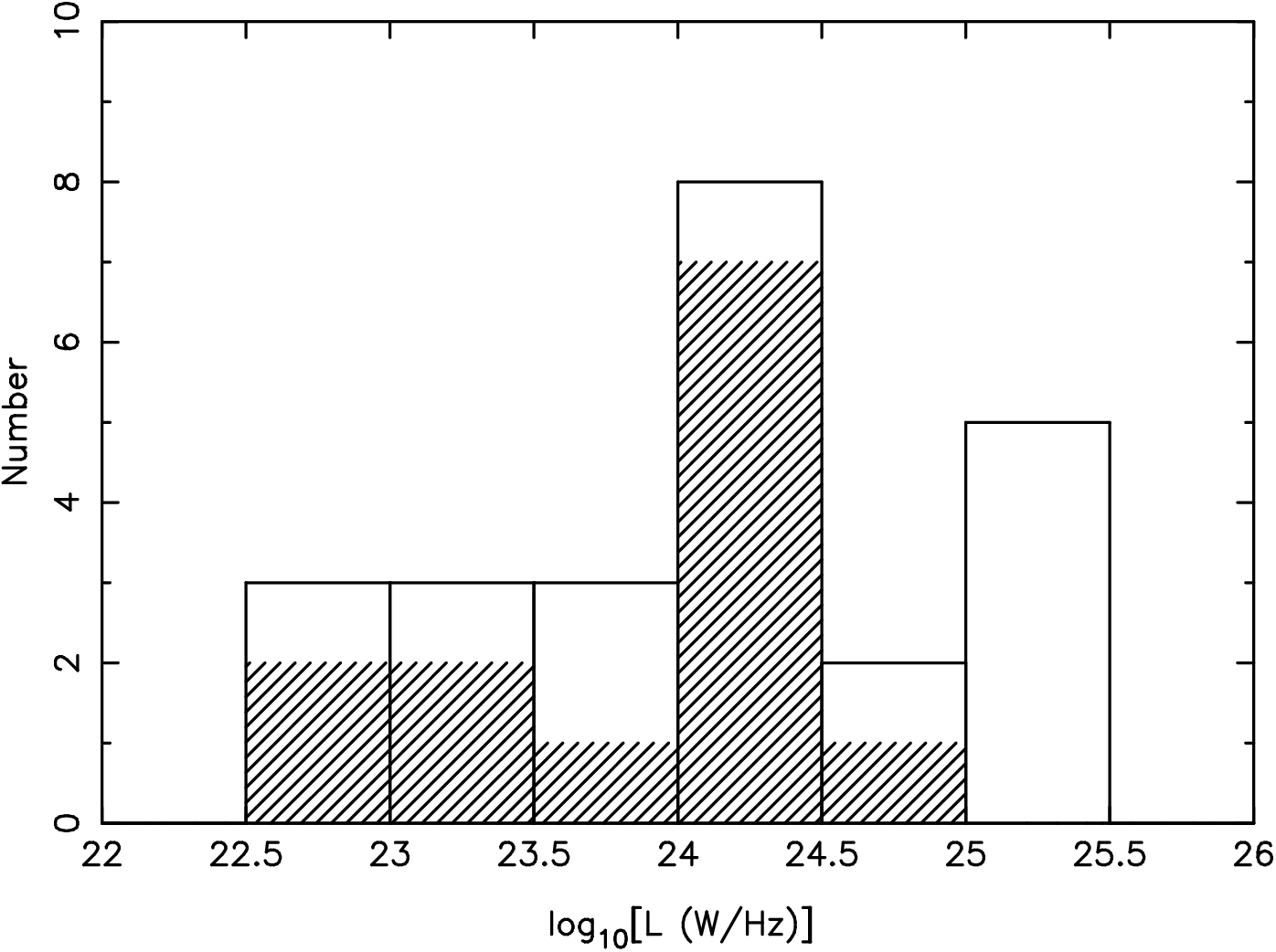}
\caption{Distribution of radio luminosities for the sources in Table
  \ref{tab:SDSS} and \ref{tab:cosmos} with either photometric or
  spectroscopic redshifts. The hatched histogram is for those with a
  spectroscopic redshift.}
  \label{fig:rlum}
\end{figure*}

For those sources with measured redshifts, total radio luminosities
have been determined from the NVSS integrated flux densities.
Figure~\ref{fig:rlum} shows the distribution of radio luminosity for
the 13 sources with spectroscopic redshifts as well as a further 11
sources with SDSS photometric redshift estimates. The radio
luminosities range from $\sim 5\times10^{22}$ to $\sim 2.5\times
10^{25}$\,W\,Hz$^{-1}$, which is consistent with the sources being
radio-loud active galactic nuclei (AGNs). The 1.4-GHz radio
luminosity function of radio-loud AGNs and star-forming galaxies
crosses over at around $10^{23}$\,W\,Hz$^{-1}$ (Sadler et al. 2002;
Best et al. 2005; Mauch \& Sadler 2007).

\begin{figure*}
  \centering
  \includegraphics[angle=-90.0, width=80mm]{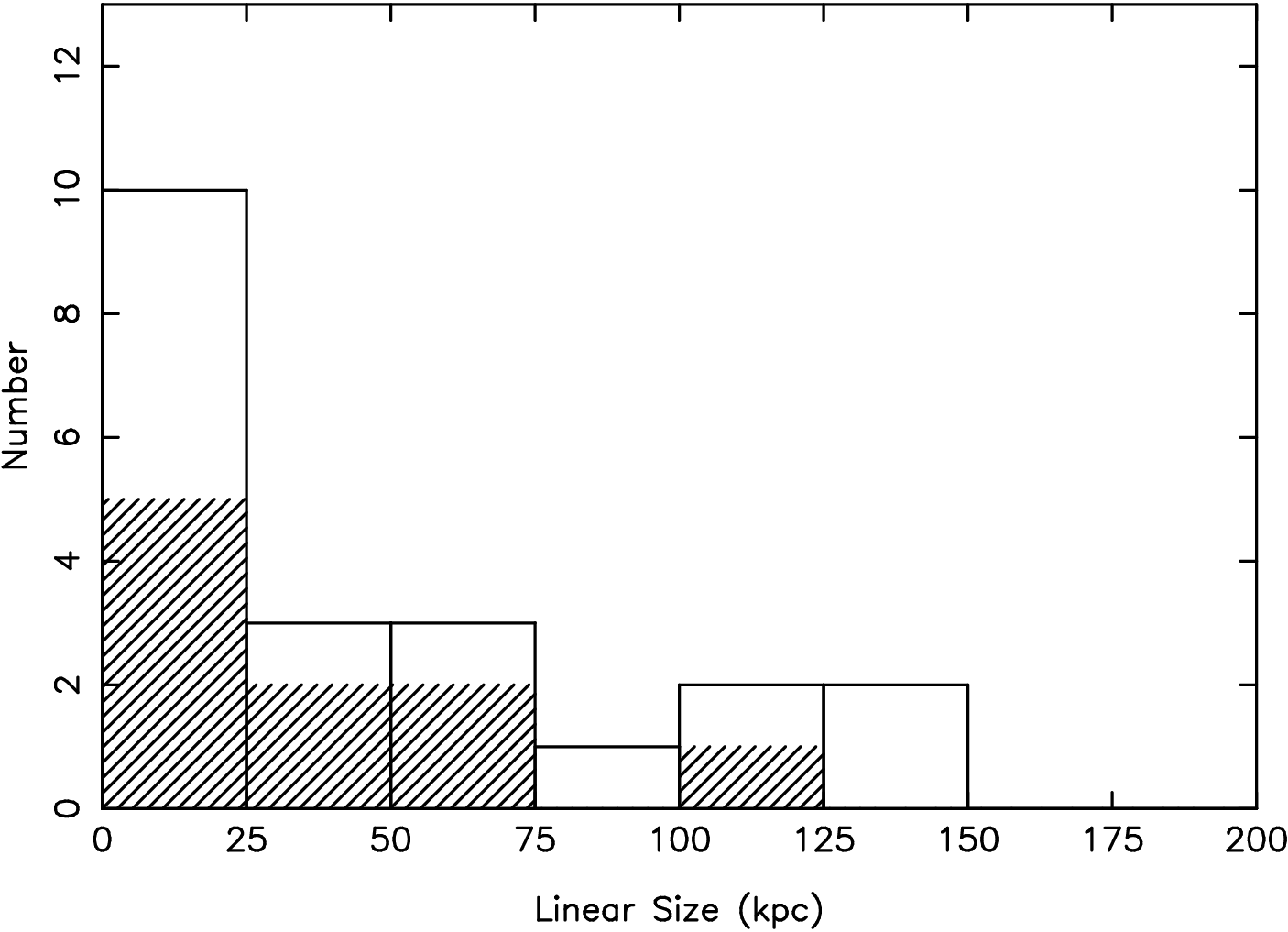}
\caption{Distribution of linear sizes (FWHM of deconvolved major axis)
  for the sources in Table \ref{tab:SDSS} and \ref{tab:cosmos} with
  either photometric or spectroscopic redshifts. The hatched histogram
  is for those with spectroscopic redshift. The unresolved sources,
  J163221+155147, J035815$-$303129 and J094615$-$311406, are
  excluded. }
  \label{fig:rlin}
\end{figure*}

The high resolution radio observations with ATCA and VLA found that
most sources observed with the VLA are extended (beam size 5\,arcsec)
and one of the sources observed with the ATCA is extended (beam size
$\sim$\,9\,arcsec\,$\times$\,18\,arcsec).  In addition, the majority
of sources (30 out of 36) detected in the FIRST survey (beam size
$\sim$\,5\,arcsec) are extended, with deconvolved sizes greater than
2\,arcsec (Table \ref{tab:highres}). The two known pulsars are both
found to be unresolved in the FIRST survey, as expected. The LINER/OH
megamaser, J163221+155147, is also found to be unresolved in the FIRST
survey. The projected linear size distribution (based on the FWHM of
the deconvolved major-axis) of those extended radio sources with
spectroscopic or SDSS photometric redshifts (Tables \ref{tab:SDSS} and
\ref{tab:cosmos}) is shown in Fig.~\ref{fig:rlin}.

\section{Radio spectral index}

\begin{figure*}
  \centering
  \includegraphics[angle=-90.0, width=80mm]{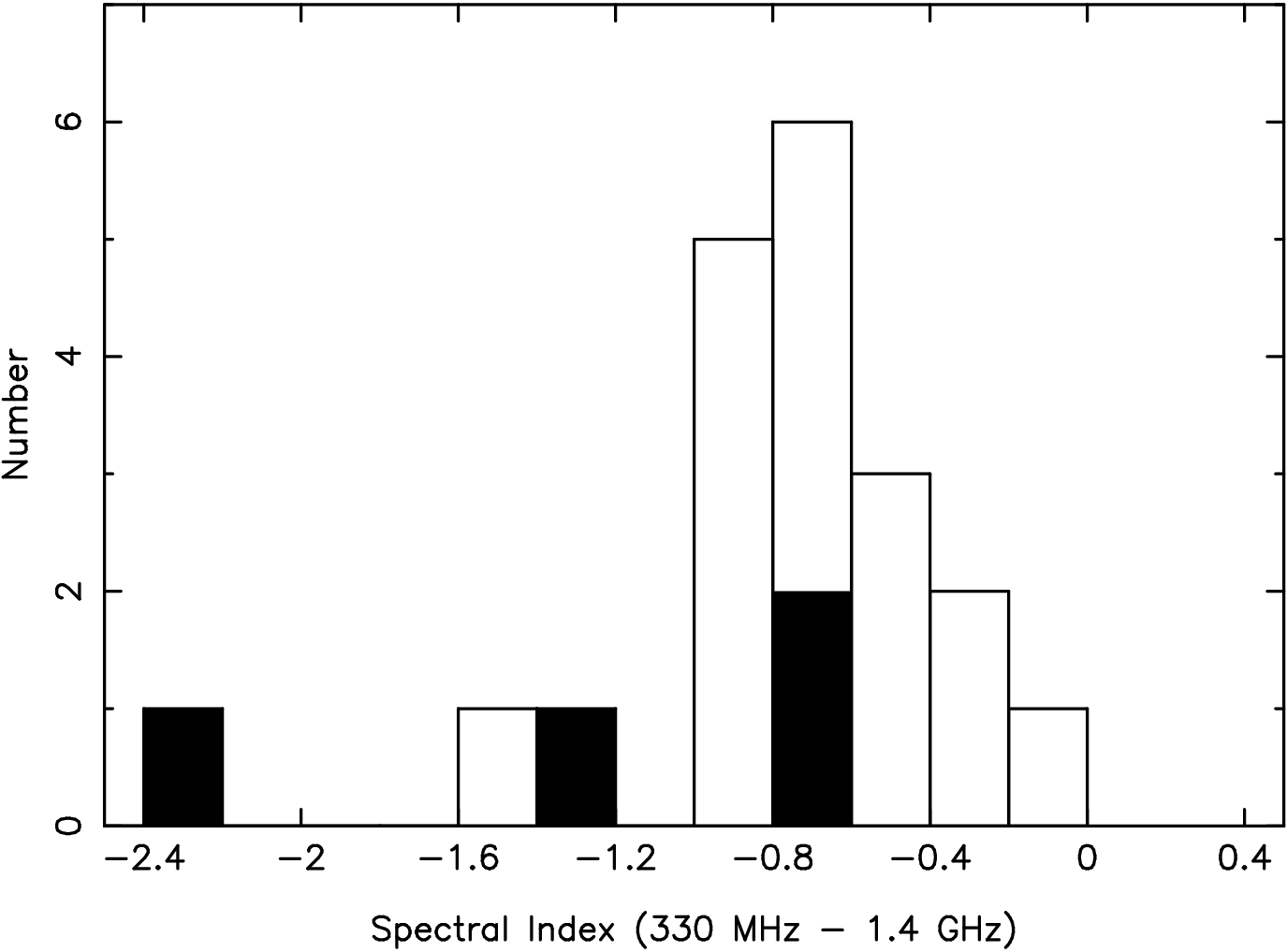}
\caption{Spectral index distribution between 330\,MHz and 1.4\,GHz for
  the 20 sources in the sample that are also detected in WENSS. The
  pulsars are shown in solid black.}
  \label{fig:spin}
\end{figure*}

Spectral indices of the sources have been estimated by comparing NVSS
flux densities with the Westerbork Northern Sky Survey (WENSS) at
330\,MHz (Rengelink et al. 1997) and with Sydney University Molonglo
Sky Survey (SUMSS) at 843MHz (Bock, Large \& Sadler 1999; Mauch et
al. 2003, Murphy et al. 2007).  Figure~\ref{fig:spin} shows the
330\,MHz to 1.4\,GHz spectral index histogram of the sample of 20
sources with WENSS flux densities. The median spectral index of this
sample is $-0.76\pm 0.48$ (or $-0.73\pm 0.30$ excluding the
pulsars). In comparison, for the 10 sources with SUMSS flux densities,
the median spectral index between 843\,MHz and 1.4\,GHz is $-0.95\pm
0.50$ which is consistent with the NVSS-WENSS median spectral index;
there are no neighbouring NVSS sources within the SUMSS beam for these
10 sources.

Kimball and Ivezi{\'c} (2008) estimated the distribution of the
spectral index between 330\,MHz and 1.4\,GHz for radio sources
detected in all three surveys (NVSS, FIRST and WENSS) and found that
the median spectral index of `compact' sources to be $-0.58$, of
`resolved' sources to be $-0.80$ and of `complex' sources to be
$-0.79$\footnote{These numbers are slightly different from those given
  by Kimball and Ivezi{\'c} (2008), due to a slight error in the NVSS
  flux densities quoted in their original paper, and correspond to the
  updated version of their online catalogue.}.  The radio morphologies
of our sample mostly include sources defined in their paper as
`compact' and `resolved' (see fig.~8 in Kimball \& Ivezi{\'c}
2008). An overall median spectral index of `compact' and `resolved'
sources was obtained from their table of median spectral indices for
the two types of sources, weighted by the number of sources in each
category. It was found to be $-0.67$, similar to the median spectral
index of our sample. This shows that the low frequency radio spectral
index distribution of our sample of highly polarized NVSS sources is
similar to a sample of NVSS sources with similar radio morphology but
without polarization selection.

\section{Nature of the sources}

The radio and optical properties of these highly polarized radio
sources discussed above show that at least the majority of optically
bright objects are elliptical radio galaxies, based on their radio
luminosity and their optical morphology and spectra. While radio
galaxies are known to have such high linear polarization in isolated
regions, such as part of a radio jet, there are no known examples of
radio galaxies with such high overall linear polarization at
1.4\,GHz. Even the highly polarized source J0658.7$-$5559 (55 per cent
at 8.8\,GHz) is only $\sim 1.4$ per cent polarized at 1.4\,GHz (Liang
et al. 2001). Faraday depolarization is much stronger at lower
frequencies, as a result of thermal plasma lying either within or in
front of the radio-emitting region.

In general, strong linear polarization implies that a substantial
fraction of the magnetic field in the emission region is regular or
compressed into a plane containing the line of sight (Laing 1981),
which for some of the sources in our sample must extend for over a
hundred kpc. It is difficult to generate ordered magnetic fields on
such a large scale (Kulsrud \& Zweibel 2008). Any theoretical model
for the origin of magnetic fields will have to explain how such
ordered large-scale fields can be created.

Why are these sources so highly polarized compared with known radio
galaxies? Either they have instrinsically more ordered magnetic fields
than average, or they have the same intrinsic properties but suffer
less Faraday depolarization. We therefore consider four possibilities:
(i) the intrinsic average polarization is higher because of a more
ordered large-scale source magnetic field; (ii) the internal Faraday
depolarization at the source is much lower due to a source magnetic
field orientation that is mostly perpendicular to the line of sight
(i.e. $B_{||}\sim 0$) and/or a low and uniform thermal electron
density at the source emission region; (iii) these sources inhabit a
much sparser environment than radio galaxies showing low polarization;
and (iv) the lines of sight to these sources pass through lower
density regions than those of an average radio galaxy (i.e. lack of
foreground magnetised plasma). The fact that these sources have very
high polarization already suggest that the large scale magnetic field
at the source must be well ordered. However, low polarization radio
sources may also have well-ordered magnetic fields at the source, but
suffer depolarization along the line of sight to the
observer. Internal Faraday depolarization can be significant even if
the magnetic field and electron density are both uniform. Foreground
magnetised plasma with small-scale magnetic field and electron density
inhomogeneity can also cause Faraday depolarization. Note that the
magnetised plasma in the interstellar medium of our own Galaxy is
known to cause Faraday rotation, but the amount of Faraday
depolarization due to the Galaxy is not sufficient to be considered a
major cause for the depolarization of extragalactic sources (Burn
1966).

We can address the third and fourth possibility by examining the
immediate and line-of-sight environments of these ultrahigh
polarization radio galaxies in comparison with a sample of similar
radio sources with low linear polarization.

\subsection{Environments of ultra-high polarization sources}

In order to compare the environments of these highly polarized sources
with similar radio sources of low polarization, we need to select
those ultrahigh polarization sources with spectroscopic redshifts and
a comparison sample of low-polarization sources. For the ultrahigh
polarization subsample, only the eight sources with spectroscopically
determined redshifts in the range of 0.1 - 0.4 are selected from
Table~\ref{tab:SDSS} and \ref{tab:cosmos} (excluding the OH
megamaser). This subsample of high-polarization sources spans a
1.4-GHz radio luminosity range of $\sim 1.5\times 10^{23}$ to $\sim
2.5\times 10^{24}$\,W~Hz$^{-1}$, and radio projected linear size range
of $\sim 8-65$\,kpc. All of the sources have optical spectra typical
of low-power radio-loud AGNs, that is, elliptical galaxies with no
obvious emission lines.

For comparison, we select a low-polarization sample of radio-loud AGNs
with similar radio luminosity range, linear size range, redshift range
and optical spectra, but with linear polarization less than 2 per cent
at 1.4\,GHz. Such a sample can be readily selected from the Best et
al.  (2005) catalogue, where they correlated the spectroscopic sample
of the second data release (DR2) of the SDSS with NVSS and FIRST
survey to obtain a catalogue of 2712 radio-luminous galaxies. These
radio sources were then classified as radio-loud AGNs and star-forming
galaxies, according to their optical spectra and radio luminosity. The
radio sources found in the NVSS were also classified according to
their source morphology, where `class 1' are single-component NVSS
sources with single-component matches in the FIRST survey and `class
2' comprises single-component NVSS sources resolved into multiple
FIRST sources. For our comparison sample, only the radio-loud AGNs
with radio morphology of `class 1' and `class 2' are selected so that
the low- and high-polarization samples are morphologically similar.
The linear polarization values were obtained from the NVSS. The median
percentage polarization is only 5.4 per cent, with 70 per cent of the
sources having polarization less than 10 per cent. The
low-polarization comparison sample of 121 sources is then selected by
choosing only those with NVSS linear polarization less than 2 per
cent.

A direct observation of the thermal plasma can be obtained through
X-ray observations, but current X-ray surveys are not sensitive enough
to detect anything other than rich clusters in this redshift range. In
addition, sufficient resolution and brightness sensitivity are needed
to distinguish X-ray emission from the central engine of an AGN and
the surrounding medium. The ROSAT all-sky survey is the deepest
all-sky X-ray catalogue available. Out of the 129 highly polarized
sources in Table~\ref{tab:129nvss}, only J103601+050714 was detected
in the ROSAT all-sky survey catalogue and just four sources out of the
121 low-polarization sample from Best et al. (2005) have counterparts
in the ROSAT catalogue.

Another indirect method of assessing the thermal plasma density of the
environment is by counting the number of galaxies within a given
radius. Thermal plasma is known to exist between galaxies in groups
and clusters, with richer clusters having higher thermal electron
densities and temperature (Mulchaey \& Zabludoff 1998; Ledlow et
al. 2003).

\begin{figure*}
    \centering
    \subfigure[]{\includegraphics[height=7cm,angle=-90.0]{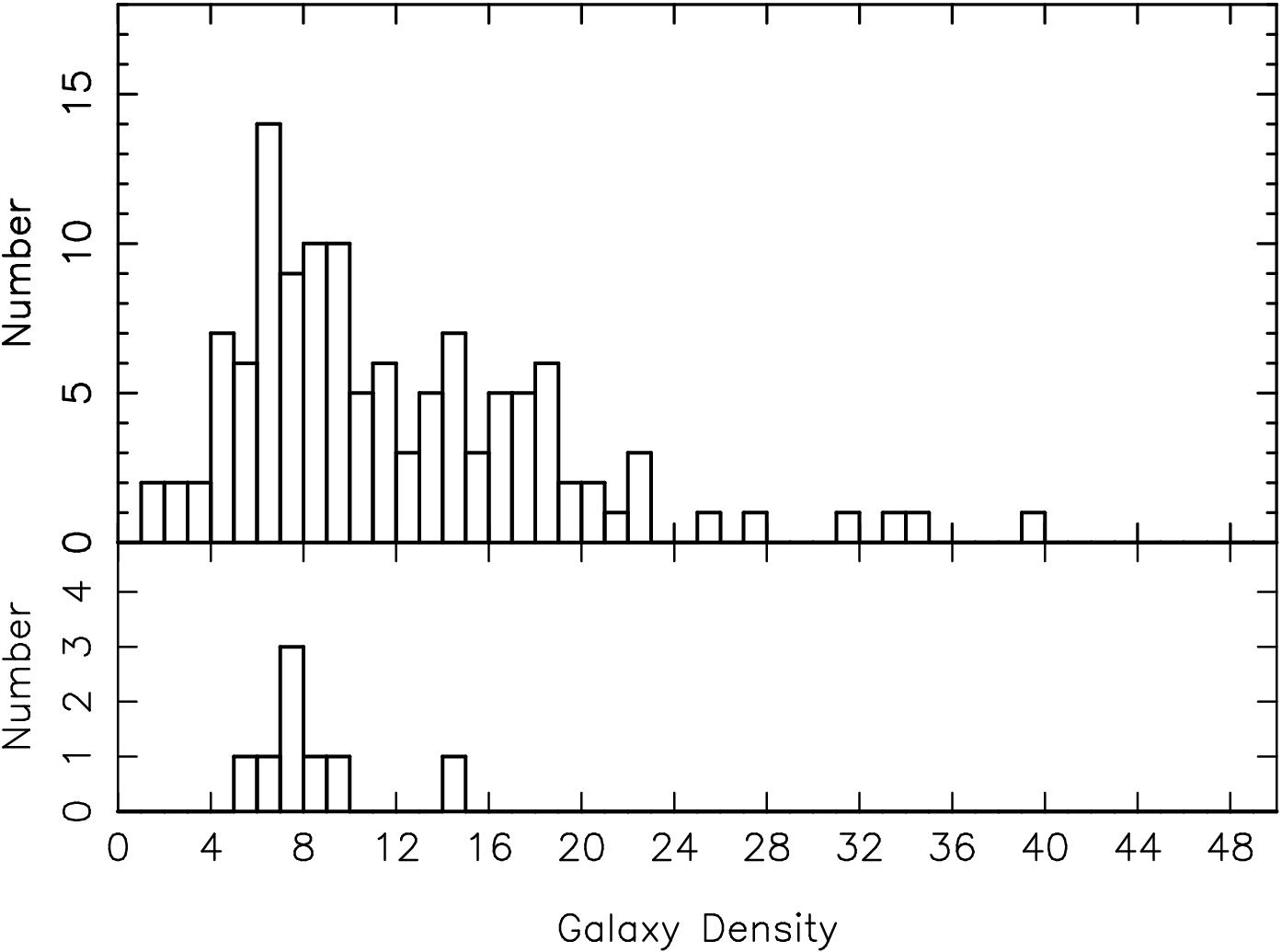}
                }
    \subfigure[]{\includegraphics[height=7cm,angle=-90.0]{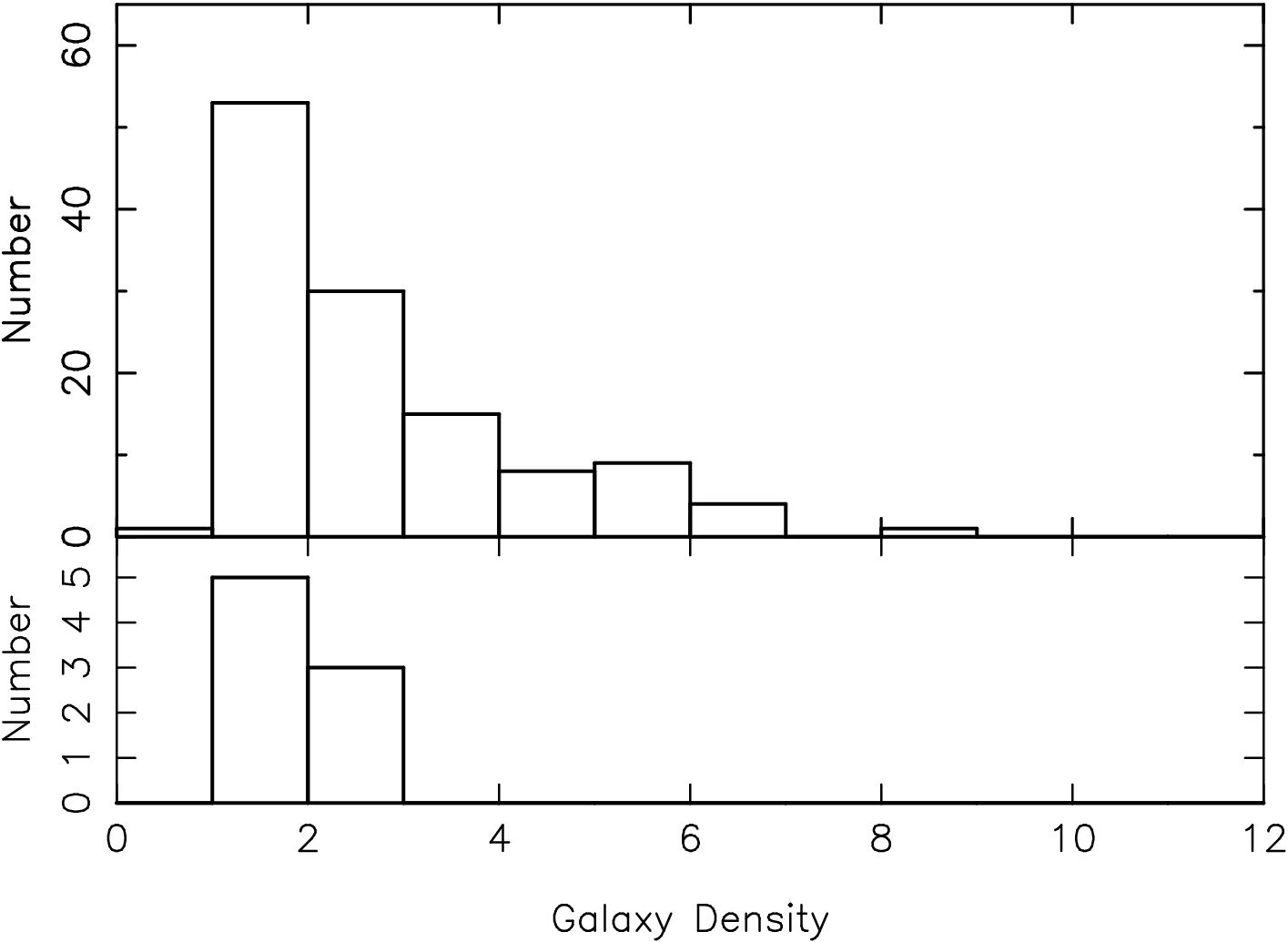}
                }
\caption{(a) Galaxy counts within a 1\,Mpc radius of a radio-loud AGN
  for the low-polarization comparison sample of 121 sources (upper
  part) and the ultrahigh polarization sample of eight sources (lower
  part); and (b) the same as (a) but within a 200\,kpc radius.}
  \label{fig:counts}
\end{figure*}

Galaxy number counts within a 1\,Mpc, 500\,kpc and 200\,kpc radius (in
the plane of the sky) for the ultrahigh polarization subsample are
compared with that of the low-polarization sample. The galaxy count
statistics within the 200\,kpc radius give an indication of the more
immediate environment around a radio source. A galaxy is counted if
its SDSS photometric redshift is consistent with the radio-source
redshift within the photometric redshift error, which is given by
$\sigma_{0}(1+z)$, where $\sigma_{0}\sim 0.04$ is appropriate for the
current redshift range (Abdalla et al. 2008; Wen, Han \& Liu
2009). Note that the photometric redshift error is much greater than
the velocity dispersion of a typical rich cluster of galaxies. In
addition, only galaxies brighter than an apparent magnitude limit of
$r=21.5$ and an absolute $r$ magnitude limit of $-21$ were
selected. The apparent magnitude limit was selected for a reliable
star/galaxy separation (Lupton et al. 2001). The absolute magnitude
limit was chosen above the completeness limit for the redshift range
concerned to avoid Malmquist bias.

Table~\ref{tab:counts} gives a summary of the results of the galaxy
counts, and the galaxy count distributions within 1\,Mpc and 200\,kpc
are shown in Fig.~\ref{fig:counts}.  The difference in the
distributions is not significant. A Kolmogorov-Smirnov two-sample test
was applied and the null hypothesis of the samples being drawn from
the same distribution could not be rejected even at the 10 per cent
level. The median galaxy count within a 1\,Mpc radius is $9\pm 7$ for
the low-polarization sample and $7\pm 3$ for the ultrahigh
polarization sample.  In comparison, using the same galaxy counting
criteria, rich clusters of galaxies in the same redshift range, such
as A773 and A2219, have 83 and 61 galaxies, respectively. On the other
hand, the mean galaxy counts within a radius of 1\,Mpc on the plane of
the sky centred on a random position between the RA range of
$\sim$180$\degr$ to 200$\degr$ and Dec. range of 0$\degr$ -- 10$\degr$
within a redshift range of 0.1--0.4 is $4.5\pm 0.3$. It appears,
therefore, that these compact, ultrahigh polarization radio galaxies
are in environments similar to their low-polarization counterparts,
which may be up to twice the average galaxy density in the same
redshift range, but nearly an order of magnitude less dense than the
richest clusters.

\subsection{Line-of-sight environments of ultrahigh polarization 
sources}

\begin{figure*}
    \centering \includegraphics[angle=-90.0, width=80mm]{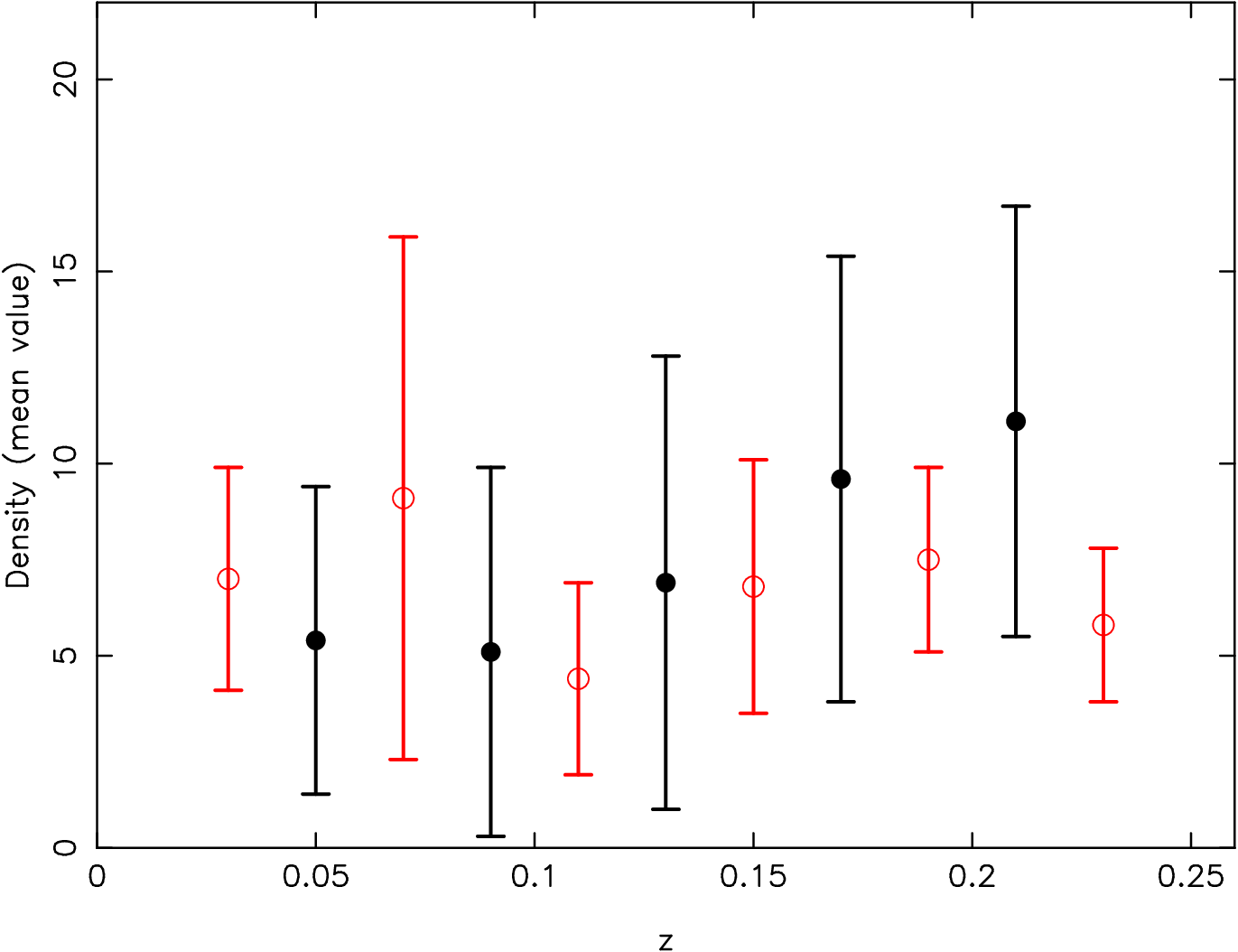}
 \caption{Mean galaxy counts versus redshift within a 1\,Mpc radius in
   the plane of the sky (redshift range given by the photometric
   redshift error) along the line of sight towards a radio-loud AGN
   for the low-polarization comparison sample (black filled circles)
   and the ultrahigh polarization sample (red open circles). The error
   bars indicate the 1$\sigma$ spread of the line-of-sight galaxy
   densities for each sample at a particular
   redshift. } \label{fig:linecounts} \end{figure*}

The above section compared the immediate environments of the ultrahigh
polarization radio sources and the low-polarization radio
sources. There remains the possibility that the lines of sight to
these ultrahigh polarization sources pass through relatively low
density environments compared with low-polarization sources. The same
galaxy counting method can be applied to regions along the line of
sight in the redshift range of $0.03<z<0.25$. A 1\,Mpc search radius
in the plane of the sky and a redshift range determined by the
photometric redshift error as defined in the above section are used
for galaxy counting. The redshift range was selected such that more
than half of the high-polarization sources are at redshifts greater
than 0.25 and the lower bound of the redshift range was selected so
that the angular size corresponding to 1\,Mpc is a manageable size.
The mean galaxy count for the ultrahigh polarization and the
low-polarization samples are compared at each redshift in
Fig.~\ref{fig:linecounts}, which shows that the line-of-sight galaxy
densities towards the ultrahigh and low-polarization sources are
indistinguishable.

Our results suggest that the difference in polarization between the
ultrahigh polarization sources and those with low polarization is not
because of differences in depolarization due to the external
environment along the line of sight.  The reasons for the extremely
high polarization must therefore be due to the intrinsic properties of
the radio sources.  The high polarization requires a more ordered
source magnetic field, low $B_{||}$ and thermal plasma density in the
radio emission region.  Any mixing of a large amount of thermal
electrons with the synchrotron emitting relativistic electrons will
cause intrinsic Faraday depolarization.

From the sources with polarization measurements at both 1.4 and
2.5\,GHz, it can be deduced that there is little internal Faraday
depolarization, as the fractional polarization values are similar at
the two frequencies within the limits of uncertainty
(Fig.~\ref{fig:polmap2} and Table~\ref{tab:obs_nvss}). This means that
either $B_{||}$ and/or the density of the thermal plasma within the
radio-emitting region is also very low.

The rotation of the polarization angle between 1.4 and 2.5\,GHz can be
accounted for by the rotation measure due to the foreground plasma in
our own Galaxy.  Table~\ref{tab:obs_nvss} shows the apparent rotation
of the polarization position angle between the two frequencies and
that expected from the Galactic rotation measure along the line of
sight (Simard-Normandin, Kronberg \& Button 1981). The apparent
rotation of the polarization angle estimated from just two frequencies
suffers from $n\pi$ ambiguity and hence the values given represent the
minimum rotation. Table~\ref{tab:obs_nvss} shows that this apparent
minimum rotation is always similar to or less than the rotation of the
polarization angle expected from the Galactic interstellar medium.

Higher resolution, higher frequency and multifrequency radio
polarization observations of a larger sample of ultrahigh polarization
sources are necessary to obtain the intrinsic magnetic field structure
and the intrinsic Faraday rotation measure of these sources. These
goals can be realised in the near future by current facilities (ATCA,
EVLA), and ultimately by the SKA.

\subsection{Optically faint radio sources}

The above discussions were centred around optically bright sources
with spectroscopic measurements, as those are the ones with optically
bright counterparts. Of the FIRST detected sources, 61 per cent have
an optical counterpart detected in the SDSS DR7 and 30 per cent have
optical spectral data. Hence, we can only confirm that the optically
brightest 30 per cent of high-polarization sources are radio-loud
AGNs. The optically fainter objects may not be the same type of
objects. Since we know, at least eight of the optically faint
high-polarization sources are pulsars, the nature of these sources
remains uncertain. However, the majority of the optically faint
sources are unlikely to be pulsars, since pulsars tend to be
unresolved. For example, only six sources are unresolved amongst the
36 FIRST detected sources in the high-polarization sample and the only
two known pulsars observed in FIRST were found to be unresolved. These
optically faint sources are likely to be higher redshift sources, but
their nature still needs to be investigated with deeper imaging and
follow-up spectroscopy.

\section{Conclusions}

High-resolution radio follow-up observations at both the ATCA and the
VLA show that the NVSS linear polarization values for the sample of
ultrahigh polarization sources listed in Table~\ref{tab:129nvss} are
reliable at the stated levels. The optically bright ultrahigh
polarization sources are identified with radio-loud AGNs. The radio
properties, such as luminosity and spectral index of these compact,
ultrahigh linear polarization sources, are unremarkable and typical of
lower power radio-loud AGNs. Deeper optical observations are necessary
to determine the identity of the optically faint sources.

The ultrahigh polarization sources appear to be in local environments
similar to their low-polarization counterparts with similar radio
power, linear size, redshift range and optical morphology. The
line-of-sight environments also appear indistinguishable. The
ultrahigh polarization must be due to the intrinsic properties of the
sources that favour highly ordered magnetic fields and low thermal
plasma density or low $B_{||}$ in the radio-emitting region.
Preliminary observations at two frequencies show that Faraday
depolarization is low and hence confirm that the internal Faraday
depolarization is low. A simple explanation could be that the magnetic
fields in these sources are aligned preferentially in the plane of the
sky giving a low $B_{||}$ and hence low internal Faraday
depolarization.

\section*{Acknowledgements}

Support from the National Natural Science Foundation (NNSF) of China
(10773016, 10821061, and 1083303), and the National Key Basic Research
Science Foundation of China (2007CB815403) is gratefully acknowledged.
HL acknowledges the hospitality of the National Astronomical
Observatory of China and the University of Sydney, School of
Physics. We thank Zhong Lue Wen, Alfonso Aragon-Salamanca and Ron
Ekers for useful discussions and Olivier Guiller, Rebecca Lange for
assistance. KARMA was used for the radio-optical overlays and MIRIAD
was used for imaging the high-resolution data from the ATCA. The ATCA
is part of the Australia Telescope, funded by the Commonwealth of
Australia for operation as a National Facility and managed by the
CSIRO.  The National Radio Astronomy Observatory is a facility of the
National Science Foundation operated under cooperative agreement by
Associated Universities, Inc. The research made use of the SDSS
archive provided by the Alfred P. Sloan Foundation, the Participating
Institutions, the National Science Foundation, the U.S. Department of
Energy, the National Aeronautics and Space Administration, the
Japanese Monbukagakusho, the Max Planck Society and the Higher
Education Funding Council for England. The SSS material is based on
photographic data originating from the UK, Palomar and ESO Schmidt
telescopes and was provided by the Wide-Field Astronomy Unit,
Institute for Astronomy, University of Edinburgh. The research made
use of the NVSS and FIRST radio surveys, which were carried out using
the VLA of the National Radio Astronomy Observatory.

\input table1.tex
\clearpage
\input table2.tex
\input table3.tex
\clearpage
\input table4.tex

\clearpage
\input table5.tex
\clearpage
\input table6.tex
\clearpage

\end{document}

%% file: sdss_final_22.tex
\begin{figure*}
\begin{tabular}{rrr}
\mbox{\includegraphics[bb=72 99 557 713,angle=270,width=63mm,clip]{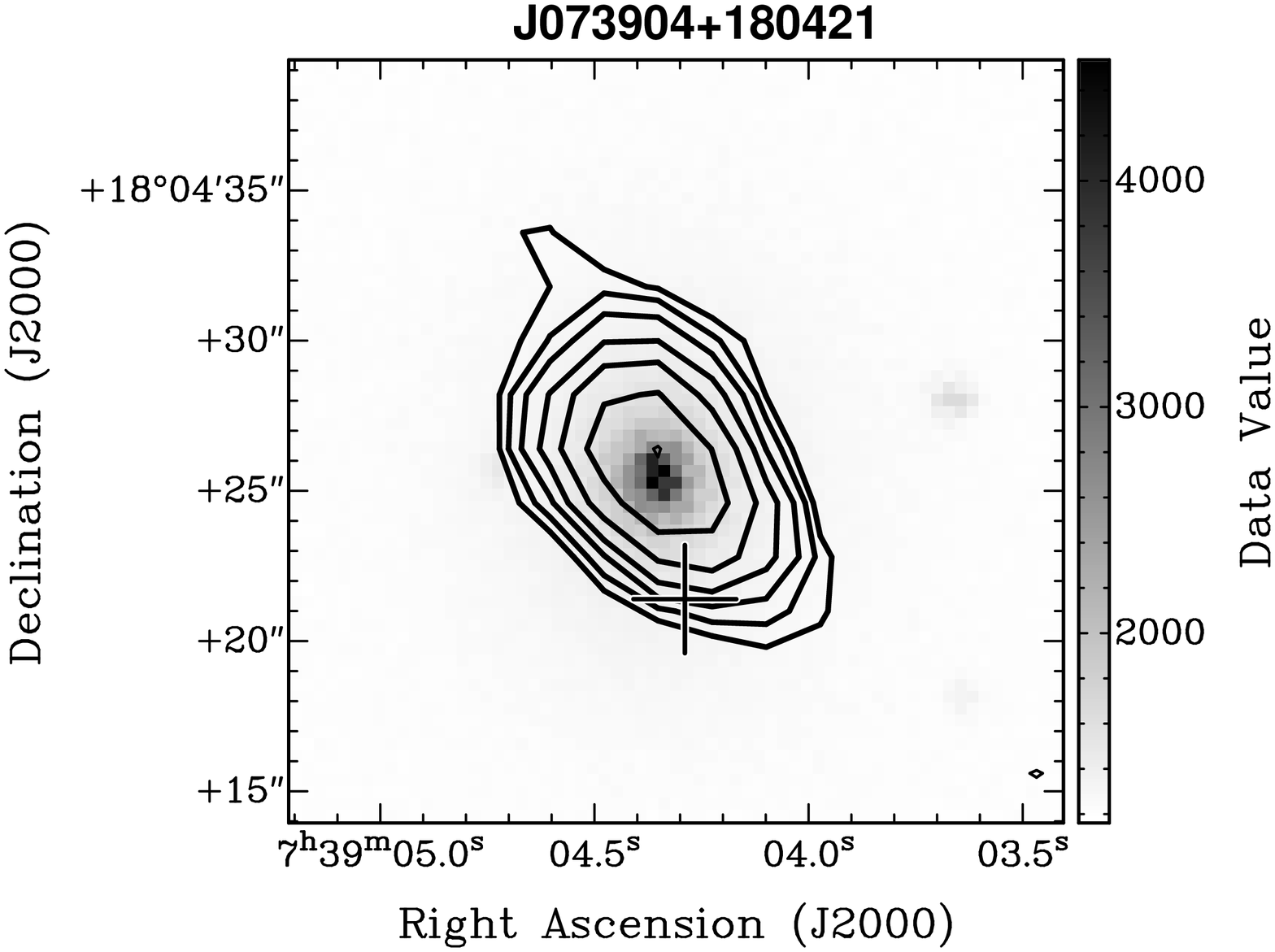}}&
\mbox{\includegraphics[bb=72 126 557 716,angle=270,width=60mm,clip]{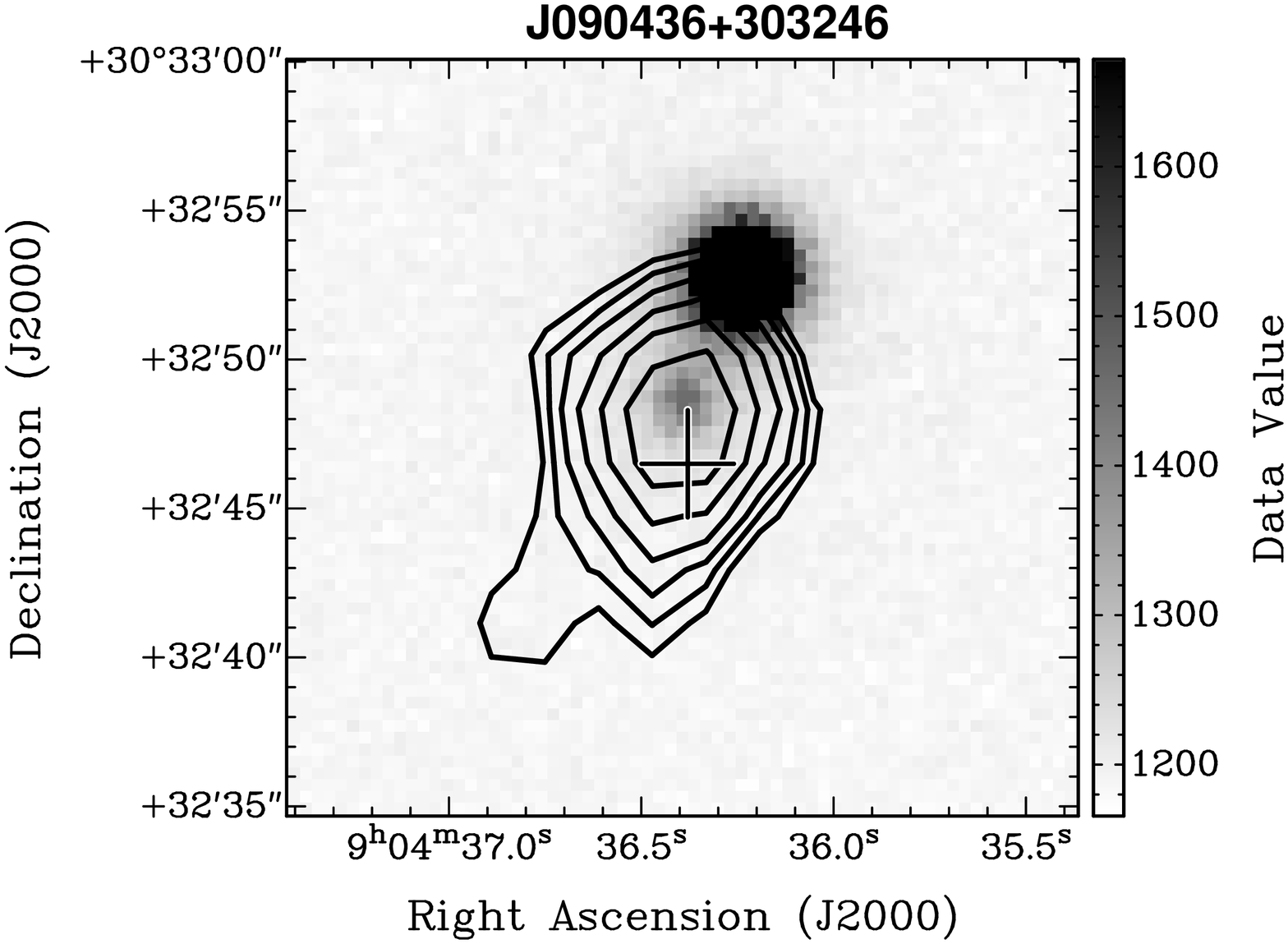}}&
\mbox{\includegraphics[bb=82 118 546 688,angle=270,width=60mm,clip]{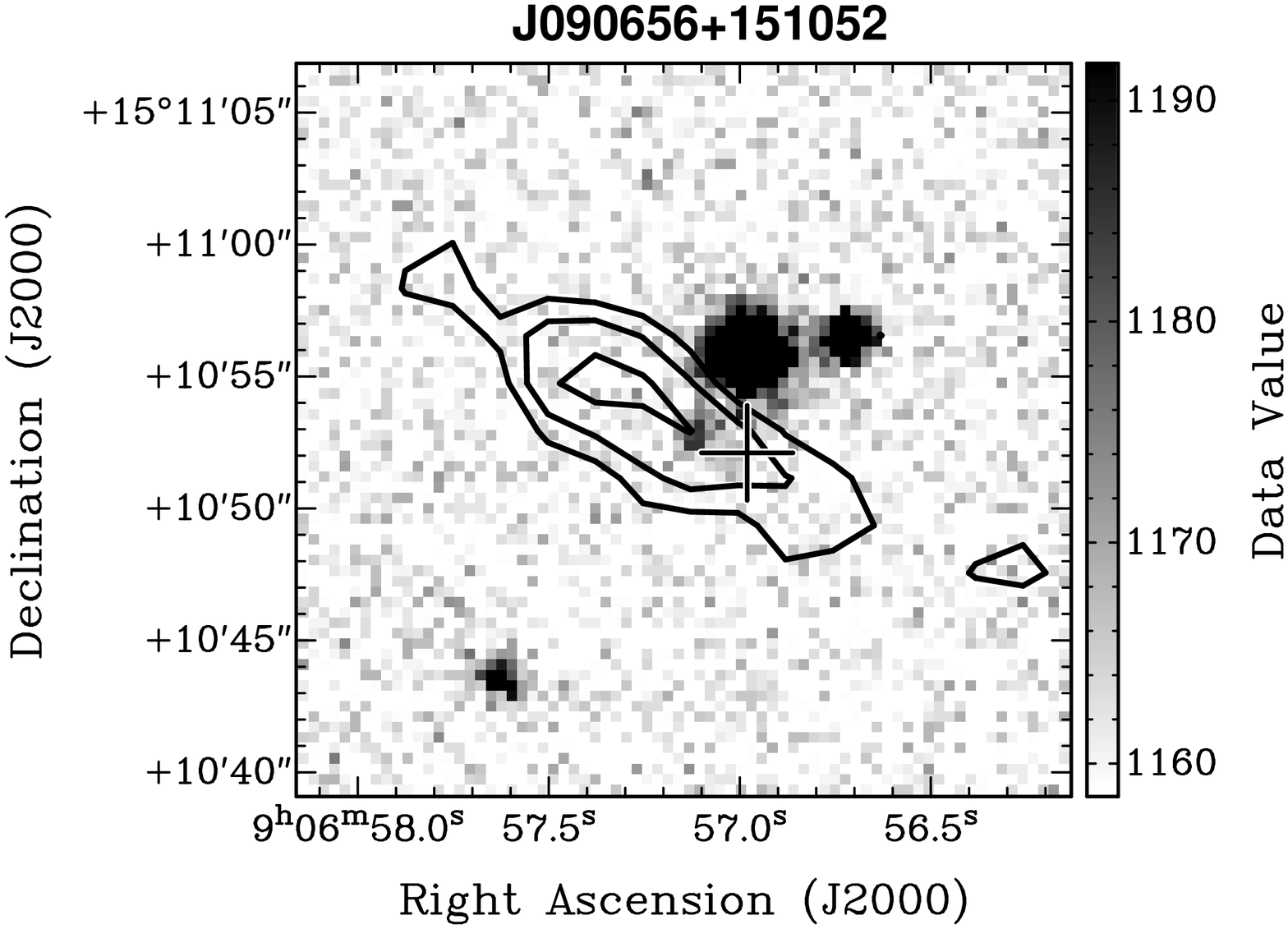}}\\[2mm]
\mbox{\includegraphics[bb=73 86 555 718,angle=270,width=63mm,clip]{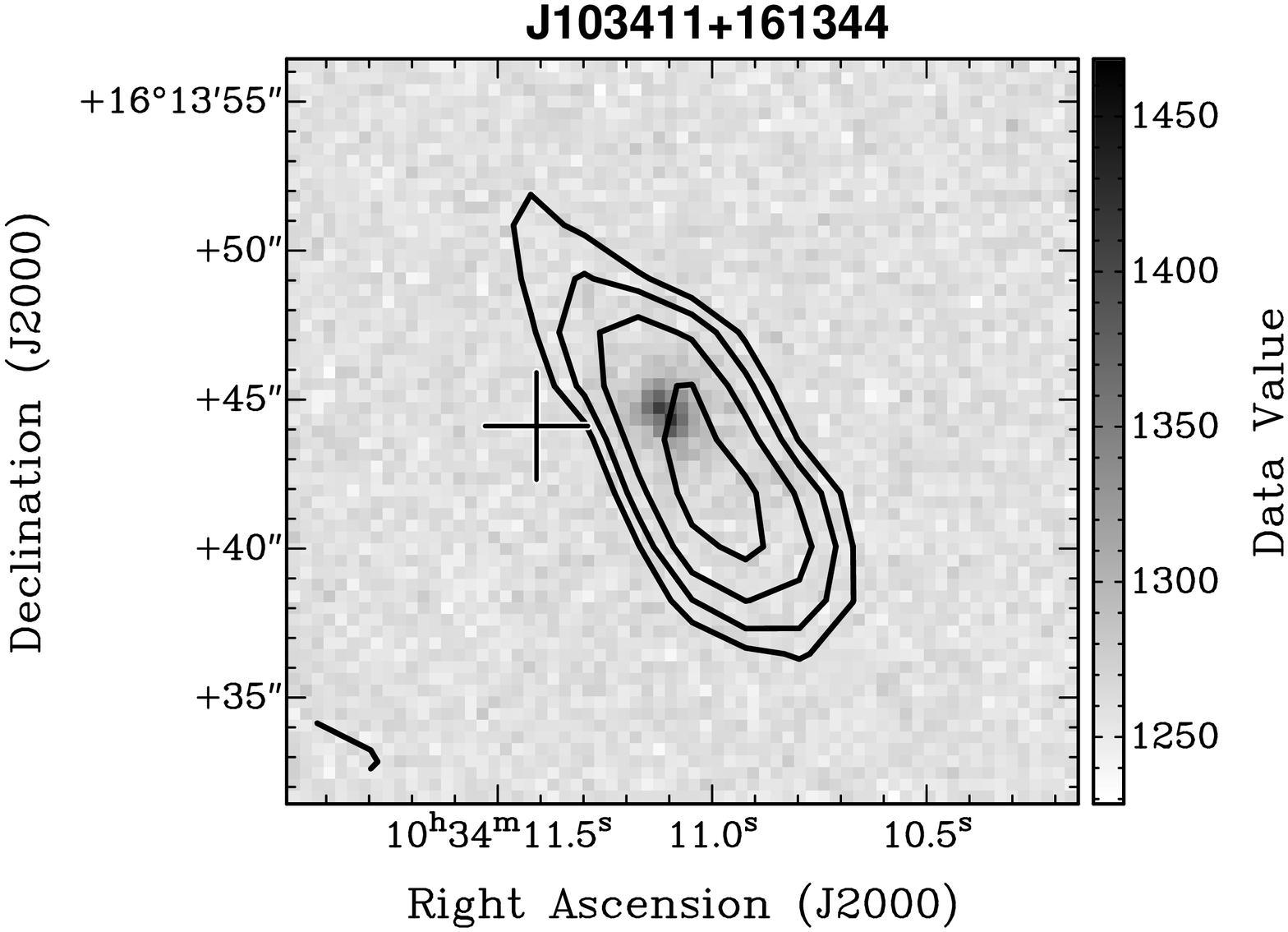}}&
\mbox{\includegraphics[bb=82 125 546 682,angle=270,width=60mm,clip]{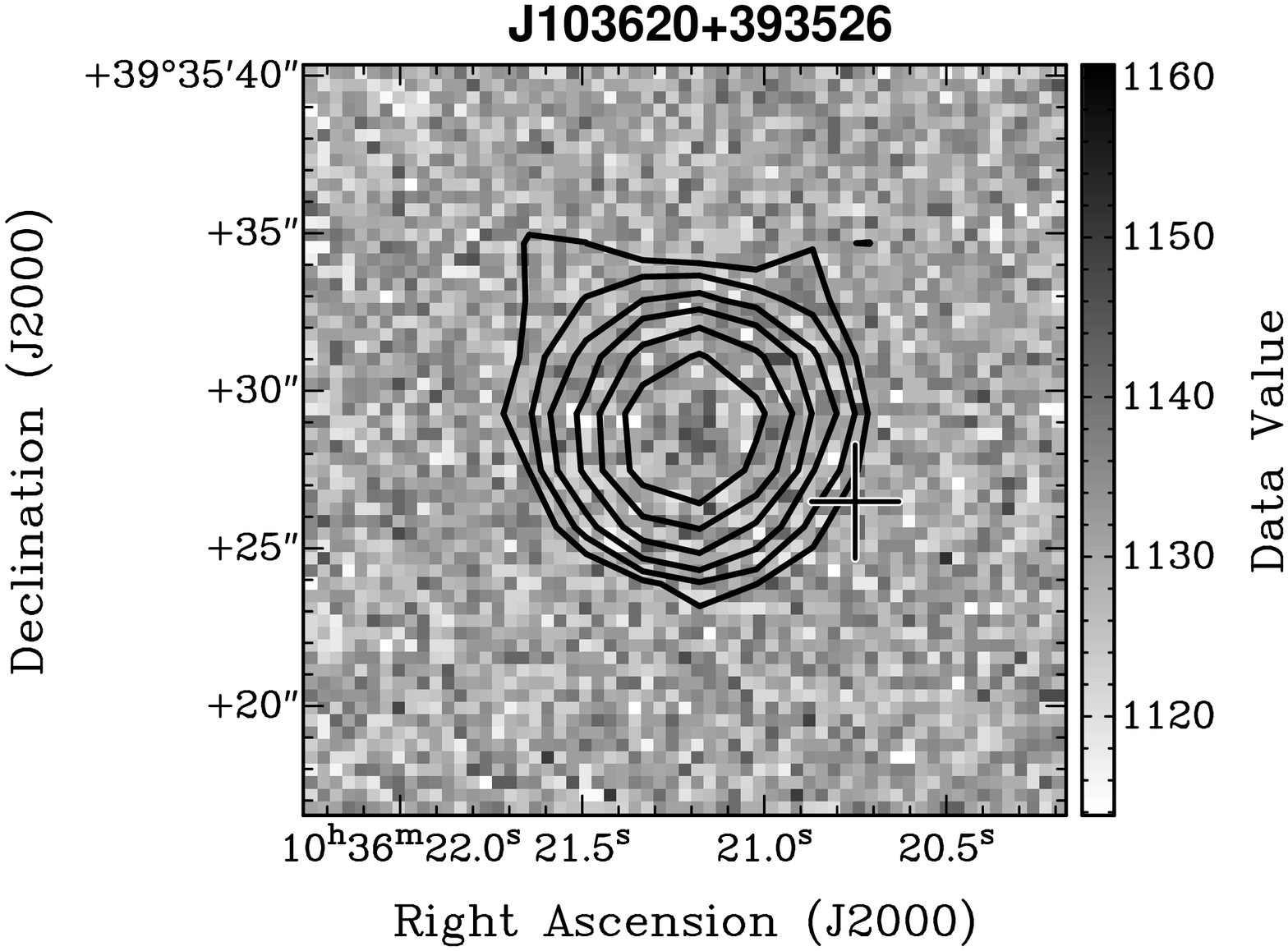}}&
\mbox{\includegraphics[bb=69 125 560 717,angle=270,width=60mm,clip]{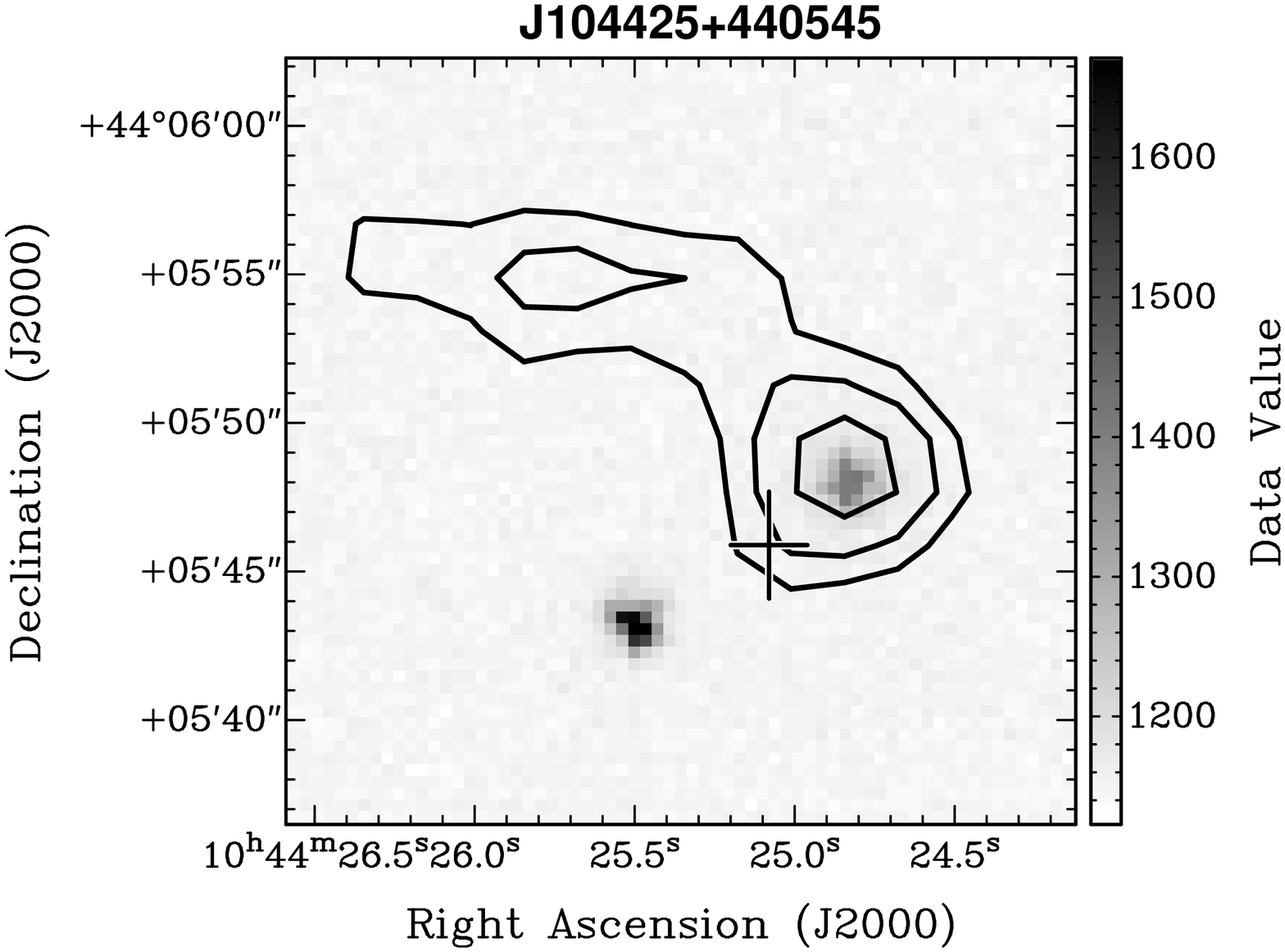}}\\[2mm] 
\mbox{\includegraphics[bb=67 80 560 723,angle=270,width=63mm,clip]{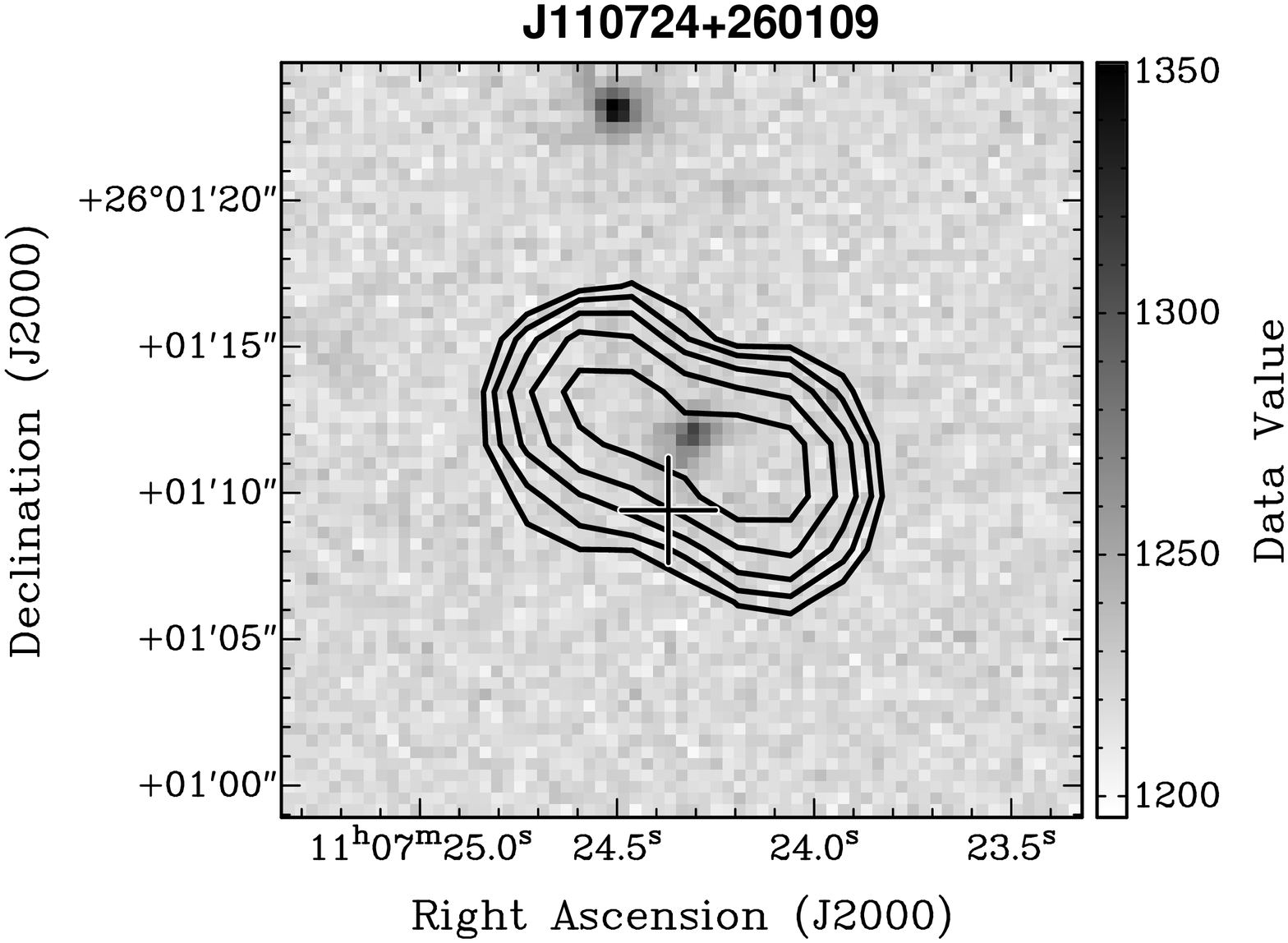}}&
\mbox{\includegraphics[bb=80 128 548 713,angle=270,width=60mm,clip]{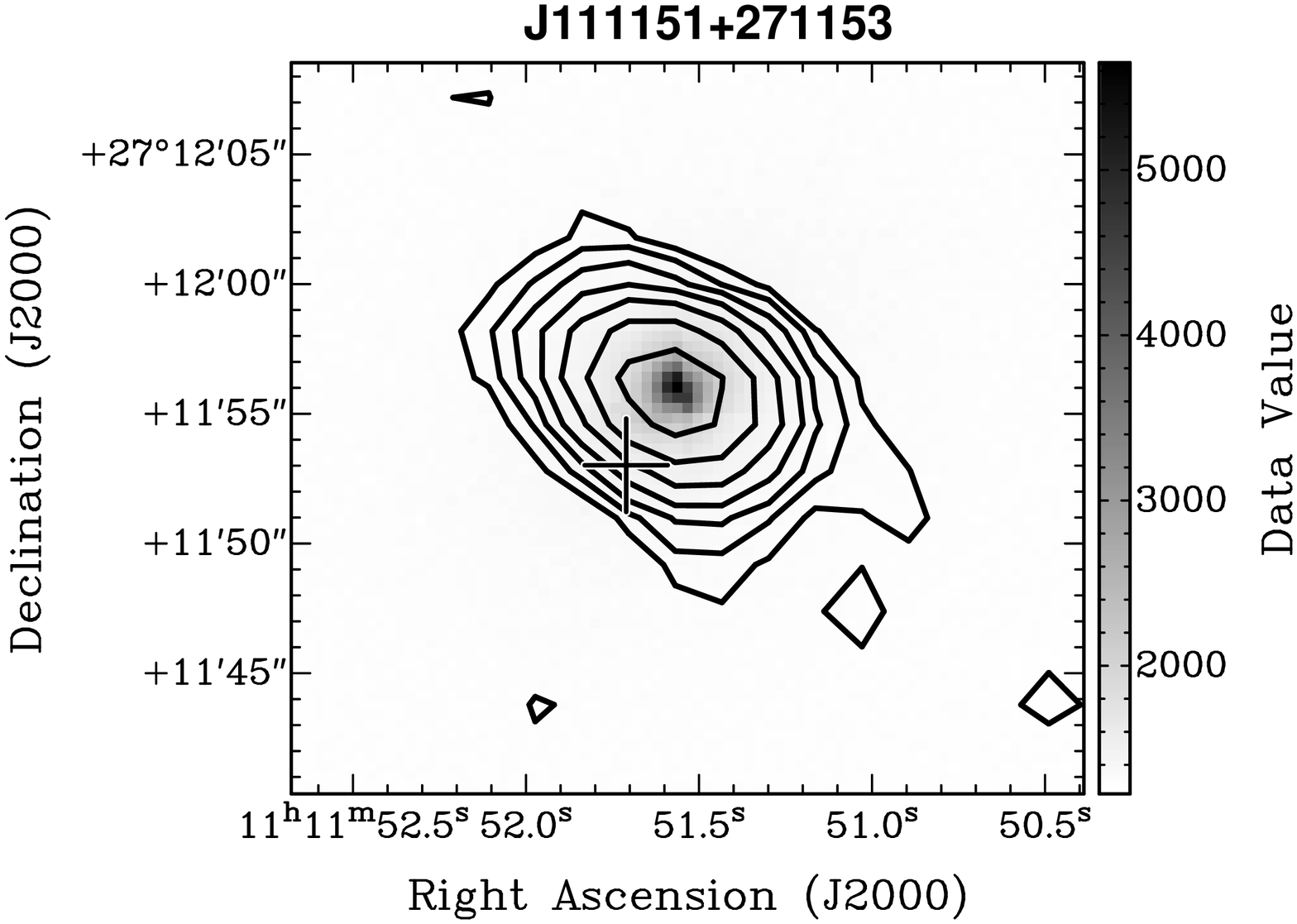}}& 
\mbox{\includegraphics[bb=74 123 554 719,angle=270,width=60mm,clip]{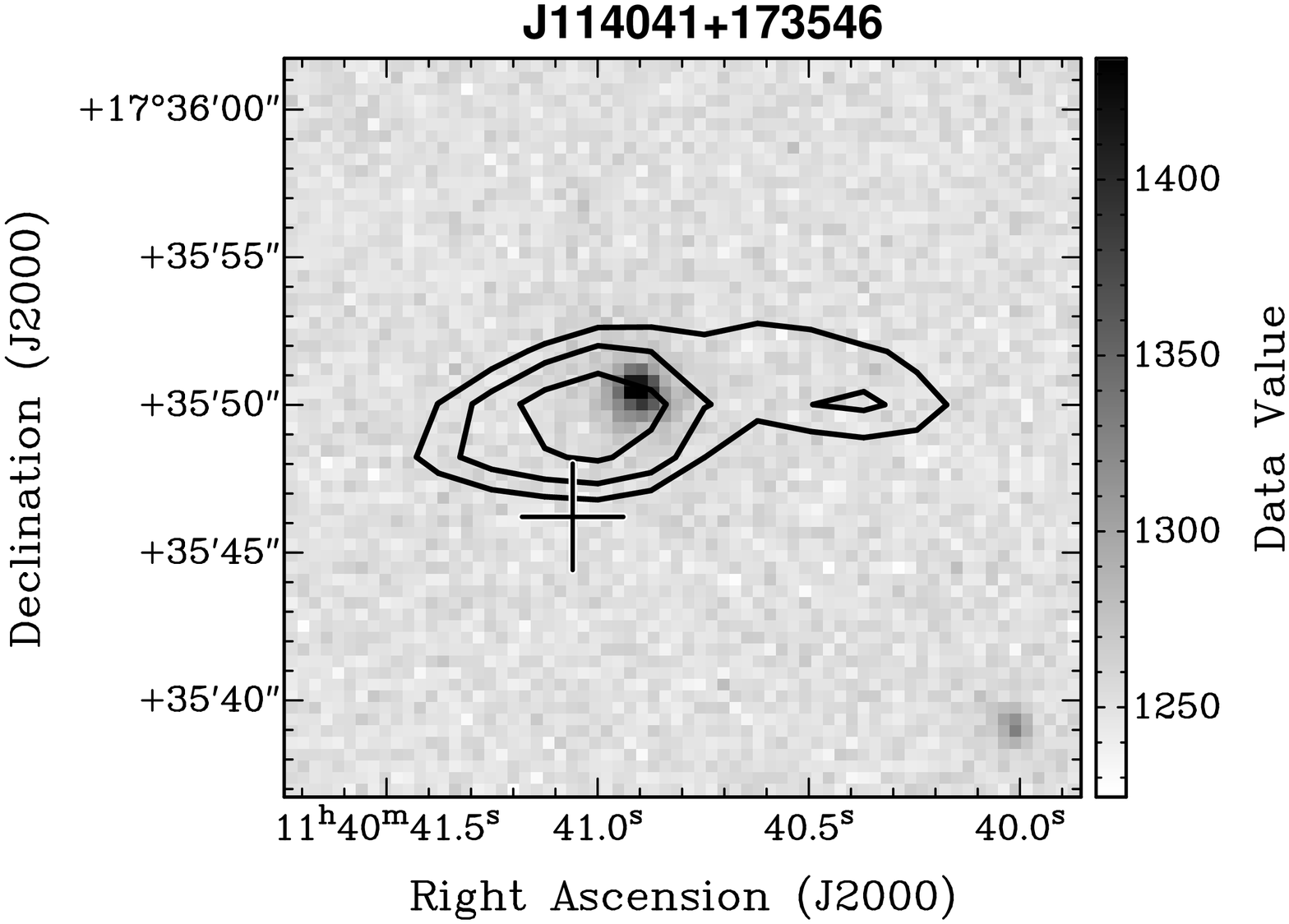}}\\[2mm]
\mbox{\includegraphics[bb=72 85 558 718,angle=270,width=63mm,clip]{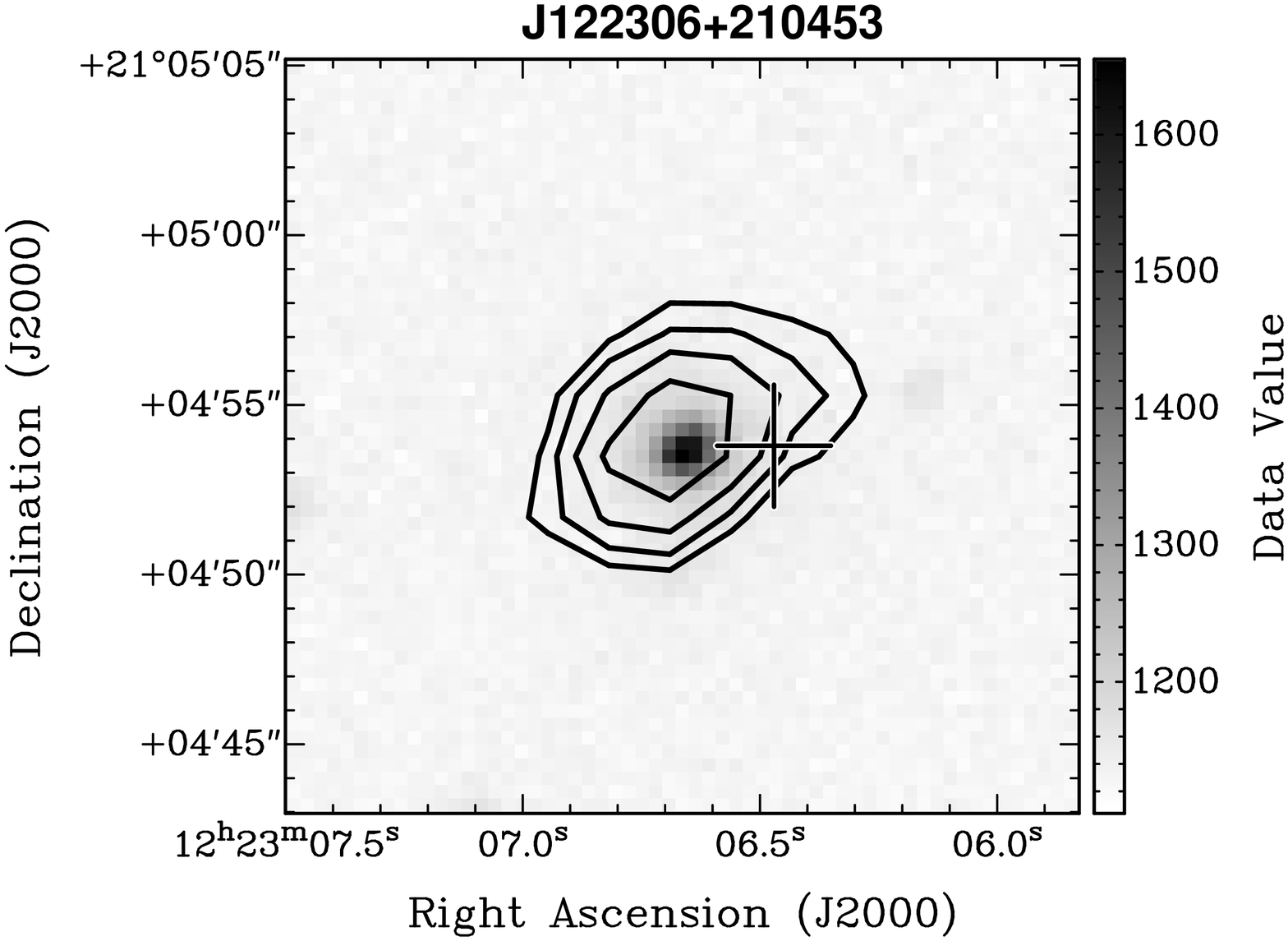}}&
\mbox{\includegraphics[bb=66 122 563 719,angle=270,width=60mm,clip]{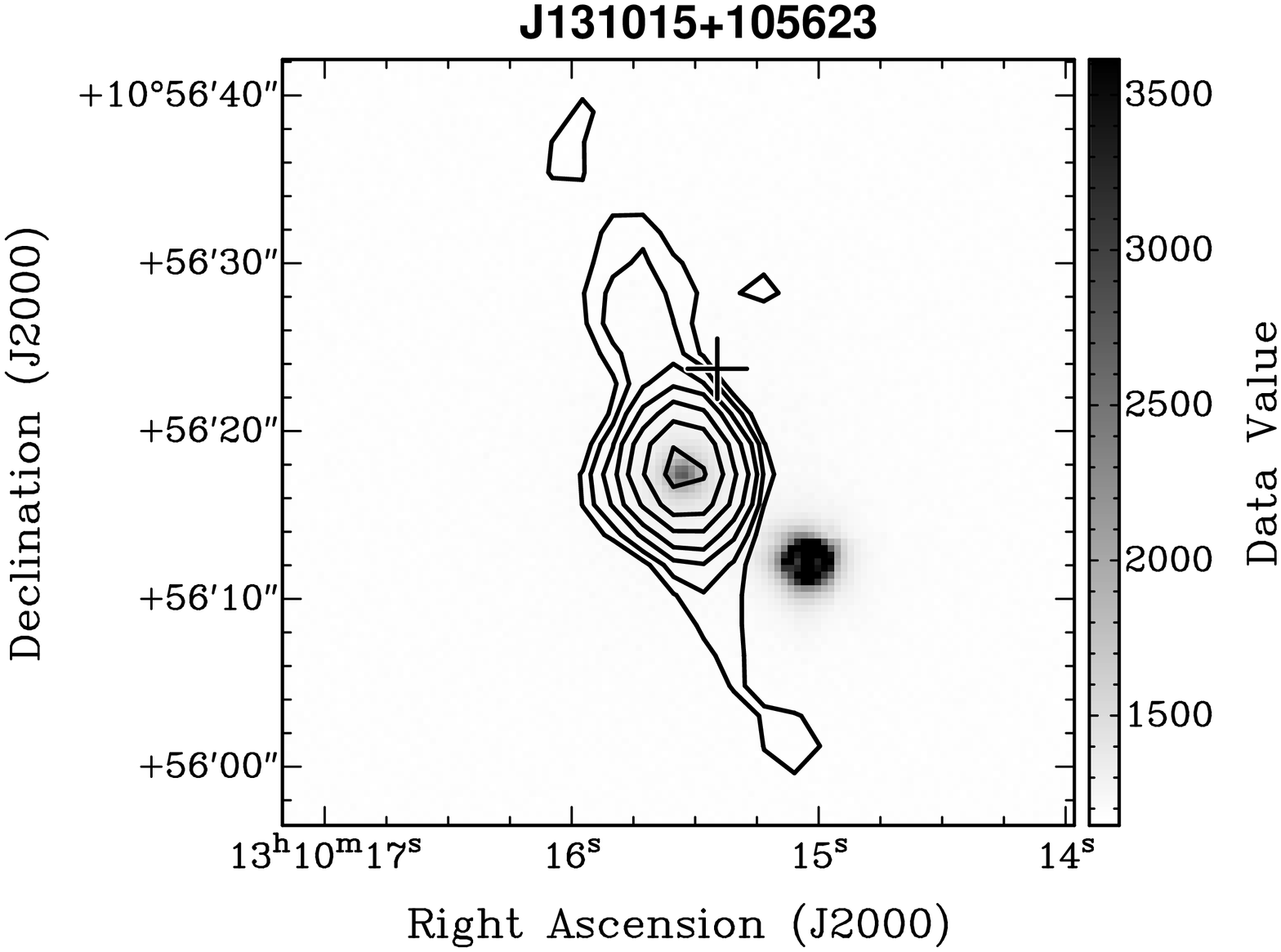}}&
\mbox{\includegraphics[bb=68 119 561 723,angle=270,width=60mm,clip]{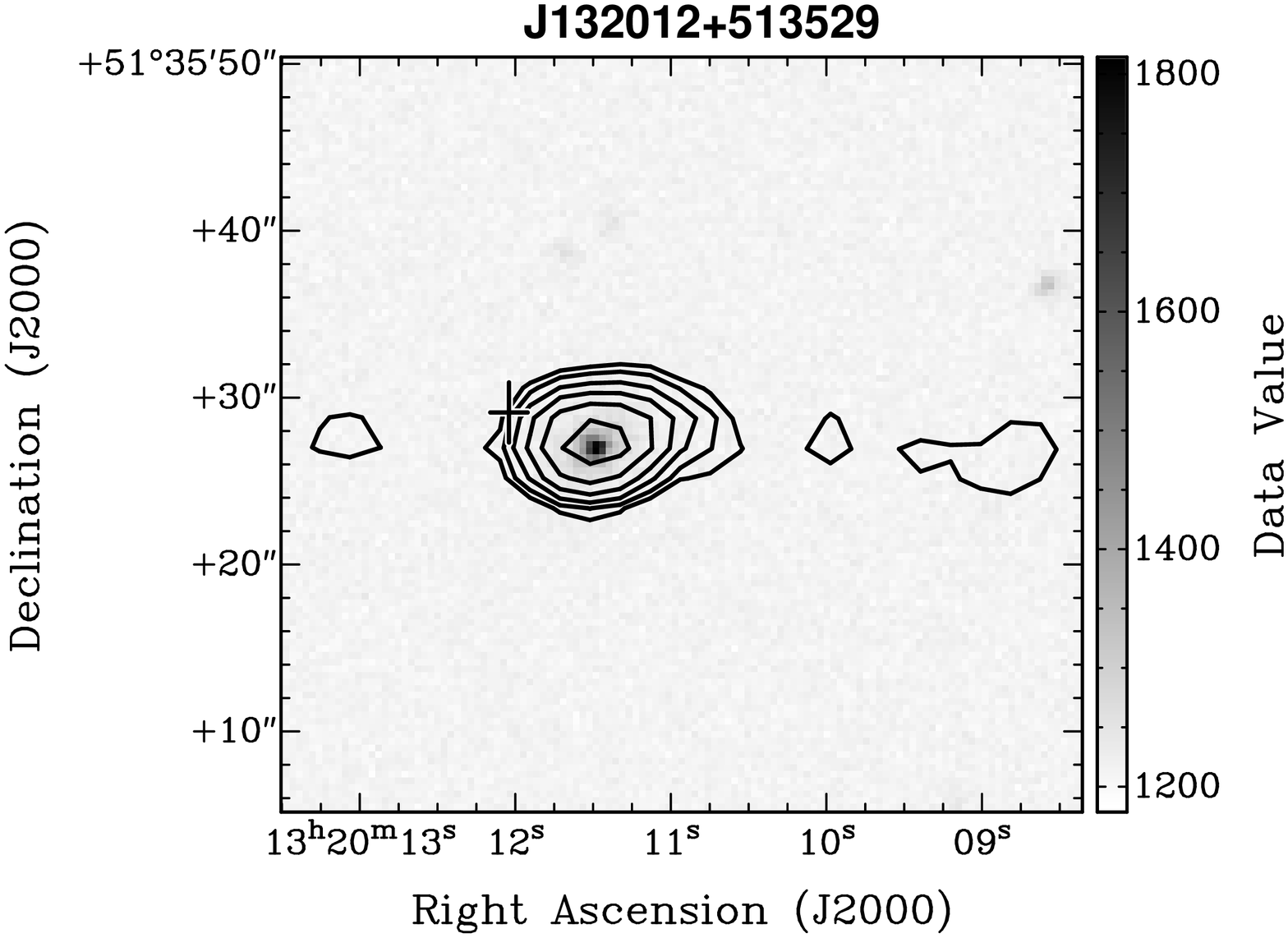}}\\
\end{tabular}%
\caption{FIRST radio contours overlaid on SDSS $i$-band images for the
  22 sources with SDSS identifications (Table~\ref{tab:SDSS}).
  Contours are $\pm 3\sigma \times 2^{n/2}$ ($n=0, 1, 2, 3,$
  \dots\ and the value of $\sigma$ is from the 8th column in
  Table~\ref{tab:highres}). The `+' symbol denotes the position of the
  source detected in the NVSS.}\label{fig:optid}
\end{figure*}%
\begin{figure*}
\begin{tabular}{rrr}
\mbox{\includegraphics[bb=74 84 554 719,angle=270,width=63mm,clip]{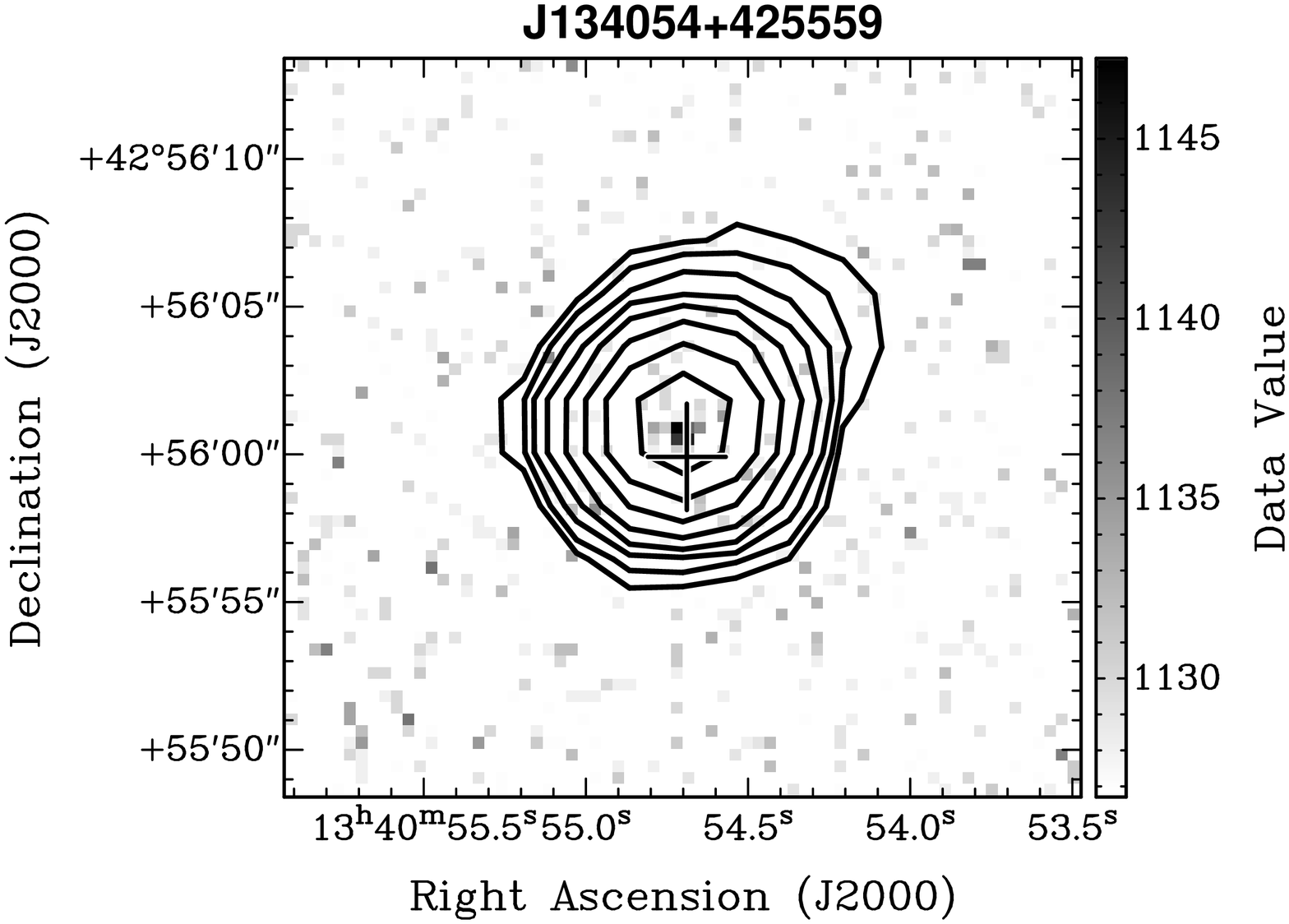}}&
\mbox{\includegraphics[bb=73 129 556 713,angle=270,width=60mm,clip]{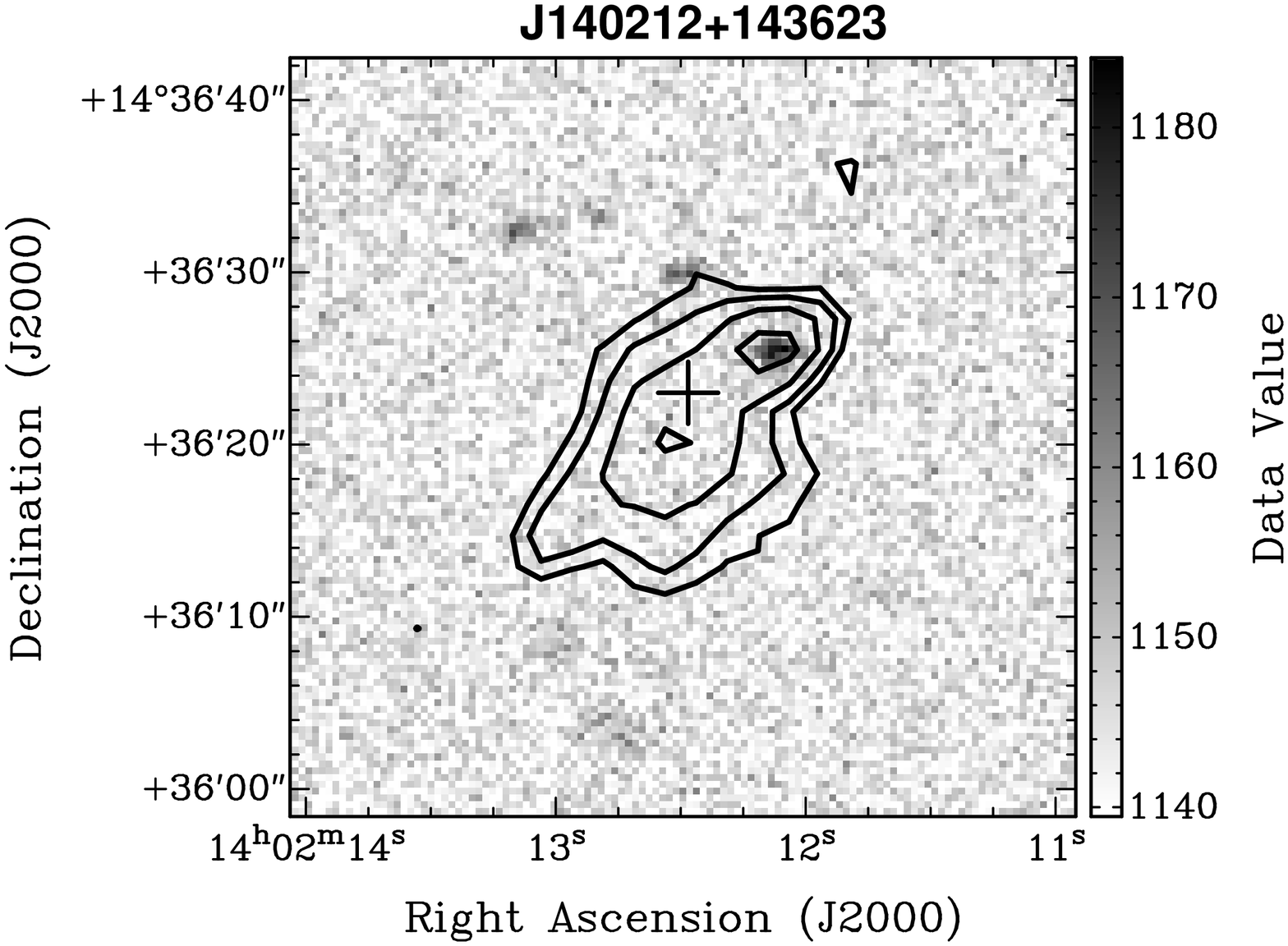}}& 
\mbox{\includegraphics[bb=99 138 532 669,angle=270,width=60mm,clip]{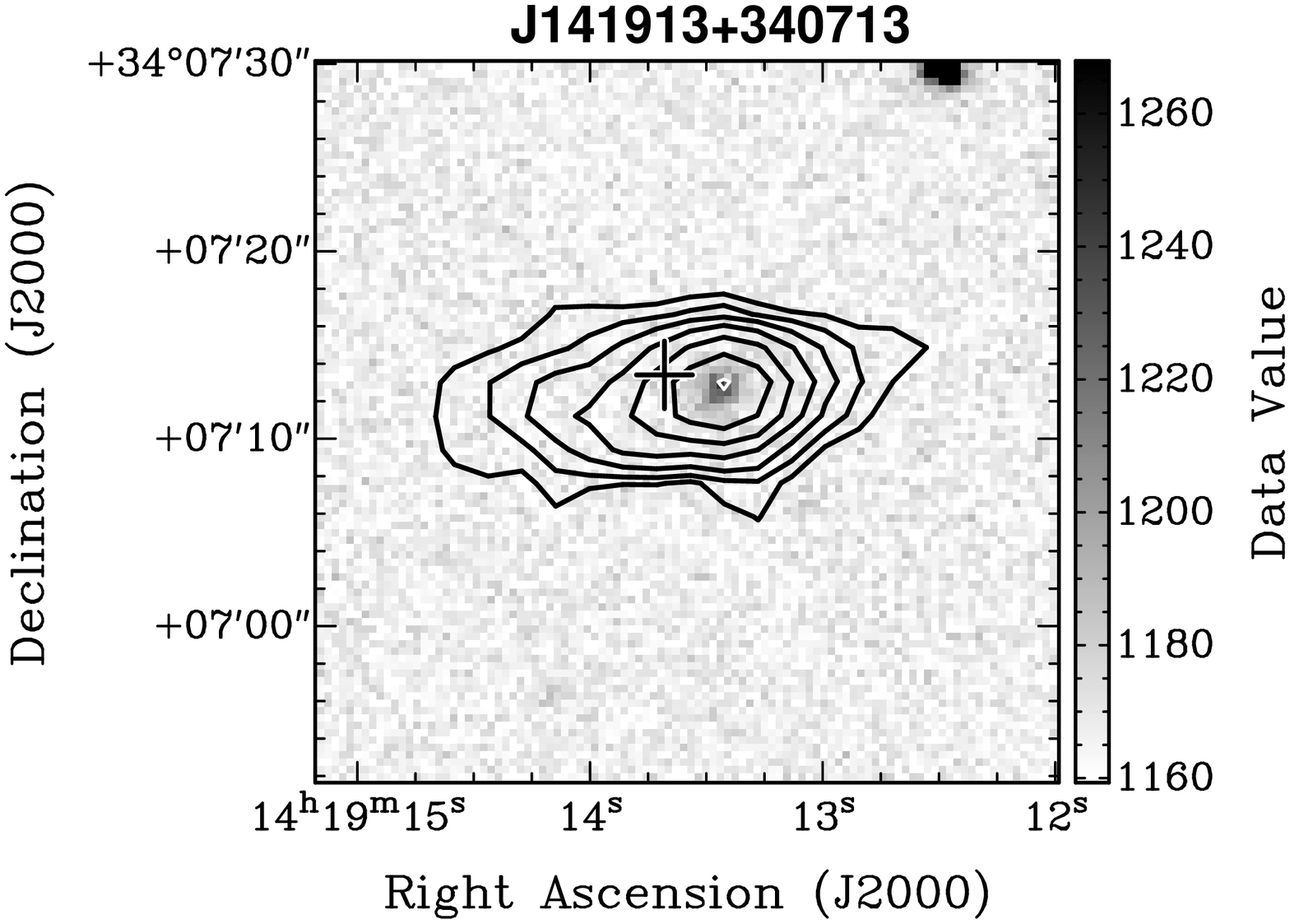}}\\[2mm]
\mbox{\includegraphics[bb=80 95 549 708,angle=270,width=63mm,clip]{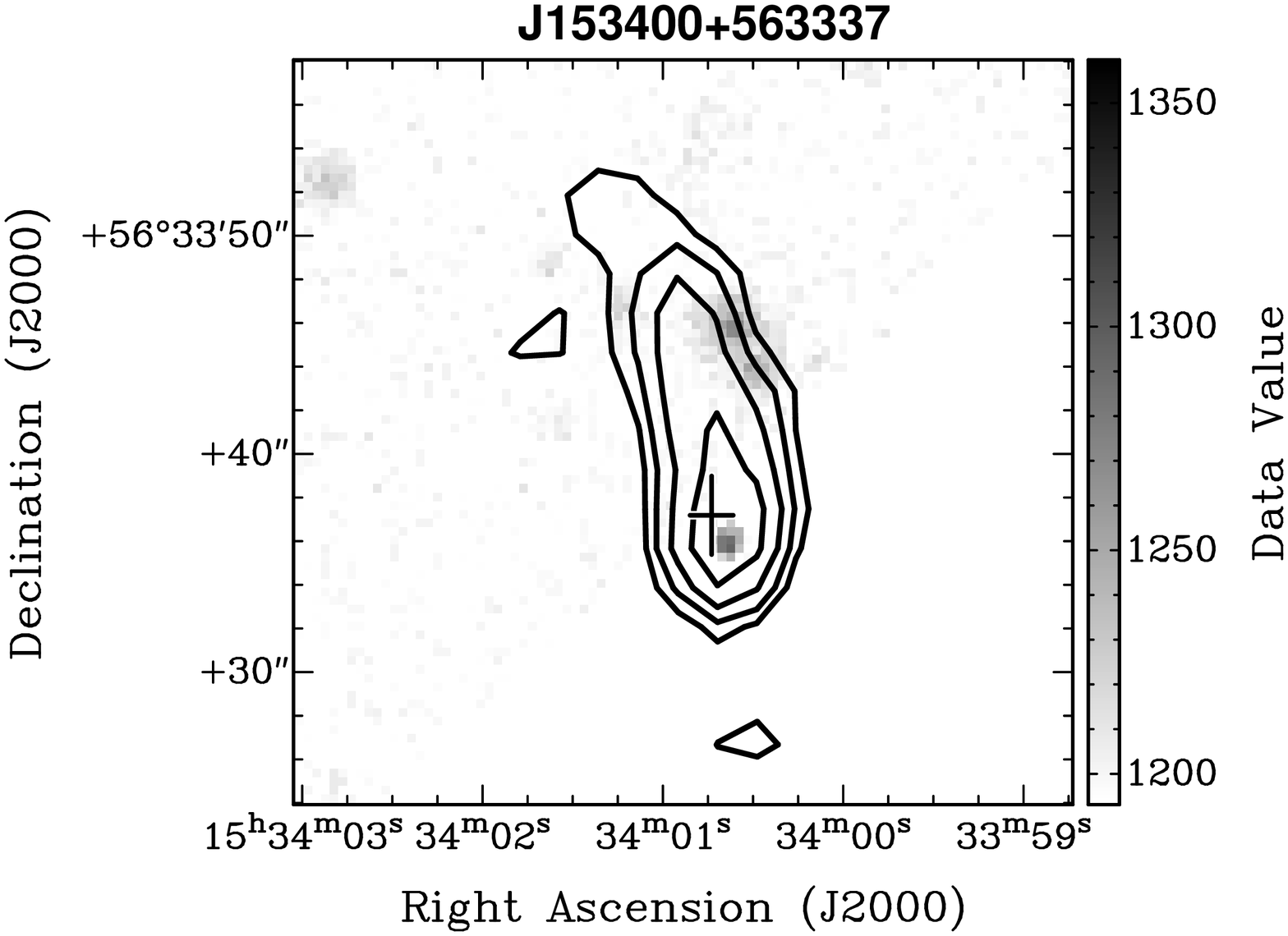}}&
\mbox{\includegraphics[bb=86 123 542 684,angle=270,width=60mm,clip]{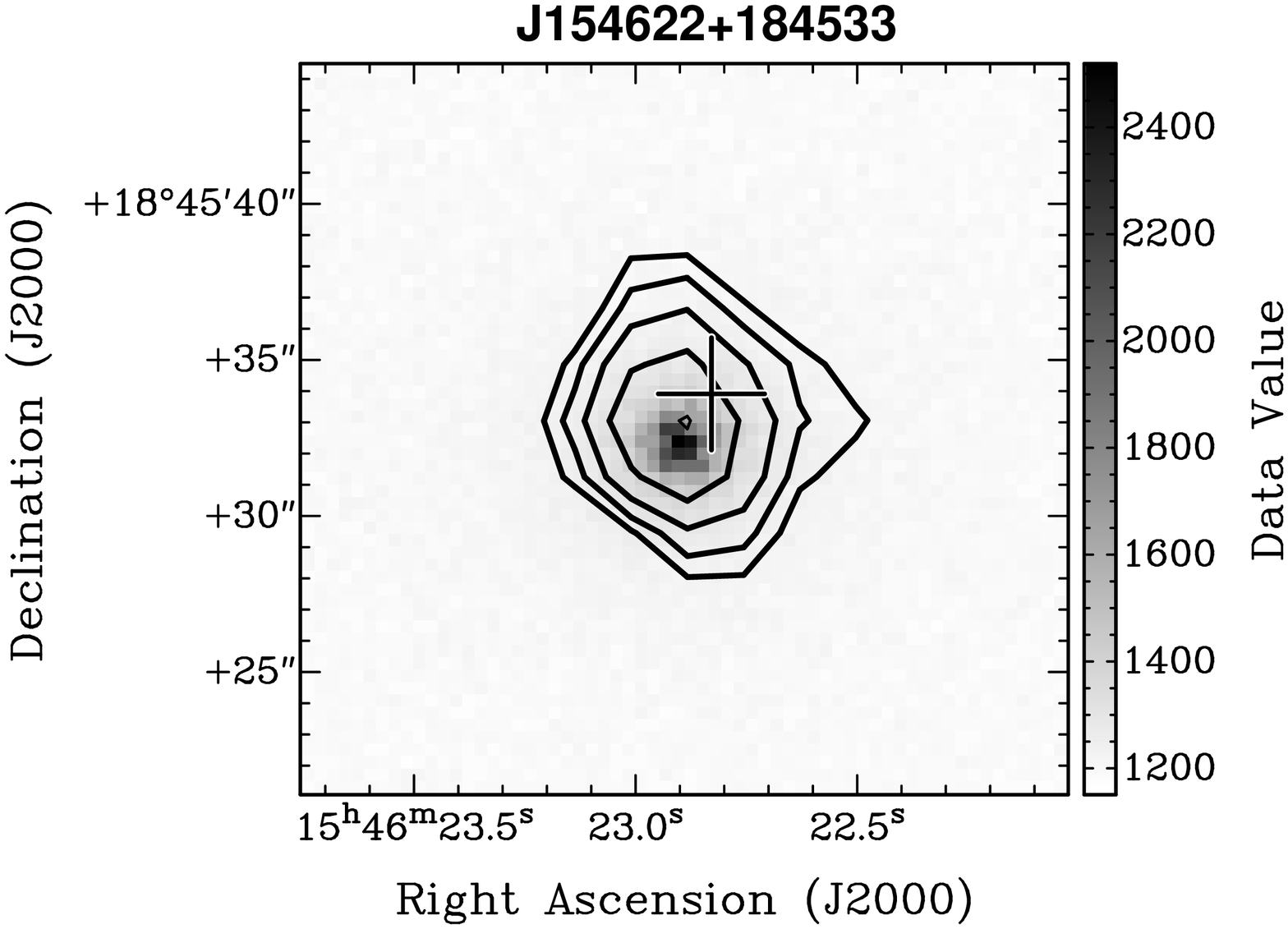}}& 
\mbox{\includegraphics[bb=68 120 560 723,angle=270,width=60mm,clip]{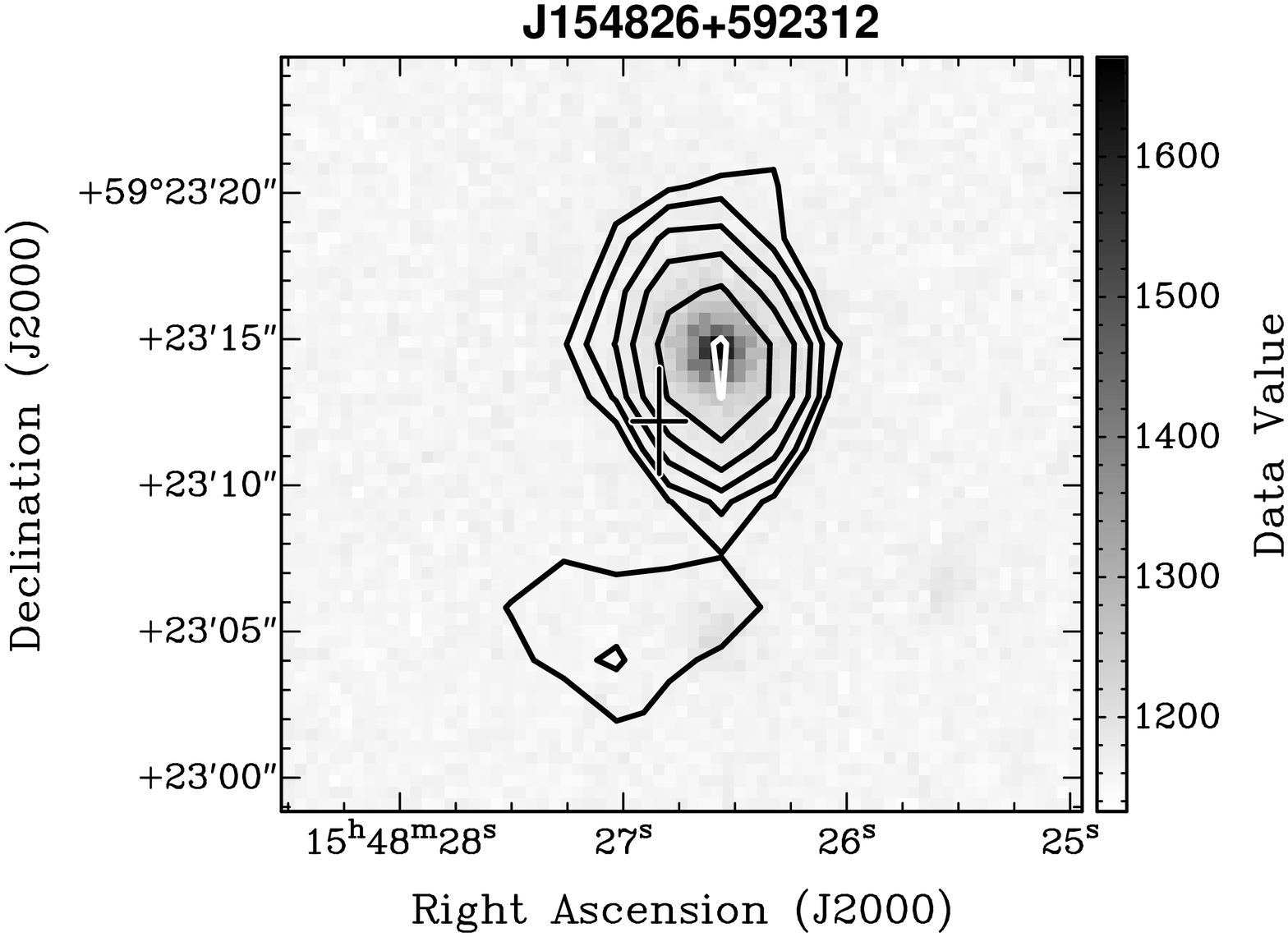}}\\[2mm]
\mbox{\includegraphics[bb=74 89 554 714,angle=270,width=63mm,clip]{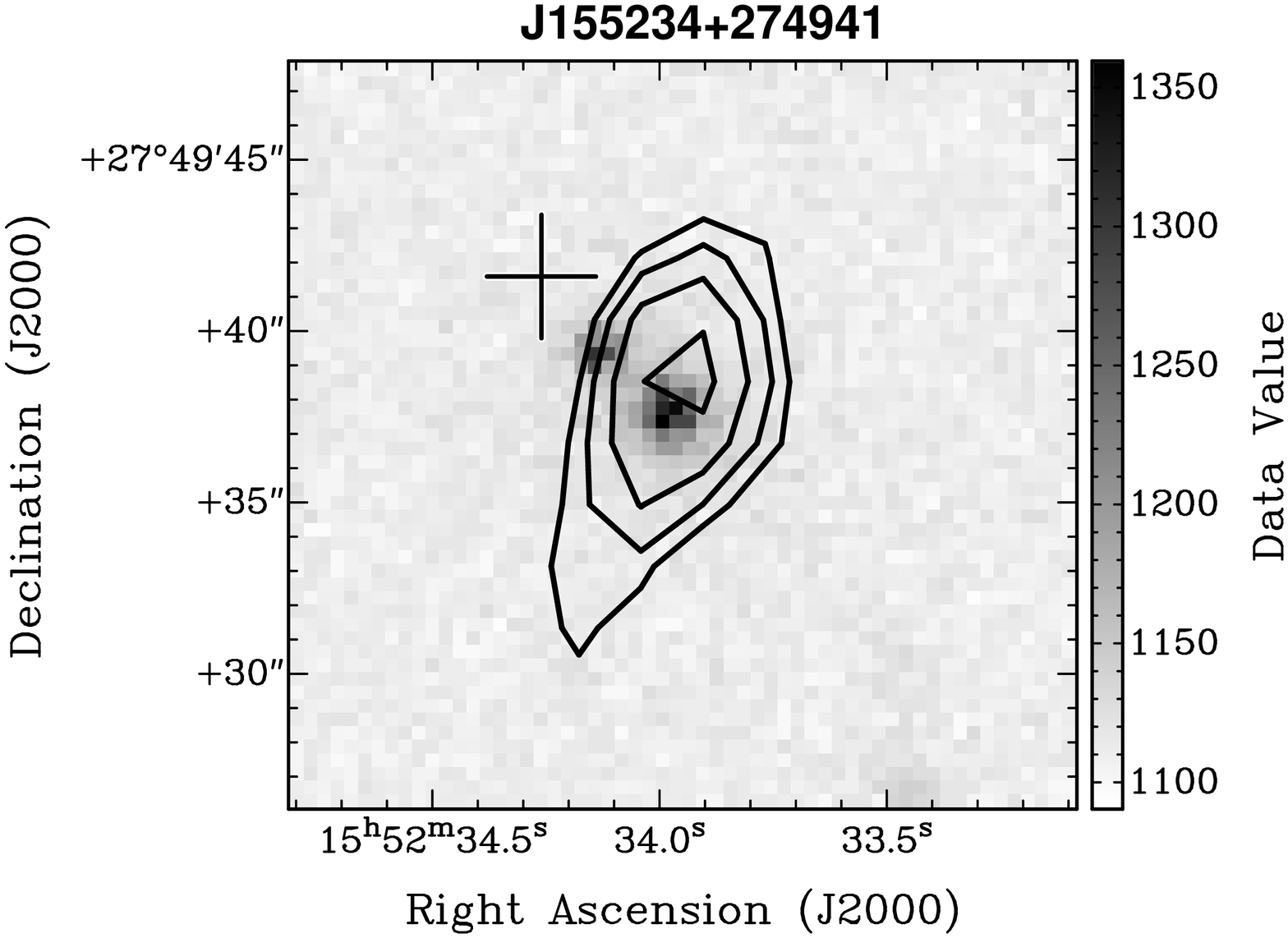}}&
\mbox{\includegraphics[bb=80 121 548 687,angle=270,width=60mm,clip]{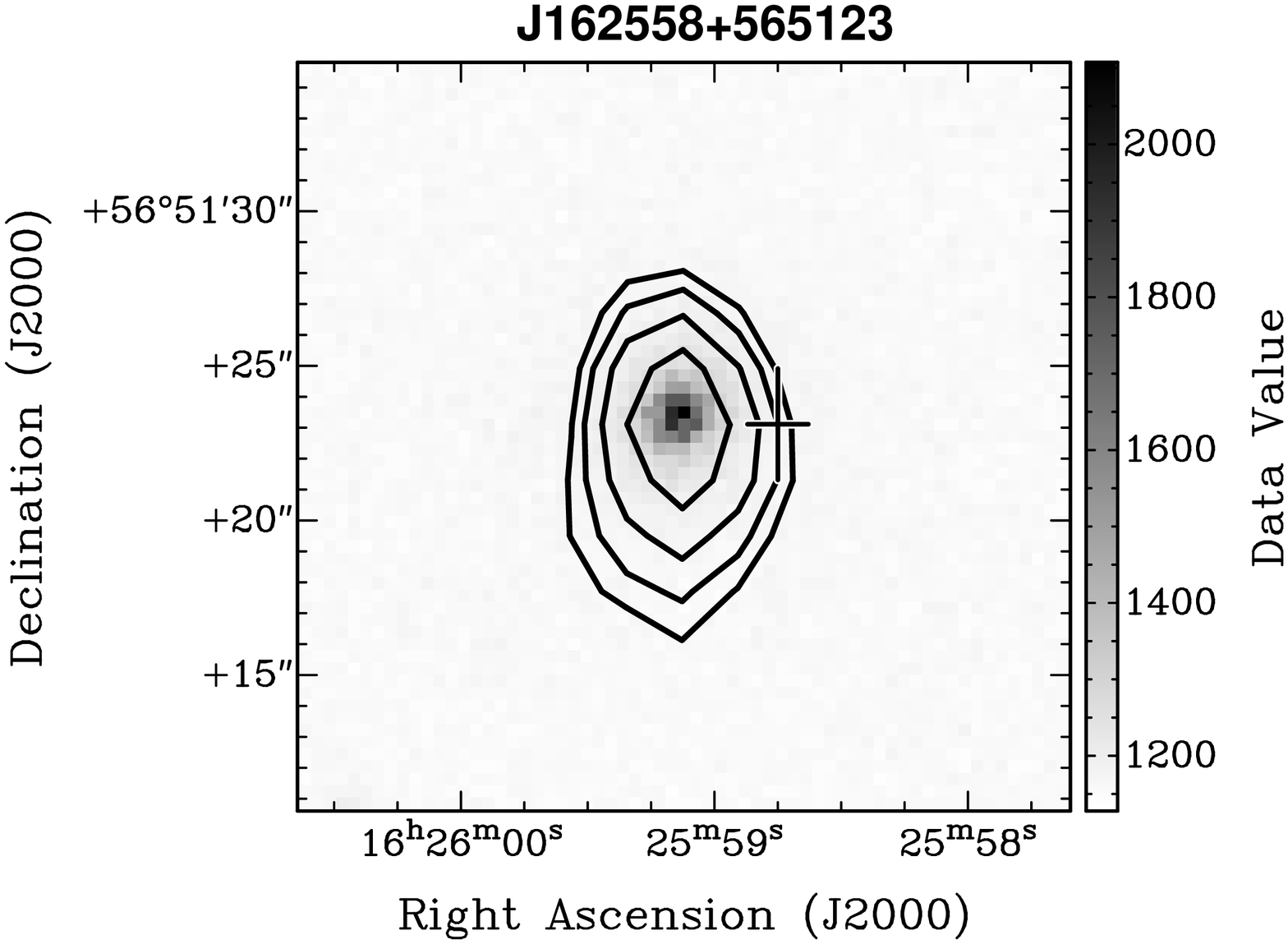}}&
\mbox{\includegraphics[bb=68 122 560 719,angle=270,width=60mm,clip]{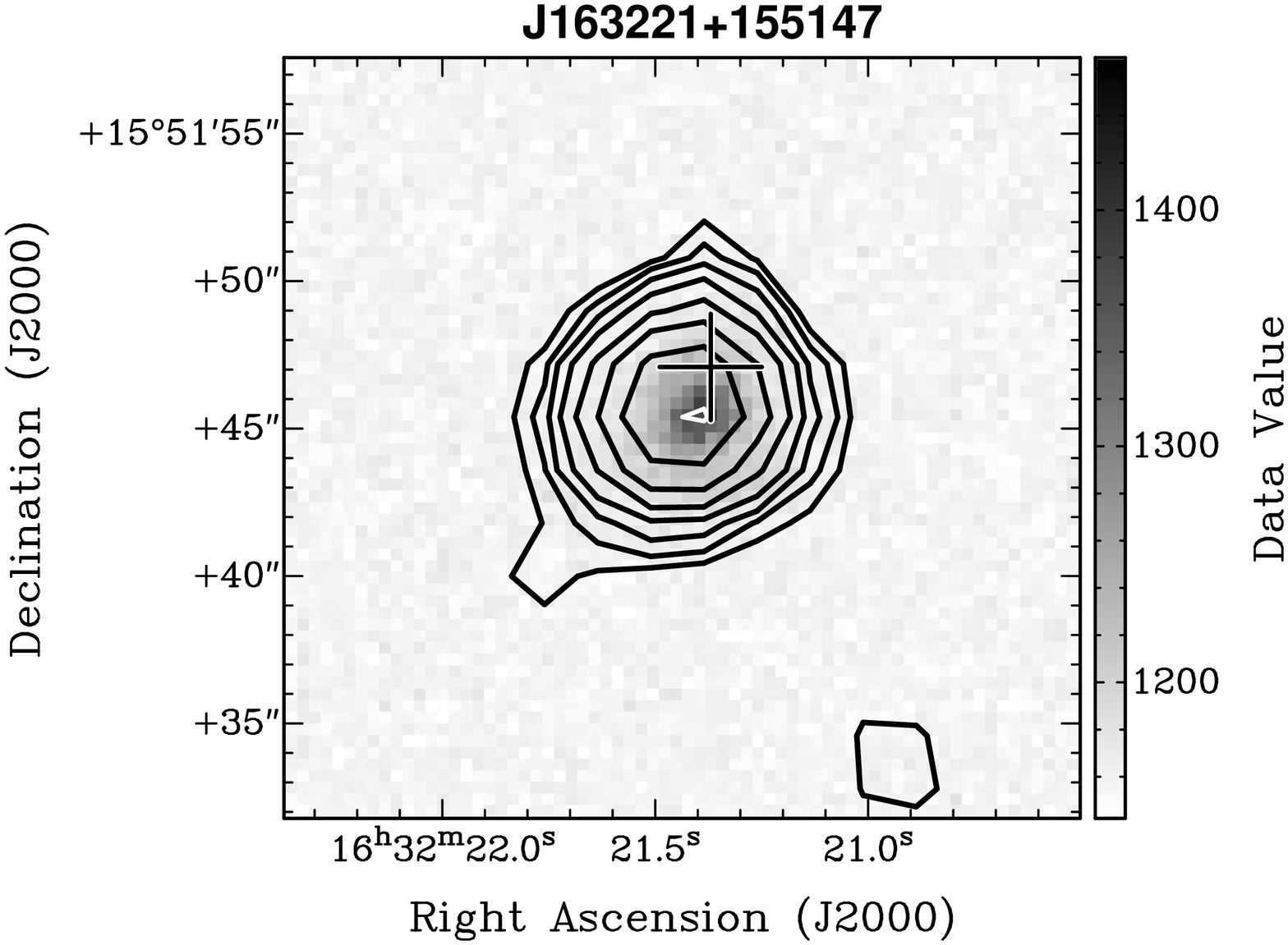}}\\[2mm] 
\mbox{\includegraphics[bb=80 82 548 686,angle=270,width=63mm,clip]{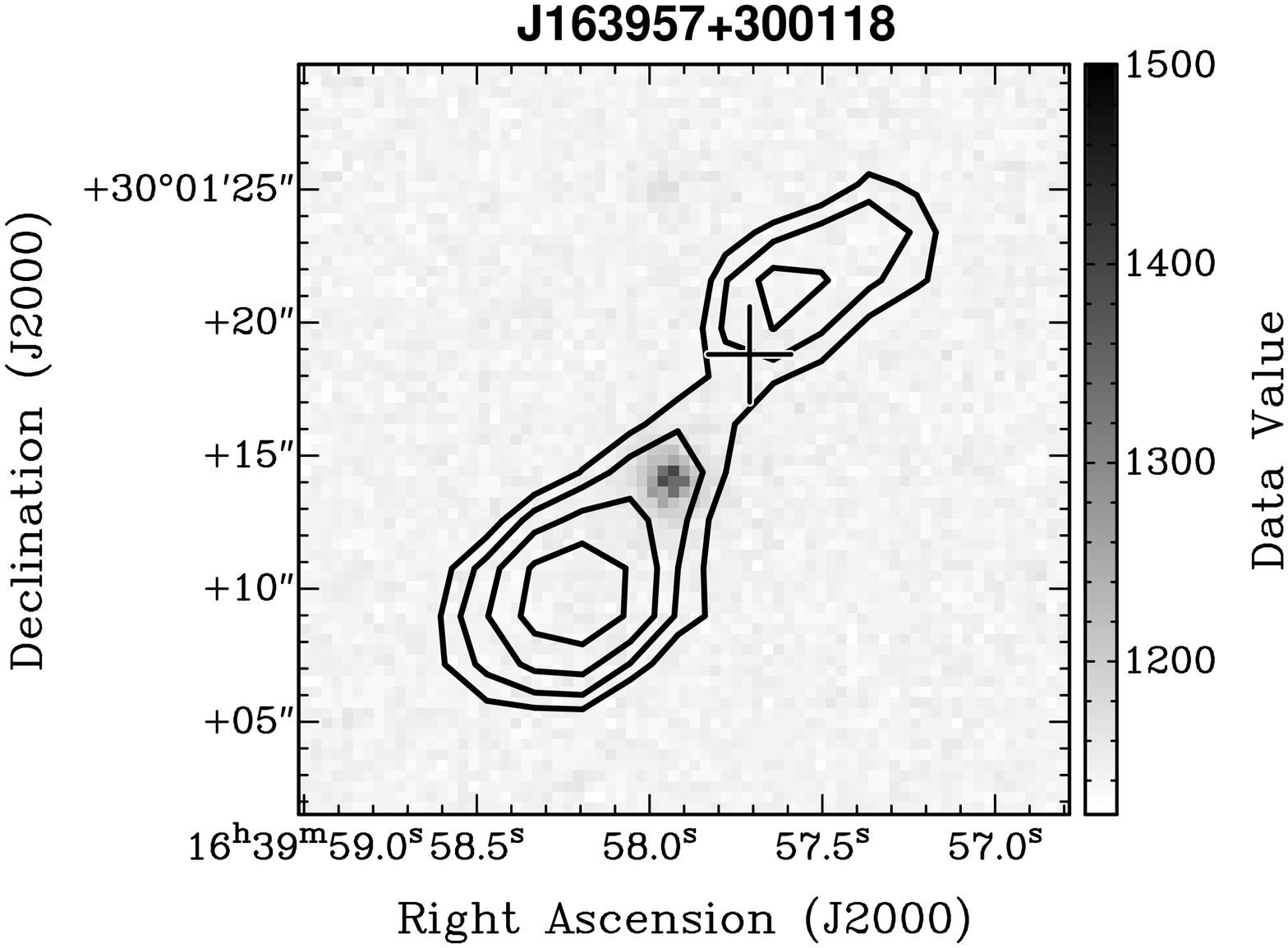}}&
\end{tabular}%
\contcaption{}
\end{figure*}%

%% file: table1.tex
\begin{table*}
\caption{The sample of 129 ultrahigh polarisation radio sources
  selected from the NVSS.}\label{tab:129nvss}
\begin{tabular}{lrrrrlr}
\hline
\hline
\multicolumn{1}{c}{Source}  & 
\multicolumn{1}{c}{$l$} & 
\multicolumn{1}{c}{$b$} &
\multicolumn{1}{c}{$I$} &
\multicolumn{1}{c}{$P$} &
\multicolumn{1}{c}{$\Pi$} &
\multicolumn{1}{c}{PPA} \\
 & 
\multicolumn{1}{c}{$(^\circ)$} & 
\multicolumn{1}{c}{$(^\circ)$} & 
\multicolumn{1}{c}{(mJy)} & 
\multicolumn{1}{c}{(mJy)} & 
\multicolumn{1}{c}{(per cent)} &
\multicolumn{1}{c}{$(^\circ)$} \\
\hline
 J000518+280145&  $110.7$&  $-33.7$&  $ 3.6\pm0.4$&  $ 1.9 \pm0.5 $&  $ 51\pm14$&  $ 12\pm5$ \\
 J000953+003956&  $101.5$&  $-60.5$&  $ 6.8\pm0.5$&  $ 2.7 \pm0.6 $&  $ 40\pm10$&  $-75\pm4$ \\
 J001615$-$143722&  $ 87.9$&  $-75.0$&  $ 5.0\pm0.5$&  $ 2.2 \pm0.5 $&  $ 44\pm11$&  $-73\pm4$ \\
 J003706+515737&  $120.7$&  $-10.8$&  $ 7.3\pm0.5$&  $ 3.0 \pm0.4 $&  $ 41\pm6$&  $ -2\pm3$ \\
 J004742+191412&  $121.7$&  $-43.6$&  $ 6.9\pm0.5$&  $ 2.2 \pm0.5 $&  $ 32\pm7$&  $ 79\pm4$ \\
 J010829+280828&  $127.5$&  $-34.6$&  $ 9.2\pm0.5$&  $ 2.8 \pm0.6 $&  $ 31\pm7$&  $  2\pm4$ \\
 J010907+253341&  $127.9$&  $-37.1$&  $ 5.8\pm0.5$&  $ 1.9 \pm0.5 $&  $ 33\pm9$&  $-17\pm5$ \\
 J013051$-$065547&  $149.6$&  $-67.7$&  $ 5.4\pm0.5$&  $ 1.7 \pm0.4 $&  $ 31\pm8$&  $ 53\pm5$ \\
 J013558$-$022406&  $148.1$&  $-63.1$&  $ 5.6\pm0.5$&  $ 1.8 \pm0.4 $&  $ 32\pm8$&  $-51\pm5$ \\
 J014646$-$255558&  $210.5$&  $-77.6$&  $ 3.8\pm0.5$&  $ 2.0 \pm0.5 $&  $ 52\pm14$&  $ 23\pm5$ \\
 J020338+385942&  $138.0$&  $-21.8$&  $ 6.7\pm0.5$&  $ 2.1 \pm0.5 $&  $ 31\pm8$&  $ 47\pm4$ \\
 J021908+121836&  $154.1$&  $-45.2$&  $ 6.6\pm0.5$&  $ 2.4 \pm0.4 $&  $ 36\pm7$&  $-16\pm4$ \\
 J021933+425512&  $139.6$&  $-17.1$&  $ 3.9\pm0.4$&  $ 1.6 \pm0.4 $&  $ 40\pm11$&  $ 84\pm5$ \\
 J024746+452407&  $143.4$&  $-12.8$&  $ 7.2\pm0.5$&  $ 2.3 \pm0.6 $&  $ 32\pm9$&  $-41\pm4$ \\
 J025936+285952&  $154.0$&  $-26.0$&  $ 7.4\pm0.4$&  $ 2.3 \pm0.5 $&  $ 31\pm6$&  $-51\pm4$ \\
 J030212+153752&  $163.4$&  $-36.8$&  $13.5\pm0.6$&  $ 4.2 \pm0.4 $&  $ 31\pm3$&  $-60\pm2$ \\
 J032305+231722&  $162.5$&  $-27.7$&  $ 6.7\pm0.5$&  $ 2.1 \pm0.5 $&  $ 32\pm7$&  $ 37\pm4$ \\
 J033331$-$242300&  $217.9$&  $-53.6$&  $ 5.5\pm0.5$&  $ 2.4 \pm0.6 $&  $ 44\pm11$&  $ 57\pm4$ \\
 J034925+071637&  $180.9$&  $-34.9$&  $ 5.4\pm0.5$&  $ 1.7 \pm0.5 $&  $ 32\pm9$&  $ 45\pm5$ \\
 J035815$-$303129&  $229.2$&  $-49.2$&  $ 6.6\pm0.5$&  $ 2.1 \pm0.4 $&  $ 32\pm7$&  $-41\pm4$ \\
 J035853+541315{*}& $148.2$&$ 0.8$&  $10.0\pm0.5$&  $ 3.5 \pm0.5 $&  $ 35\pm5$&  $ 19\pm3$ \\
 J040245+431320&  $155.9$&  $ -7.1$&  $ 4.0\pm0.4$&  $ 1.6 \pm0.4 $&  $ 39\pm11$&  $-29\pm5$ \\
 J042204+363943&  $163.2$&  $ -9.3$&  $ 9.3\pm0.5$&  $ 2.9 \pm0.6 $&  $ 31\pm6$&  $ 58\pm4$ \\
 J043945+500319&  $155.6$&  $  2.2$&  $18.4\pm0.7$&  $ 6.4 \pm0.5 $&  $ 35\pm3$&  $ 37\pm1$ \\
 J054826+621008&  $150.9$&  $ 16.9$&  $ 6.8\pm0.5$&  $ 2.1 \pm0.5 $&  $ 31\pm8$&  $ 74\pm4$ \\
 J062619$-$260934&  $234.2$&  $-16.8$&  $ 9.8\pm0.6$&  $ 3.0 \pm0.6 $&  $ 31\pm6$&  $ 10\pm3$ \\
 J064226+452259&  $170.3$&  $ 17.5$&  $16.4\pm0.6$&  $ 5.1 \pm0.5 $&  $ 31\pm3$&  $-54\pm2$ \\
 J072803$-$030521&  $219.9$&  $  6.7$&  $ 8.3\pm0.5$&  $ 2.6 \pm0.5 $&  $ 31\pm7$&  $ 51\pm4$ \\
 J073904+180421&  $201.6$&  $ 18.4$&  $ 4.4\pm0.5$&  $ 2.7 \pm0.7 $&  $ 60\pm17$&  $ 86\pm4$ \\
 J074213+605759&  $155.9$&  $ 29.6$&  $12.1\pm0.6$&  $ 3.7 \pm0.5 $&  $ 30\pm4$&  $ 55\pm2$ \\
 J080351+820954&  $131.5$&  $ 29.3$&  $ 5.1\pm0.5$&  $ 1.7 \pm0.4 $&  $ 33\pm9$&  $ 88\pm5$ \\
 J080840+212610&  $201.0$&  $ 26.1$&  $ 5.3\pm0.4$&  $ 1.7 \pm0.4 $&  $ 32\pm8$&  $-22\pm5$ \\
 J081130$-$014548&  $224.0$&  $ 16.9$&  $ 5.7\pm0.5$&  $ 2.2 \pm0.4 $&  $ 38\pm8$&  $-68\pm4$ \\
 J081458+742905{*}& $140.0$&$ 31.6$&  $ 7.2\pm0.5$&  $ 2.7 \pm0.5 $&  $ 37\pm8$&  $-45\pm4$ \\
 J082124+175736&  $205.9$&  $ 27.7$&  $ 3.7\pm0.4$&  $ 1.7 \pm0.4 $&  $ 46\pm13$&  $-90\pm5$ \\
 J083035$-$312745&  $252.0$&  $  4.6$&  $ 6.5\pm0.5$&  $ 2.1 \pm0.4 $&  $ 33\pm7$&  $ 53\pm4$ \\
 J085104$-$142446&  $240.7$&  $ 18.4$&  $11.4\pm0.6$&  $ 4.0 \pm0.7 $&  $ 35\pm6$&  $-37\pm3$ \\
 J085930+510912&  $167.4$&  $ 40.6$&  $ 5.8\pm0.5$&  $ 2.0 \pm0.5 $&  $ 34\pm9$&  $-26\pm4$ \\
 J090436+303246&  $194.7$&  $ 40.7$&  $ 5.6\pm0.5$&  $ 2.1 \pm0.5 $&  $ 37\pm9$&  $ 16\pm4$ \\
 J090522+271728&  $199.0$&  $ 40.2$&  $ 8.0\pm0.5$&  $ 2.8 \pm0.5 $&  $ 36\pm7$&  $ 45\pm3$ \\
 J090656+151052&  $213.8$&  $ 36.7$&  $ 5.1\pm0.5$&  $ 2.0 \pm0.5 $&  $ 40\pm11$&  $-25\pm4$ \\
 J091048+312725&  $193.8$&  $ 42.2$&  $ 5.7\pm0.5$&  $ 3.4 \pm0.5 $&  $ 60\pm10$&  $ 55\pm3$ \\
 J091337+652628&  $148.7$&  $ 39.0$&  $ 7.2\pm0.5$&  $ 2.6 \pm0.5 $&  $ 36\pm8$&  $-78\pm3$ \\
 J091958+615934&  $152.6$&  $ 40.9$&  $ 4.6\pm0.4$&  $ 1.4 \pm0.4 $&  $ 31\pm9$&  $ 43\pm5$ \\
 J092214+063824{*}& $225.4$&$ 36.4$&  $ 6.4\pm0.5$&  $ 3.8 \pm0.5 $&  $ 59\pm8$&  $ 41\pm2$ \\
 J094615$-$311406&  $262.7$&  $ 16.8$&  $ 7.7\pm0.5$&  $ 2.5 \pm0.5 $&  $ 32\pm6$&  $-66\pm4$ \\
 J094818+685735&  $142.7$&  $ 40.4$&  $ 7.4\pm0.5$&  $ 2.5 \pm0.5 $&  $ 33\pm7$&  $ 36\pm4$ \\
 J100110$-$015230&  $241.3$&  $ 39.8$&  $ 7.2\pm0.5$&  $ 2.2 \pm0.6 $&  $ 30\pm8$&  $-52\pm4$ \\
 J100133$-$191624&  $256.7$&  $ 28.0$&  $ 8.3\pm0.6$&  $ 2.6 \pm0.6 $&  $ 31\pm7$&  $  3\pm4$ \\
 J100556+655946&  $144.7$&  $ 43.5$&  $11.3\pm0.5$&  $ 3.7 \pm0.4 $&  $ 33\pm4$&  $-45\pm2$ \\
 J102758$-$122116&  $256.9$&  $ 37.4$&  $ 6.9\pm0.5$&  $ 3.1 \pm0.5 $&  $ 45\pm8$&  $-45\pm3$ \\
 J103411+161344&  $224.7$&  $ 56.4$&  $ 4.6\pm0.5$&  $ 1.8 \pm0.5 $&  $ 39\pm11$&  $ 47\pm5$ \\
 J103601+050714&  $241.1$&  $ 51.0$&  $ 7.0\pm0.5$&  $ 3.0 \pm0.4 $&  $ 43\pm7$&  $ 48\pm3$ \\
 J103620+393526&  $180.2$&  $ 59.4$&  $ 5.5\pm0.4$&  $ 1.7 \pm0.5 $&  $ 31\pm9$&  $ 16\pm5$ \\
 J104425+440545&  $170.9$&  $ 59.4$&  $ 4.5\pm0.4$&  $ 1.8 \pm0.5 $&  $ 40\pm11$&  $-84\pm5$ \\
 J110724+260109&  $209.7$&  $ 66.8$&  $ 5.2\pm0.4$&  $ 1.9 \pm0.4 $&  $ 37\pm8$&  $ 70\pm4$ \\
 J111151+271153&  $207.0$&  $ 67.9$&  $ 9.2\pm0.5$&  $ 2.8 \pm0.6 $&  $ 31\pm6$&  $ 64\pm3$ \\
 J112158+232928&  $218.2$&  $ 69.5$&  $ 3.3\pm0.4$&  $ 1.7 \pm0.4 $&  $ 52\pm13$&  $-59\pm4$ \\
\hline
\end{tabular}
\end{table*}

\begin{table*}
\contcaption{}
\begin{tabular}{lrrrrlr}
\hline
\hline
\multicolumn{1}{c}{Source}  & 
\multicolumn{1}{c}{$l$} & 
\multicolumn{1}{c}{$b$} &
\multicolumn{1}{c}{$I$} &
\multicolumn{1}{c}{$P$} &
\multicolumn{1}{c}{$\Pi$} &
\multicolumn{1}{c}{PPA} \\
 & 
\multicolumn{1}{c}{$(^\circ)$} & 
\multicolumn{1}{c}{$(^\circ)$} & 
\multicolumn{1}{c}{(mJy)} & 
\multicolumn{1}{c}{(mJy)} & 
\multicolumn{1}{c}{(per cent)} &
\multicolumn{1}{c}{$(^\circ)$} \\
\hline
 J113347$-$360537&  $285.8$&  $ 24.2$&  $ 4.9\pm0.5$&  $ 1.7 \pm0.4 $&  $ 35\pm10$&  $-29\pm5$ \\
 J114041+173546&  $239.5$&  $ 71.1$&  $ 4.8\pm0.4$&  $ 1.6 \pm0.4 $&  $ 33\pm8$&  $-34\pm5$ \\
 J122306+210453&  $254.5$&  $ 81.2$&  $ 6.9\pm0.5$&  $ 2.3 \pm0.5 $&  $ 34\pm8$&  $ 77\pm4$ \\
 J122451+224653&  $247.3$&  $ 82.6$&  $ 5.5\pm0.5$&  $ 2.6 \pm0.5 $&  $ 47\pm11$&  $ 23\pm3$ \\
 J123940+245348{*}& $252.5$&$86.5$&  $20.0\pm0.7$&  $ 7.2 \pm0.4 $&  $ 36\pm2$&  $ 84\pm1$ \\
 J124234$-$274914&  $300.5$&  $ 35.0$&  $ 5.2\pm0.5$&  $ 2.1 \pm0.5 $&  $ 39\pm11$&  $ 34\pm5$ \\
 J125225$-$195823&  $303.2$&  $ 42.9$&  $ 6.5\pm0.5$&  $ 2.2 \pm0.4 $&  $ 34\pm7$&  $ -6\pm4$ \\
 J130519$-$160020&  $307.8$&  $ 46.7$&  $ 7.4\pm0.6$&  $ 2.5 \pm0.7 $&  $ 34\pm9$&  $-87\pm4$ \\
 J130922$-$094732&  $310.3$&  $ 52.8$&  $ 3.9\pm0.4$&  $ 2.1 \pm0.5 $&  $ 54\pm13$&  $-77\pm4$ \\
 J131015+105623&  $319.1$&  $ 73.2$&  $ 9.3\pm0.5$&  $ 2.9 \pm0.6 $&  $ 32\pm6$&  $ 32\pm3$ \\
 J131551+312114&  $ 72.8$&  $ 83.2$&  $ 6.8\pm0.5$&  $ 2.2 \pm0.6 $&  $ 32\pm9$&  $ 62\pm4$ \\
 J132012+513529&  $112.4$&  $ 64.9$&  $ 5.2\pm0.4$&  $ 1.6 \pm0.4 $&  $ 30\pm7$&  $-34\pm5$ \\
 J132145$-$310259&  $310.5$&  $ 31.4$&  $ 6.0\pm0.5$&  $ 2.2 \pm0.6 $&  $ 36\pm10$&  $-45\pm5$ \\
 J133819$-$122952&  $320.4$&  $ 48.8$&  $ 3.0\pm0.5$&  $ 1.8 \pm0.4 $&  $ 58\pm16$&  $ -0\pm5$ \\
 J134054+425559&  $ 93.7$&  $ 71.3$&  $ 7.0\pm0.4$&  $ 2.1 \pm0.4 $&  $ 30\pm5$&  $-47\pm3$ \\
 J134452+421159&  $ 90.5$&  $ 71.4$&  $ 4.8\pm0.4$&  $ 1.4 \pm0.4 $&  $ 30\pm8$&  $-89\pm5$ \\
 J135319+164437&  $  0.2$&  $ 72.3$&  $ 4.8\pm0.4$&  $ 1.9 \pm0.5 $&  $ 39\pm11$&  $-15\pm5$ \\
 J135347$-$314453&  $318.1$&  $ 29.3$&  $ 5.5\pm0.5$&  $ 1.8 \pm0.5 $&  $ 33\pm9$&  $ 49\pm5$ \\
 J135531$-$291930&  $319.3$&  $ 31.5$&  $ 5.1\pm0.5$&  $ 1.8 \pm0.4 $&  $ 36\pm9$&  $-76\pm5$ \\
 J135727+005817&  $336.9$&  $ 59.4$&  $ 4.7\pm0.5$&  $ 2.0 \pm0.5 $&  $ 43\pm12$&  $-10\pm5$ \\
 J140212+143623&  $359.3$&  $ 69.3$&  $ 9.3\pm0.5$&  $ 3.3 \pm0.5 $&  $ 35\pm6$&  $-49\pm3$ \\
 J141913+340713&  $ 58.7$&  $ 69.9$&  $ 9.2\pm0.5$&  $ 2.9 \pm0.6 $&  $ 32\pm6$&  $-61\pm3$ \\
 J143016+200120&  $ 20.8$&  $ 66.3$&  $ 4.9\pm0.4$&  $ 1.6 \pm0.4 $&  $ 32\pm8$&  $-72\pm5$ \\
 J143017+303111&  $ 47.7$&  $ 68.1$&  $ 6.0\pm0.5$&  $ 2.2 \pm0.6 $&  $ 36\pm10$&  $-57\pm4$ \\
 J143511$-$260245&  $330.3$&  $ 31.3$&  $ 6.2\pm0.5$&  $ 2.2 \pm0.6 $&  $ 36\pm9$&  $-90\pm4$ \\
 J151129$-$204120&  $341.9$&  $ 31.4$&  $32.6\pm1.1$&  $10.1 \pm0.5 $&  $ 31\pm2$&  $ 22\pm1$ \\
 J153400+563337&  $ 89.9$&  $ 48.8$&  $ 5.4\pm0.4$&  $ 1.9 \pm0.4 $&  $ 34\pm9$&  $-30\pm4$ \\
 J154622+184533&  $ 30.7$&  $ 49.1$&  $ 5.8\pm0.5$&  $ 2.1 \pm0.5 $&  $ 37\pm9$&  $-59\pm5$ \\
 J154826+592312&  $ 92.3$&  $ 45.8$&  $ 6.4\pm0.5$&  $ 2.6 \pm0.6 $&  $ 41\pm10$&  $-18\pm4$ \\
 J155234+274941&  $ 44.8$&  $ 50.1$&  $ 5.6\pm0.4$&  $ 2.0 \pm0.5 $&  $ 36\pm10$&  $ 16\pm5$ \\
 J160047+522049&  $ 81.9$&  $ 46.8$&  $ 6.5\pm0.5$&  $ 2.4 \pm0.7 $&  $ 37\pm10$&  $-48\pm4$ \\
 J161114$-$344454&  $343.0$&  $ 12.2$&  $ 7.8\pm0.5$&  $ 2.4 \pm0.6 $&  $ 31\pm8$&  $-84\pm5$ \\
 J162226+010249&  $ 14.9$&  $ 33.2$&  $ 5.6\pm0.5$&  $ 1.8 \pm0.5 $&  $ 32\pm9$&  $ 81\pm5$ \\
 J162558+565123&  $ 86.5$&  $ 42.1$&  $ 4.2\pm0.5$&  $ 2.2 \pm0.5 $&  $ 53\pm14$&  $  1\pm4$ \\
 J162709$-$275117&  $350.5$&  $ 14.5$&  $ 4.9\pm0.5$&  $ 2.8 \pm0.6 $&  $ 57\pm14$&  $ 88\pm4$ \\
 J163221+155147&  $ 32.4$&  $ 37.8$&  $ 7.5\pm0.5$&  $ 2.7 \pm0.6 $&  $ 37\pm8$&  $-51\pm4$ \\
 J163540$-$390700&  $343.2$&  $  5.6$&  $ 5.9\pm0.5$&  $ 2.1 \pm0.6 $&  $ 35\pm10$&  $-55\pm6$ \\
 J163644+553439&  $ 84.4$&  $ 40.9$&  $ 5.1\pm0.4$&  $ 2.1 \pm0.4 $&  $ 42\pm9$&  $-49\pm4$ \\
 J163957+300118&  $ 50.6$&  $ 40.3$&  $ 5.2\pm0.4$&  $ 1.8 \pm0.4 $&  $ 35\pm8$&  $ 26\pm4$ \\
 J165952$-$220451&  $359.9$&  $ 12.4$&  $ 5.9\pm0.5$&  $ 2.8 \pm0.6 $&  $ 48\pm11$&  $  8\pm4$ \\
 J172333+782907&  $110.5$&  $ 30.9$&  $10.0\pm0.6$&  $ 3.2 \pm0.5 $&  $ 32\pm5$&  $ 32\pm3$ \\
 J175106$-$155339&  $ 11.9$&  $  5.6$&  $ 4.5\pm0.6$&  $ 3.8 \pm0.9 $&  $ 84\pm23$&  $ 40\pm4$ \\
 J175154+761537&  $107.6$&  $ 29.8$&  $ 4.8\pm0.5$&  $ 1.7 \pm0.5 $&  $ 36\pm10$&  $ 81\pm5$ \\
 J175533+375216&  $ 63.9$&  $ 26.7$&  $ 6.9\pm0.5$&  $ 2.4 \pm0.4 $&  $ 35\pm7$&  $-62\pm4$ \\
 J175808+393018&  $ 65.8$&  $ 26.6$&  $ 6.3\pm0.4$&  $ 2.2 \pm0.4 $&  $ 35\pm7$&  $-40\pm4$ \\
 J180523$-$344947&  $357.0$&  $ -6.6$&  $ 7.1\pm0.5$&  $ 2.2 \pm0.6 $&  $ 31\pm9$&  $-83\pm5$ \\
 J180625+483125&  $ 76.0$&  $ 27.2$&  $ 5.3\pm0.4$&  $ 1.7 \pm0.4 $&  $ 32\pm7$&  $ 66\pm5$ \\
 J181818$-$044404&  $ 24.9$&  $  5.2$&  $ 5.9\pm0.5$&  $ 2.7 \pm0.7 $&  $ 45\pm12$&  $-72\pm5$ \\
 J182530$-$093521{*}& $21.4$&$  1.3$&  $27.2\pm0.9$&  $ 9.0 \pm0.7 $&  $ 33\pm3$&  $-39\pm1$ \\
 J182701$-$204237&  $ 11.8$&  $ -4.2$&  $ 5.1\pm0.5$&  $ 2.4 \pm0.5 $&  $ 46\pm12$&  $  6\pm5$ \\
 J184044+564055{*}& $86.1$&$ 23.8$&  $ 8.5\pm0.5$&  $ 2.7 \pm0.6 $&  $ 32\pm8$&  $ 31\pm4$ \\
 J185501$-$050656&  $ 28.8$&  $ -3.1$&  $ 5.5\pm0.5$&  $ 2.5 \pm0.6 $&  $ 45\pm12$&  $-81\pm5$ \\
 J190411+745943&  $106.3$&  $ 25.3$&  $ 8.6\pm0.5$&  $ 3.3 \pm0.7 $&  $ 38\pm8$&  $-12\pm3$ \\
 J190907$-$171033&  $ 19.4$&  $-11.5$&  $ 3.6\pm0.5$&  $ 2.2 \pm0.5 $&  $ 61\pm16$&  $ -9\pm5$ \\
 J193214+105931{*}& $47.4$&$ -3.9$&  $27.0\pm0.9$&  $17.9 \pm0.6 $&  $ 66\pm3$&  $ 46\pm1$ \\
 J193835+025903&  $ 41.0$&  $ -9.1$&  $14.2\pm0.6$&  $ 4.5 \pm0.8 $&  $ 32\pm5$&  $ 52\pm3$ \\
 J202219$-$291546&  $ 13.7$&  $-31.5$&  $ 9.4\pm0.6$&  $ 3.1 \pm0.6 $&  $ 33\pm6$&  $-84\pm3$ \\
 J202249+515449{*}& $87.9$&$  8.4$&  $17.1\pm0.7$&  $ 6.1 \pm0.5 $&  $ 35\pm3$&  $ 11\pm2$ \\
\hline
\end{tabular}
\end{table*}

\begin{table*}
\contcaption{}
\begin{tabular}{lrrrrlr}
\hline
\hline
\multicolumn{1}{c}{Source}  & 
\multicolumn{1}{c}{$l$} & 
\multicolumn{1}{c}{$b$} &
\multicolumn{1}{c}{$I$} &
\multicolumn{1}{c}{$P$} &
\multicolumn{1}{c}{$\Pi$} &
\multicolumn{1}{c}{PPA} \\
 & 
\multicolumn{1}{c}{$(^\circ)$} & 
\multicolumn{1}{c}{$(^\circ)$} & 
\multicolumn{1}{c}{(mJy)} & 
\multicolumn{1}{c}{(mJy)} & 
\multicolumn{1}{c}{(per cent)} &
\multicolumn{1}{c}{$(^\circ)$} \\
\hline
J203451+505746&  $ 88.2$&  $  6.3$&  $ 5.1\pm0.4$&  $ 1.8 \pm0.5 $&  $ 34\pm9$&  $  4\pm5$ \\
J203453+214620&  $ 64.6$&  $-11.0$&  $ 4.6\pm0.5$&  $ 4.1 \pm0.9 $&  $ 89\pm21$&  $  2\pm3$ \\
J205104+582930&  $ 95.6$&  $  9.0$&  $ 5.4\pm0.4$&  $ 1.8 \pm0.5 $&  $ 34\pm9$&  $-17\pm5$ \\
J210031$-$084309&  $ 40.1$&  $-32.6$&  $ 5.8\pm0.4$&  $ 1.8 \pm0.4 $&  $ 30\pm8$&  $ 20\pm5$ \\
J221157+074408&  $ 69.2$&  $-38.0$&  $ 8.3\pm0.5$&  $ 2.5 \pm0.5 $&  $ 31\pm6$&  $  2\pm4$ \\
J223609+043049&  $ 71.8$&  $-44.5$&  $ 7.7\pm0.5$&  $ 2.8 \pm0.8 $&  $ 37\pm10$&  $ 44\pm4$ \\
J224132+165029&  $ 83.6$&  $-35.9$&  $ 3.4\pm0.4$&  $ 1.9 \pm0.4 $&  $ 55\pm13$&  $-32\pm4$ \\
J224206$-$014017&  $ 66.8$&  $-49.9$&  $ 4.8\pm0.4$&  $ 1.6 \pm0.4 $&  $ 33\pm9$&  $ 22\pm5$ \\
J230348$-$365252&  $  2.9$&  $-65.3$&  $ 6.7\pm0.5$&  $ 2.0 \pm0.5 $&  $ 30\pm8$&  $-10\pm5$ \\
J231707+084904&  $ 87.2$&  $-47.4$&  $ 6.9\pm0.5$&  $ 2.5 \pm0.7 $&  $ 36\pm10$&  $-33\pm4$ \\
J232642$-$193709&  $ 49.6$&  $-69.2$&  $ 5.0\pm0.5$&  $ 1.6 \pm0.4 $&  $ 33\pm9$&  $ 53\pm5$ \\
J233825+424248&  $108.9$&  $-18.2$&  $ 9.3\pm0.5$&  $ 2.9 \pm0.5 $&  $ 32\pm6$&  $ 62\pm3$ \\
J233902+303143&  $105.0$&  $-29.8$&  $ 3.7\pm0.4$&  $ 1.8 \pm0.4 $&  $ 50\pm12$&  $ 39\pm4$ \\
\hline
\multicolumn{7}{l}{\it Column description:} \\
\multicolumn{7}{l}{{Column (1)}: NVSS source name;} \\
\multicolumn{7}{l}{{Column (2) and (3)}: Galactic coordinates;} \\
\multicolumn{7}{l}{{Column (4)}: Flux density in the NVSS;} \\
\multicolumn{7}{l}{{Column (5)}: Linearly polarised flux density in NVSS;} \\
\multicolumn{7}{l}{{Column (6)}: Percentage polarisation in the NVSS;} \\
\multicolumn{7}{l}{{Column (7)}: Polarisation position angle (N to E) in the NVSS;} \\
\multicolumn{7}{l}{{*}: Known pulsar.}
\end{tabular}
\end{table*}

%% file: table2.tex
\begin{table*}
\caption{High-resolution observations at 1.4\,GHz with ATCA, VLA and
  FIRST survey.}\label{tab:highres}
\begin{tabular}{lccccccccc}
\hline
\hline
\multicolumn{1}{c}{Source} & RA & DEC & Pos.$-$NVSS  & Size & PA &  $I_{\rm peak}$ &  rms &   $I_{\rm int}$ &  $I_{\rm NVSS}$ 
\\
       & {(~h~~m~~~s)} & {(~~~$^\circ$~~~$'$~~~$''$)} & $('')$ & ($'' \times ''$) & ($^\circ$)& (mJy/beam) & (mJy/beam) & (mJy) & (mJy)  \\
\hline
ATCA:\\
J035815$-$303129&  03 58 16.05&   $-$30 31 27.7&    2.3& $< 19\times9$&  ---& 3.9&  0.15&  5.9&   $ 6.6\pm0.5$  \\
J083035$-$312745{**}&  08 30 36.04&   $-$31 27 46.1& 1.3& ~$20\times6$&   20& 4.3&  0.19&  6.3&   $ 6.5\pm0.5$  \\
J094615$-$311406&  09 46 15.64&   $-$31 14 03.1&    4.9& $< 19\times8$&  ---& 5.6&  0.35&  7.5&   $ 7.7\pm0.5$  \\
J161114$-$344454&  16 11 14.43&   $-$34 44 55.2&    0.3& $< 15\times9$&  ---& 3.4&  0.46&  4.4&   $ 7.8\pm0.5$  \\
J180523$-$344947&  18 05 23.54&   $-$34 49 48.8&    1.4& $< 15\times9$&  ---& 7.9&  0.22&  8.2&   $ 7.1\pm0.5$  \\
J230348$-$365252&  23 03 48.19&   $-$36 52 55.2&    2.3& $< 15\times9$&  ---& 2.8&  0.38&  5.8&   $ 6.7\pm0.5$  \\
\\
\hline
VLA:\\
J030212+153752{**}&  03 02 12.95&   +15 37 51.8& 1.2 & $16\times4$&   148&  3.9&  0.05& 11.8&   $13.5\pm0.6$  \\
J043945+500319&  04 39 44.98&   +50 03 19.6&    0.2&  $ 4\times3$&  158.5&  9.7&  0.07& 13.6&   $18.4\pm0.7$  \\
J064226+452259&  06 42 26.75&   +45 22 59.7&    1.4&  $ 8\times1$&   93.5&  7.4&  0.05& 14.9&   $16.4\pm0.6$  \\
J074213+605759&  07 42 13.68&   +60 57 58.6&    0.8&  $ 5\times1$&   82.8&  5.6&  0.05&  8.5&   $12.1\pm0.6$  \\
J100556+655946&  10 05 56.17&   +65 59 45.0&    1.7&  $ 2\times1$&   21.6&  8.7&  0.08&  9.4&   $11.3\pm0.5$  \\
\\
\hline
FIRST:\\
J000953+003956 & 00 09 53.53& +00 40 00.5&          7.7& $ 7\times 6$&  72.7&   1.6&  0.10&   3.5&   $ 6.8\pm0.5$\\
J073904+180421 & 07 39 04.35& +18 04 25.9&          4.6& $ 6\times 0$&  38.3&   4.4&  0.17&   5.8&   $ 4.4\pm0.5$\\
J074213+605759 & 07 42 13.69& +60 57 58.7&          0.7& $ 5\times 0$&  89.6&   6.2&  0.16&   8.4&   $12.1\pm0.6$\\
J080840+212610 &         ---&         ---&          ---&          ---&   ---&   ---&  0.14&    ---&   $ 5.3\pm0.4$\\
J082124+175736 &         ---&         ---&          ---&          ---&   ---&   ---&  0.38&    ---&   $ 3.7\pm0.4$\\
J085930+510912 & 08 59 30.95& +51 09 10.8&          2.6& $ 10\times 1$& 170.7&  1.6&  0.23&   3.3&   $ 5.8\pm0.5$\\
J090436+303246 & 09 04 36.41& +30 32 47.8&          1.3& $ 5\times 0$& 168.2&   3.5&  0.13&   4.3&   $ 5.6\pm0.5$\\
J090522+271728 &         ---&         ---&          ---&          ---&   ---&   ---&  0.15&    ---&   $ 8.0\pm0.5$\\
J090656+151052 & 09 06 57.25& +15 10 53.8&          4.3& $18\times 2$&  55.3&   1.1&  0.15&   4.0&   $ 5.1\pm0.5$\\
J091048+312725 &         ---&         ---&          ---&          ---&   ---&   ---&  0.13&    ---&   $ 5.7\pm0.5$\\
J091958+615934 & 09 19 58.07& +61 59 46.8&         12.3& $ 0\times 0$&   8.1&   3.7&  0.18&   3.1&   $ 4.6\pm0.4$\\
J092214+063824{*}& 09 22 14.01& +06 38 22.8&       5.9& $ 1\times 0$&  150.5&  10.3&  0.14&  10.5&   $ 6.4\pm0.5$\\
J100110$-$015230 & 10 01 10.48& $-$01 52 27.5&     3.3& $10\times 2$&   26.5&   2.1&  0.14&   4.3&   $ 7.2\pm0.5$\\
J103411+161344 & 10 34 11.03& +16 13 43.0&          5.6& $12\times 0$&  33.1&   1.8&  0.15&   3.8&   $ 4.6\pm0.5$\\
J103601+050714 &         ---&         ---&          ---&          ---&   ---&   ---&  0.14&    ---&   $ 7.0\pm0.5$\\
J103620+393526 & 10 36 21.20& +39 35 28.9&          5.7& $ 3\times 3$& 179.7&   3.7&  0.13&   4.6&   $ 5.5\pm0.4$\\
J104425+440545 & 10 44 24.97& +44 05 49.6&          3.9& $10\times 7$&  34.9&   1.1&  0.14&   3.7&   $ 4.5\pm0.4$\\
J110724+260109 & 11 07 24.32& +26 01 11.6&          2.3& $ 8\times 1$&  65.1&   2.7&  0.15&   5.1&   $ 5.2\pm0.4$\\
J111151+271153 & 11 11 51.58& +27 11 55.9&          3.4& $ 6\times 2$&  61.3&   4.5&  0.15&   7.6&   $ 9.2\pm0.5$\\
J112158+232928 &         ---&         ---&          ---&          ---&   ---&   ---&  0.27&    ---&   $ 3.3\pm0.4$\\
J114041+173546 & 11 40 40.96& +17 35 49.9&          3.9& $10\times 0$& 101.1&   1.4&  0.15&   2.4&   $ 4.8\pm0.4$\\
J122306+210453 & 12 23 06.68& +21 04 54.1&          2.9& $ 5\times 0$& 122.4&   2.1&  0.17&   2.6&   $ 6.9\pm0.5$\\
J122451+224653 &         ---&         ---&          ---&          ---&   ---&   ---&  0.14&    ---&   $ 5.5\pm0.5$\\
J123940+245348{*}& 12 39 40.39& +24 53 49.9&       1.7& $ 0\times 0$&  104.6&  11.5&  0.15&  11.2&   $20.0\pm0.7$\\
J131015+105623 & 13 10 15.55& +10 56 17.6&          6.5& $ 5\times 2$&  11.7&   4.2&  0.15&   6.2&   $ 9.3\pm0.5$\\
J131551+312114 & 13 15 52.54& +31 21 14.2&         10.0& $ 6\times 5$&  87.4&   1.4&  0.13&   2.9&   $ 6.8\pm0.5$\\
J132012+513529 & 13 20 11.46& +51 35 27.4&          5.7& $ 6\times 0$&  96.8&   3.3&  0.15&   4.8&   $ 5.2\pm0.4$\\
J134054+425559 & 13 40 54.70& +42 56 01.1&          1.2& $ 2\times 1$& 126.5&   6.5&  0.14&   7.3&   $ 7.0\pm0.4$\\
J134452+421159 & 13 44 52.68& +42 11 59.2&          1.2& $10\times 5$& 144.2&   1.9&  0.13&   5.5&   $ 4.8\pm0.4$\\
J135319+164437 & 13 53 19.37& +16 44 45.3&          7.8& $ 5\times 1$&  82.3&   3.5&  0.15&   4.8&   $ 4.8\pm0.4$\\
J135727+005817 &         ---&         ---&          ---&          ---&   ---&   ---&  0.37&    ---&   $ 4.7\pm0.5$\\
J140212+143623{**} & 14 02 12.29& +14 36 23.2&      2.6& $20 \times 9 $&   131&   1.4&  0.15&  10.1&   $ 9.3\pm0.5$\\
J141913+340713 & 14 19 13.52& +34 07 12.3&          2.3& $12\times 3$&  96.4&   3.2&  0.15&   8.9&   $ 9.2\pm0.5$\\
J143016+200120 & 14 30 17.20& +20 01 19.7&          3.3& $ 1\times 0$& 139.6&   3.5&  0.14&   3.5&   $ 4.9\pm0.4$\\
J143017+303111 &         ---&         ---&          ---&          ---&   ---&   ---&  0.15&    ---&   $ 6.0\pm0.5$\\
J153400+563337 & 15 34 00.74& +56 33 40.2&          3.0& $17\times 3$&  12.8&   1.7&  0.15&   6.3&   $ 5.4\pm0.4$\\
J154622+184533 & 15 46 22.89& +18 45 33.0&          1.2& $ 4\times 2$&  20.8&   2.1&  0.15&   2.9&   $ 5.8\pm0.5$\\
J154826+592312 & 15 48 26.61& +59 23 14.4&          2.8& $ 5\times 0$&   5.4&   2.7&  0.14&   3.7&   $ 6.4\pm0.5$\\
J155234+274941 & 15 52 33.98& +27 49 38.0&          5.2& $ 8\times 0$& 158.4&   1.6&  0.14&   2.1&   $ 5.6\pm0.4$\\
J160047+522049 & 16 00 47.66& +52 20 56.3&          8.8& $ 4\times 3$&  34.2&   4.1&  0.14&   5.9&   $ 6.5\pm0.5$\\
J162558+565123 & 16 25 59.14& +56 51 22.5&          3.3& $ 6\times 0$&   0.8&   2.0&  0.16&   3.0&   $ 4.2\pm0.5$\\
\hline
\end{tabular}
\end{table*}

\begin{table*}
\contcaption{}
\begin{tabular}{lccccccccc}
\hline
\hline
\multicolumn{1}{c}{Source} & RA & DEC & Pos.$-$NVSS  & Size & PA &  $I_{\rm peak}$ &  rms &   $I_{\rm int}$ &  $I_{\rm NVSS}$  \\
       & {(~h~~m~~~s)} & {(~~~$^\circ$~~~$'$~~~$''$)} & $('')$ & ($'' \times ''$) & ($^\circ$)& (mJy/beam) & (mJy/beam) & (mJy) &  (mJy)  \\
\hline
J163221+155147 & 16 32 21.44& +15 51 45.7&          1.7& $ 2\times 0$& 131.0&   5.7&  0.15&   5.9&   $ 7.5\pm0.5$\\
J163644+553439 & 16 36 44.00& +55 34 35.8&          5.4& $ 2\times 0$&  31.6&   4.7&  0.15&   4.8&   $ 5.1\pm0.4$\\
J163957+300118{**} & 16 39 57.84& +30 01 16.1&      3.2& $23 \times 3$&    145&   1.6&  0.14&   5.6&   $ 5.2\pm0.4$\\
J224206$-$014017 & 22 42 06.42& $-$01 40 11.8&      6.7& $ 6\times 0$&   8.1&   2.6&  0.13&   3.5&   $ 4.8\pm0.4$\\
\hline
\multicolumn{10}{l}{\it Column description:}\\
\multicolumn{10}{l}{{Column (1)}: NVSS source name;} \\
\multicolumn{10}{l}{{Column (2) and (3)}: J2000 coordinates from 
high-resolution radio observations (ATCA, VLA or FIRST);} \\
\multicolumn{10}{l}{{Column (4)}: Position offset between NVSS position
and the peak position from high-resolution radio images; }\\
\multicolumn{10}{l}{{Column (5) and (6)}: Deconvolved size and position angle
  (N to E) of sources (for FIRST observations, the deconvolved sizes
  are from } \\ 
\multicolumn{10}{l}{~~~~~~~~~~~~~~~~~~~~~~~~~~~ the FIRST catalogue and
  a source is unresolved if the deconvolved size is $<2^{''}$);}\\
\multicolumn{10}{l}{{Column (7),(8) and (9)}: Peak flux density, rms noise and 
integrated flux density in the high-resolution radio observations;}\\
\multicolumn{10}{l}{{Column (10)}: Flux density in the NVSS;}\\
\multicolumn{10}{l}{`---' in columns (2)--(7) and (9): the source is not detected in the FIRST survey; }\\
\multicolumn{10}{l}{{*}: Known pulsar; }\\
\multicolumn{10}{l}{{**}: Clearly resolved into two sources in radio
  high-resolution observations (the listed coordinate is an average of the two peak positions).}\\ 
\end{tabular}
\end{table*}

%% file: table3.tex
\begin{table*}
\caption{Linear polarisation measurements from high-resolution observations with the ATCA and VLA.}\label{tab:obs_nvss}
\begin{tabular}{lrrrrrrrr}
\hline
\hline
\multicolumn{1}{c}{Source} &  $\rm \Pi_{\rm NVSS}$ & PPA$_{\rm NVSS}$ & $\Pi_{\rm 1.4}$ & PPA$_{\rm 1.4}$ & $\Pi_{\rm 2.5}$ & PPA$_{\rm 2.5}$ & $\Delta$PPA & $\Delta$PPA$_{G}$ \\
       & (per cent) & $(^{\circ})$ & (per cent) &  $(^{\circ})$ & (per cent) & $(^{\circ})$  & $(^{\circ})$ &  $(^{\circ})$ \\
\hline
ATCA:\\
J035815$-$303129 &  $32\pm7$&  $-41\pm4$& $40\pm12$&  $-55\pm7$&   $47\pm15$& $-70\pm9$ & 15 & 25 \\
J083035$-$312745{**} & $33\pm7$&  $53\pm4$&   $34\pm9$&  $41\pm6$&    $51\pm20$&  $36\pm9$ & 5 & 109 \\
J094615$-$311406&  $32\pm6$&  $-66\pm4$& $27\pm7$&  $-67\pm6$&  $25\pm7$&  $86\pm7$ & 27 & 59   \\
J161114$-$344454&  $31\pm8$&  $96\pm5$&  $19\pm7$&  $74\pm9$&    $37\pm14$&  $-64\pm9$ & 42 & $108$\\
J180523$-$344947&  $31\pm9$&  $97\pm5$&  $8\pm3$&  $58\pm10$&    $16\pm6$&   $-4\pm10$ & 62 & 118  \\
J230348$-$365252&  $30\pm8$&  $-10\pm5$&  $27\pm10$&  $-24\pm8$&   $<14$&  --- & --- & ---   \\
\hline
VLA:\\
J030212+153752{**}& $31\pm3$&  $-60\pm2$&  $40\pm7$&  $-59\pm4$&  &  & &  \\
J043945+500319&  $35\pm3$&  $37\pm1$&    $30\pm5$&  $-22\pm7$&  &  & &  \\
J064226+452259&  $31\pm3$&  $-54\pm2$&   $31\pm4$&  $-57\pm3$&  &  &  & \\
J074213+605759&  $30\pm4$&  $55\pm2$&   $49\pm8$&  $61\pm5$&  &  & &  \\
J100556+655946&  $33\pm4$&  $-45\pm2$&   $44\pm6$&  $-47\pm2$&  &  &  & \\
\hline
\multicolumn{9}{l}{\it Column description:} \\
\multicolumn{9}{l}{{Column (1)}: NVSS source name;} \\
\multicolumn{9}{l}{{Column (2) and (3)}: Percentage linear polarisation and polarisation position angle in the NVSS;} \\
\multicolumn{9}{l}{{Column (4) and (5)}: Percentage linear polarisation and polarisation position angle in high-resolution }\\ 
\multicolumn{9}{l}{~~~~~~~~~~~~~~~~~~~~~~~~~~~~observations at 1.4 GHz with ATCA and VLA;} \\
\multicolumn{9}{l}{{Column (6) and (7)}: Percentage linear polarisation and polarisation position angle in high-resolution}\\ 
\multicolumn{9}{l}{~~~~~~~~~~~~~~~~~~~~~~~~~~~~observations at 2.5 GHz with the ATCA;} \\
\multicolumn{9}{l}{{Column (8)}: Apparent rotation of the polarisation position angle between 1.4 GHz and 2.5 GHz;} \\
\multicolumn{9}{l}{{Column (9)}: Rotation of the polarisation position angle between 1.4 GHz and 2.5 GHz expected }\\
\multicolumn{9}{l}{~~~~~~~~~~~~~~~~~from the Galactic rotation measure given in Simard-Normandin, Kronberg \& Button (1981);} \\
\multicolumn{9}{l}{{**}: Clearly resolved into two sources in radio
  high-resolution observations.}\\
\end{tabular}
\end{table*}

%% file: table4.tex
\begin{table*}
\caption{Optical identifications with the SDSS.}\label{tab:SDSS}
\begin{tabular}{lccccccccc}
\hline
\hline
\multicolumn{1}{c}{Source} & RA & DEC  & radio$-$SDSS  & $g$  & $i$ & radio$-$2MASS  & $K$ & redshift & type \\
       & (~h~~m~~~s) & (~~~$^\circ$~~~$'$~~~$''$) & ($''$) &  &  & ($''$) &  &  (photometric redshift)&\\
\hline
J073904+180421& 07 39 04.35& +18 04 25.5&     0.4&    15.20& 13.90&   0.5&      12.72&  $(0.08\pm0.01)$&  \\
J090436+303246& 09 04 36.39& +30 32 48.6&     0.9&    19.16& 17.13&   1.1&      14.74&  $0.2819\pm0.0002$& E\\
J090656+151052& 09 06 57.10& +15 10 52.7&     2.5&    24.93& 20.93&    ---&         ---&  $(0.97\pm0.06)$&   \\
J103411+161344& 10 34 11.10& +16 13 44.6&     1.9&    20.63& 18.20&    ---&         ---&  $0.3808\pm0.0002$& E\\
J103620+393526& 10 36 21.17& +39 35 28.6&     0.4&    24.50& 22.25&    ---&         ---&  $(0.56\pm0.13)$&   \\
J104425+440545& 10 44 24.82& +44 05 47.8&     2.4&    19.93& 18.02&    2.5&      15.46&  $(0.24\pm0.02)$&   \\
J110724+260109& 11 07 24.31& +26 01 11.8&     0.2&    22.29& 19.34&     ---&         ---&  $(0.71\pm0.04)$&  \\
J111151+271153& 11 11 51.56& +27 11 55.9&     0.2&    15.15& 13.92&    0.2&      12.38&  $0.0471\pm0.0002$&   E\\
J114041+173546& 11 40 40.90& +17 35 50.2&     0.9&    20.94& 18.44&    0.9&      15.84&  $0.3857\pm0.0002$&   E\\
J122306+210453& 12 23 06.64& +21 04 53.8&     0.6&    18.92& 17.03&    0.6&      14.66&  $0.2369\pm0.0002$&  E\\
J131015+105623& 13 10 15.54& +10 56 17.2&     0.4&    16.58& 15.17&    0.3&      13.56&  $0.1025\pm0.0002$&   E\\
J132012+513529& 13 20 11.49& +51 35 27.1&     0.4&    19.38& 17.22&    0.5&      14.73&  $0.2773\pm0.0002$&   E\\
J134054+425559& 13 40 54.71& +42 56 00.9&     0.3&    24.19& 21.88&     ---&          ---&  $(0.64\pm0.07)$&   \\
J140212+143623& 14 02 12.10& +14 36 25.4&     0.2\rlap{**} &    23.49& 20.41&   ---&  ---&  $(0.57\pm0.03)$&   \\
J141913+340713& 14 19 13.44& +34 07 12.7&     1.0&    22.34& 19.11&     ---&          ---&  $(0.72\pm0.04)$&   \\
J153400+563337& 15 34 00.64& +56 33 36.0&     0.1\rlap{**} &  21.15&  18.57&    ---&   ---&  $0.4628\pm0.0003$&   E\\
J154622+184533& 15 46 22.90& +18 45 32.3&     0.7&    16.68& 15.26&    0.6&      13.46&  $0.1069\pm0.0002$&   E\\
J154826+592312& 15 48 26.59& +59 23 14.6&     0.3&    18.69& 16.86&    0.4&      14.61&  $(0.21\pm0.01)$&   \\
J155234+274941& 15 52 33.98& +27 49 37.7&     0.3&    20.29& 17.85&    0.5&      15.21&  $0.3313\pm0.0002$&   E\\
J162558+565123& 16 25 59.13& +56 51 23.4&     0.9&    18.08& 16.51&    0.9&      14.28&  $(0.15\pm0.01)$&   \\
J163221+155147& 16 32 21.38& +15 51 45.5&     0.8&    18.30& 17.15&    0.3&      14.30&  $0.2418\pm0.0001$&   L\\
J163957+300118& 16 39 57.94& +30 01 14.2&     0.6\rlap{**} & 20.70& 18.35& 0.3\rlap{**} &  15.44& $(0.34\pm0.02)$&   \\
\hline
\multicolumn{10}{l}{\it Column description:} \\
\multicolumn{10}{l}{{Column (1)}: NVSS source name; } \\
\multicolumn{10}{l}{{Column (2) and (3)}: J2000 coordinates of optical identification in the SDSS;} \\
\multicolumn{10}{l}{{Column (4)}: Position offset between the FIRST survey and SDSS; } \\
\multicolumn{10}{l}{{Column (5)}: {\it g}-band magnitude from the SDSS corrected for extinction; } \\
\multicolumn{10}{l}{{Column (6)}: {\it i}-band magnitude from the SDSS corrected for extinction; } \\
\multicolumn{10}{l}{{Column (7)}: Position offset between the FIRST survey and 2MASS; } \\
\multicolumn{10}{l}{{Column (8)}: {\it K}-band magnitude from the 2MASS;} \\
\multicolumn{10}{l}{{Column (9)}: Redshift from the SDSS (photometric redshifts are given in brackets); } \\
\multicolumn{10}{l}{{Column (10)}: Type of identifications (`E' for elliptical galaxy and `L' for LINER); } \\
\multicolumn{10}{l}{'---' in columns (7) and (8): not detected in the 2MASS ;}\\
\multicolumn{10}{l}{{**}: Distance from optical identification to central axis of the radio source.} \\
\end{tabular}
\end{table*}

%% file: table5.tex
\begin{table*}
\caption{Optical identifications with SuperCOSMOS.}\label{tab:cosmos}
\begin{tabular}{lccccccccc}
\hline
\hline
\multicolumn{1}{c}{Source} & RA & DEC  & radio$-$SSA  & $B_J$ &$I_N$ & radio$-$2MASS  & $K$ & redshift & 
type \\
       & (~h~~m~~~s) & (~~~$^\circ$~~~$'$~~~$''$) & ($''$) &  &  & ($''$) &  &  &\\
\hline
J030212+153752& 03 02 12.98&   +15 37 51.5&    0.3\rlap{**} & 22.3$\sharp$&   19.53&  0.4\rlap{$\dagger$**} &  
15.5$\sharp$&    
--- &  \\
J035815$-$303129& 03 58 16.01& $-$30 31 28.2&    0.7&    18.26&     16.79&      0.3&    14.35&    0.183 & E      \\
J064226+452259& 06 42 26.73&   +45 22 59.8&    0.2&    17.49&     15.65&      0.9&    13.53&        ---   &     \\
J074213+605759& 07 42 13.73&   +60 57 58.7&    0.4&    17.58&     15.77&      0.1&    13.62&        ---   &     \\
J083035$-$312745& 08 30 36.12& $-$31 27 44.8&    0.2\rlap{$\ddagger$**} & 20.7$\sharp$& 18.5$\sharp$ &      --- &     --- &          
--- &  \\
J094615$-$311406& 09 46 15.66& $-$31 14 03.7&    0.7&    16.10&     14.49&      0.9&    12.89&    0.0565&    E \\
J230348$-$365252& 23 03 48.13& $-$36 52 53.4&    0.7&    19.88&     17.83&      1.9&    14.97&         ---   &      
\\
\hline
\multicolumn{10}{l}{\it Column description: } \\
\multicolumn{10}{l}{{Column (1)}: NVSS source name; } \\
\multicolumn{10}{l}{{Column (2) and (3)}: J2000 coordinates from SuperCOSMOS Sky Archive (SSA);} \\
\multicolumn{10}{l}{{Column (4)}: Position offset between radio high-resolution observations (ATCA or VLA) and the SSA; } \\
\multicolumn{10}{l}{{Column (5)}: $B_J$-band magnitude from the SSA;} \\ 
\multicolumn{10}{l}{{Column (6)}: $I_N$-band magnitude from the SSA;} \\
\multicolumn{10}{l}{{Column (7)}: Position offset between radio high-resolution
observations (ATCA or VLA) and the 2MASS; } \\
\multicolumn{10}{l}{{Column (8)}: $K$-band magnitude from the 2MASS;} \\
\multicolumn{10}{l}{{Column (9)}: Spectroscopic redshift; } \\
\multicolumn{10}{l}{{Column (10)}: Type of identification (`E' for elliptical galaxy); } \\
\multicolumn{10}{l}{'---' in columns (5),(6),(7),(8) and (9): not detected in the SSA and 2MASS or no spectroscopic redshift available;}\\
\multicolumn{10}{l}{$^{**}$: Distance from optical identification to the central axis of the radio source;} \\
\multicolumn{10}{l}{{$\dagger$}: Not in the 2MASS catalogue; position estimated from the peak pixel in the $K$-band image;} \\  
\multicolumn{10}{l}{{$\ddagger$}: Not in the SSA catalogue; optical position estimated from the $B_J$-band image;}\\
\multicolumn{10}{l}{{$\sharp$}: Approximate magnitude estimated by comparing with catalogued objects with similar brightness. } \\  
\end{tabular}
\end{table*}

%% file: table6.tex
\begin{table}
\setlength{\tabcolsep}{1.8mm}
\caption{Comparison of galaxy counts between low-polarisation and
  ultrahigh polarisation samples.}\label{tab:counts}
\begin{tabular}{cccccccc}
\hline
\hline
\multicolumn{1}{c}{radius} & \multicolumn{3}{c}{low polarisation sample}& 
\multicolumn{3}{c}{high polarisation sample}  & \multicolumn{1}{c}{D} \\
\multicolumn{1}{c}{(Mpc)}&\multicolumn{1}{c}{mean} & 
\multicolumn{1}{c}{median}  & \multicolumn{1}{c}{sigma}  & 
\multicolumn{1}{c}{mean} & \multicolumn{1}{c}{median}  & 
\multicolumn{1}{c}{sigma}  & \multicolumn{1}{c}{(KS test)} \\
\hline
1.0&   11.4&    9.0&      6.9&     7.9&    7.0&      2.6&  0.36 \\
0.5&   ~5.4&    4.0&      3.8&     4.0&    3.0&      1.9&  0.25 \\
0.2&   ~2.2&    2.0&      1.5&     1.4&    1.0&      0.5&  0.31 \\
\hline
\multicolumn{8}{l}{\it Column description:} \\
\multicolumn{8}{l}{{Column (1)}: Search radius for galaxy counts;} \\
\multicolumn{8}{l}{{Column (2),(3) and (4)}: mean, median and dispersion of galaxy count distribution}\\
\multicolumn{8}{l}{~~~~~ for the 121 low-polarisation sources;} \\
\multicolumn{8}{l}{{Column (5),(6) and (7)}: mean, median and dispersion of galaxy count distribution}\\ 
\multicolumn{8}{l}{~~~~~ for the eight ultrahigh polarisation sources;} \\
\multicolumn{8}{l}{{Column (8)}: Kolmogorov--Smirnov test statistic D, giving the maximum distance}\\
\multicolumn{8}{l}{~~~~~ between the normalised cumulative distributions of the two samples;} \\
\multicolumn{8}{l}{~~~~~ if D$>0.445$, then the null hypothesis of the two distribution being the}\\
\multicolumn{8}{l}{~~~~~ same can be rejected at the 10 per cent level.}\\
\end{tabular}
\end{table}

%% file: ms.bbl
\begin{thebibliography}{99}

\bibitem[Abdalla et al.(2008)]{2008arXiv0812.3831A} Abdalla F.~B.,
  Banerji M., Lahav O.,  Rashkov V.\ 2008, preprint (arXiv:0812.3831)

\bibitem[Becker et al.(1995)]{1995ApJ...450..559B} Becker R.~H.,
  White R.~L., Helfand D.~J.\ 1995, ApJ, 450, 559

\bibitem[Bernet et al.(2008)]{2008Natur.454..302B} Bernet M.~L.,
  Miniati F., Lilly S.~J., Kronberg P.~P., Dessauges-Zavadsky,
  M.\ 2008, Nature, 454, 302

\bibitem[Best et al.(2005)]{2005MNRAS.362....9B} Best P.~N.,
  Kauffmann G., Heckman T.~M., Ivezi{\'c}, {\v Z}.\ 2005, MNRAS,
  362, 9

\bibitem[Bock et al.(1999)]{1999AJ....117.1578B} Bock D.~C.-J.,
  Large M.~I., Sadler E.~M.\ 1999, AJ, 117, 1578

\bibitem[Burn (1966)]{1966MNRAS.133,67-83} Burn B.~J.\ 1966, MNRAS,
  133, 67-83

\bibitem[Condon et al.(1998)]{1998AJ....115.1693C} Condon J.~J.,
  Cotton W.~D., Greisen E.~W., Yin Q.~F., Perley R.~A., Taylor
  G.~B., Broderick J.~J.\ 1998, AJ, 115, 1693

\bibitem[Crawford et al.(2000)]{2000AJ....119.2376C} Crawford F.,
  Kaspi V.~M., Bell J.~F.\ 2000, AJ, 119, 2376

\bibitem[Darling \& Giovanelli(2000)]{2000AJ....119.3003D} Darling
  J., Giovanelli R.\ 2000, AJ, 119, 3003

\bibitem[Fletcher et al. (2004)]{2004A&A...414,53F} Fletcher A.,
  Berkhuijsen E.~M., Beck R., Shukurov A.\ 2004, A\&A, 414, 53

\bibitem[Fukugita et al. (1996)]{1996AJ....111,1748F} Fukugita M.,
  Ichikawa T., Gunn J.~E., Doi, M., Shimasaku K., Schneider
  D.~P.\ 1996, AJ, 111, 1748

\bibitem[Goodlet \& Kaiser(2005)]{2005MNRAS.359.1456G} Goodlet J.~A.,
  Kaiser C.~R.\ 2005, MNRAS, 359, 1456

\bibitem[Han et al.(2004)]{2004IAUS..218..135H} Han J.~L., Manchester
  R.~N., Lyne A.~G., Qiao G.~J.\ 2004, in Camilo F., Gaensler B.~M.,
  eds, IAU Symp. 218, Young Neutron Stars and Their
  Environments, p.135

\bibitem[Han \& Tian(1999)]{1999A&AS..136..571H} Han J.~L., Tian,
  W.~W.\ 1999, A\&AS, 136, 571

\bibitem[Kimball \& Ivezi{\'c}(2008)]{2008AJ....136..684K} Kimball
  A.~E., Ivezi{\'c} {\v Z}.\ 2008, AJ, 136, 684

\bibitem[Kronberg(1994)]{1994RPPh...57..325K} Kronberg P.~P.\ 1994,
   Rep. Prog. Phys., 57, 325

\bibitem[Kulsrud \& Zweibel( 2008)]{2008RPPH...71..046901} Kulsrud R.~M., Zwei
bel E.~G.\ 2008, Rep. Prog. Phys., 71, 046901

\bibitem[Laing et al.(1981)]{1981ApJ...248..87L}Laing R.~A.\ 1981, ApJ, 248, 87

\bibitem[Ledlow et al. (2003)]{2003AJ....126..2740} Ledlow, M.~J.,
  Voges, W., Owen, F.~N., Burns, J.~O.\ 2003, AJ, 126, 2740

\bibitem[Liang et al.(2000)]{2000ApJ...544..686L} Liang H., Hunstead
  R.~W., Birkinshaw M., Andreani P.\ 2000, ApJ, 544, 686

\bibitem[Liang et al.(2001)]{2001MNRAS.328L..21L} Liang H., Ekers
  R.~D., Hunstead R.~W., Falco E.~E., Shaver P.\ 2001, MNRAS,
  328, L21

\bibitem[Lupton et al.(2001)]{2001ASPC..238..269L} Lupton R., Gunn
  J.~E., Ivezi{\'c} Z., Knapp G.~R., Kent S.\ 2001, in Harnden F.~R.,
  Jr., Primini F.~A., Payne H.~E., eds, ASP Conf. Ser. Vol.238,
  Astronomical Data Analysis Software and Systems X,
  Astron. Soc. Pac., San Francisco, p.269

\bibitem[Mauch et al.(2003)]{2003MNRAS.342.1117M} Mauch T., Murphy
  T., Buttery H.~J., Curran J., Hunstead R.~W., Piestrzynski B.,
  Robertson J.~G., Sadler E.~M.\ 2003, MNRAS, 342, 1117

\bibitem[Mauch \& Sadler(2007)]{2007MNRAS.375..931M} Mauch T., Sadler
  E.~M.\ 2007, MNRAS, 375, 931

\bibitem[Mulchaey \& Zabludoff(1998)]{1998ApJ..496..73} Mulchaey,
  J.~S, Zabludoff, A.~I.\ 1998, ApJ, 496, 73

\bibitem[Murphy et al.(2007)]{2007MNRAS.382..382M} Murphy T., Mauch
  T., Green A., Hunstead R.~W., Piestrzynska B., Kels A.~P.,
  Sztajer P.\ 2007, MNRAS, 382, 382

\bibitem[Rengelink et al.(1997)]{1997A&AS..124..259R} Rengelink
  R.~B., Tang Y., de Bruyn A.~G., Miley G.~K., Bremer M.~N.,
  Roettgering H.~J.~A., \& Bremer M.~A.~R.\ 1997, A\&AS, 124, 259

\bibitem[Ruzmaikin et al.(1988)]{1989Natur.336..341R} Ruzmaikin A.,
 Sokoloff D.,  Shukurov A., \ 1988, Nature, 336, 341

\bibitem[Sadler et al.(2002)]{2002MNRAS.329..227S} Sadler E.~M., et
  al.\ 2002, MNRAS, 329, 227

\bibitem[Saikia \& Salter (1988)]{1988ARA&A..26...93S} Saikia D.~J.,
  \& Salter C.~J.\ 1988, Ann. Rev. Astr. Ap., 26, 93

\bibitem[Simard-Normandin et al. (1981)]{1981ApJS..45...97S}
  Simard-Normandin M.,Kronberg P.~P., Button S.\ 1981, ApJS, 45, 97

\bibitem[Sokoloff et al. (1998)]{1998MNRAS.299..189S} Sokoloff D.~D.,
  Bykov A.~A., Shukurov A., Berkhuijsen E.~M., Beck R., Poezd
  A.~D.\ 1998, MNRAS, 299, 189

\bibitem[Taylor et al. (2007)]{2007ApJ...666..201T} Taylor A.~R.,Stil
  J.~M., Grant J.~K., Landecker T.~L., Kothes R., Reid R.~I., Gray
  A.~D., Scott D., Martin P.~G. et al.\ 2007, ApJ, 666, 201

\bibitem[Wen et al.(2009)]{2009ApJS..183..197W} Wen Z.~L., Han
  J.~L., Liu F.~S.\ 2009, ApJS, 183, 197

\bibitem[Wolfe et al.(2008)]{2008Natur.455..638W} Wolfe A.~M.,
  Jorgenson R.~A., Robishaw T., Heiles C., Prochaska
  J.~X.\ 2008, Nature, 455, 638

\bibitem[York et al.(2000)]{2000AJ....120.1579Y} York D.~G., et
  al.\ 2000, AJ, 120, 1579

\end{thebibliography}
